\documentclass[11pt]{article}
\usepackage[english]{babel}
\usepackage[protrusion=true,expansion=true]{microtype}
\usepackage[hang, small,labelfont=bf,up,textfont=it,up]{caption}
\usepackage{booktabs}
\usepackage{mathrsfs}
\usepackage{lastpage}
\usepackage{float}
\usepackage{varioref}

\usepackage[colorlinks]{hyperref}
\usepackage{amsmath,amsfonts,amsthm}
\usepackage[svgnames]{xcolor}
\usepackage{subcaption}
\expandafter\def\csname ver@subfig.sty\endcsname{}
\usepackage{mathtools, amssymb}
\usepackage[margin=1in]{geometry} 
\usepackage{color, colortbl}
\usepackage{multirow}
\usepackage{graphics}
\usepackage{graphicx}
\usepackage{setspace}
\usepackage{tikz,titlesec}
\usepackage{ulem}
\usepackage{cite}

\theoremstyle{plain}

\providecommand{\Ro}{\mathcal{R}_0}

\newcommand{\beginsupplement}{%
        \setcounter{table}{0}
        \renewcommand{\thetable}{S\arabic{table}}%
        \setcounter{figure}{0}
        \renewcommand{\thefigure}{S\arabic{figure}}%
}

\newcommand{\captionfonts}{\small}
\makeatletter  
\long\def\@makecaption#1#2{%
  \vskip\abovecaptionskip
  \sbox\@tempboxa{{\captionfonts #1: #2}}%
  \ifdim \wd\@tempboxa >\hsize
    {\captionfonts #1: #2\par}
  \else
    \hbox to\hsize{\hfil\box\@tempboxa\hfil}%
  \fi
  \vskip\belowcaptionskip}
\makeatother   

\newcommand{\figpath}{figures}
\newcommand{\fitfigs}{figures/modelcomparisonplots}
\newcommand{\forecastingfigs}{figures/forecastingplots}
\newcommand{\paramfigs}{figures/paramestplots}

\newcommand{\comment}[1]{}

\begin{document}
\title{\bfseries Model distinguishability and inference robustness in mechanisms of cholera transmission and loss of immunity}
\author{Elizabeth C. Lee,$^{1}$ Michael R. Kelly, Jr.,$^{2}$ Brad M. Ochocki,$^{3}$ Segun M. Akinwumi,$^{4}$ \\
Karen E. S. Hamre,$^{5}$ Joseph H. Tien,$^{2}$ and Marisa C. Eisenberg$^{6}$\thanks{Corresponding author. Email addresses: ecl48@georgetown.edu, kelly.1156@osu.edu, brad.ochocki@rice.edu, segunmic@ualberta.ca, hamr0091@umn.edu, jtien@math.ohio-state.edu, marisae@umich.edu.}}
\maketitle

\vspace{-1cm}
\begin{center}
{\tiny\noindent$^{1}$Department of Biology, Georgetown University, 37th and O Streets, NW, Washington, DC 20057, United States of America\\
$^{2}$Department of Mathematics, The Ohio State University, 231 West 18th Ave, Columbus, OH 43210, United States of America\\
$^{3}$Department of BioSciences, Program in Ecology and Evolutionary Biology, Rice University, 6100 Main Street, MS-170, Houston, TX 77005-1892, United States of America\\
$^{4}$Department of Mathematical and Statistical Sciences, University of Alberta, Edmonton, AB, T6G 2G1, Canada\\
$^{5}$Division of Global Pediatrics and Division of Epidemiology and Community Health, University of Minnesota, 717 Delaware Street SE, 3rd Floor, Minneapolis, MN 55414, United States of America\\
$^{6}$Departments of Epidemiology and Mathematics, University of Michigan, Ann Arbor, 1415 Washington Heights, Ann Arbor, 48109, USA\\}
\end{center}

\noindent \textbf{Abstract}\\ 
{\small Mathematical models of cholera and waterborne disease vary widely in their structures, in terms of transmission pathways, loss of immunity, and a range of other features. These differences can affect model dynamics, with different models potentially yielding different predictions and parameter estimates from the same data. Given the increasing use of mathematical models to inform public health decision-making, it is important to assess model distinguishability (whether models can be distinguished based on fit to data) and inference robustness (whether inferences from the model are robust to realistic variations in model structure).

In this paper, we examined the effects of uncertainty in model structure in the context of epidemic cholera, testing a range of models with differences in transmission and loss of immunity structure, based on known features of cholera epidemiology. We fit these models to simulated epidemic and long-term data, as well as data from the 2006 Angola epidemic. We evaluated model distinguishability based on fit to data, and whether the parameter values, model behavior, and forecasting ability can accurately be inferred from incidence data. 

In general, all models were able to successfully fit to all data sets, both real and simulated, regardless of whether the model generating the simulated data matched the fitted model. However, in the long-term data, the best model fits were achieved when the loss of immunity structures matched those of the model that simulated the data. Two transmission and reporting parameters were accurately estimated across all models, while the remaining parameters showed broad variation across the different models and data sets. Forecasting efforts were not successful early in the outbreaks, but once the epidemic peak had been achieved, most models were able to capture the downward incidence trajectory with similar accuracy.

Our results suggest that we are unlikely to be able to infer mechanistic details from epidemic case data alone, underscoring the need for broader data collection, such as immunity/serology status, pathogen dose response curves, and environmental pathogen data. Nonetheless, with sufficient data, conclusions from forecasting and some parameter estimates were robust to variations in the model structure, and comparative modeling can help to determine how realistic variations in model structure may affect the conclusions drawn from models and data.
}

\section{Introduction}
Cholera is a waterborne disease caused by the bacterium \textit{Vibrio cholerae}, which manifests as severe diarrhea and vomiting leading to dehydration. Left untreated, cholera can be up to $50\%$ fatal, but rehydration treatment can greatly reduce case fatality rates to as low as 1\% \cite{sack2004, factsheet}. Worldwide, cholera causes three to five million cases and over 100,000 deaths per year \cite{WHOfactsheet2010}. Numerous mathematical models of cholera transmission have been proposed to investigate factors that impact the dynamics and transmission of waterborne diseases \cite{codeco2001, hartley2006, king2008, Joh2009, tien2010, mwasa2011mathematical, shuai2012, tian2011global, AndrewsBasu2011, Sanches2011}, and the ongoing cholera epidemic in Haiti has spurred additional interest in the subject \cite{tuite2011, Chao2011, Abrams2012, AndrewsBasu2011, bertuzzo2011,Eisenberg2013b}.

Due to the range of indirect transmission pathways and timescales, which may be represented by environmental water sources, household water containers, foodborne transmission, and more \cite{swerdlow1992, Eisenberg2013}, commonly used mathematical models for cholera vary widely in their level of detail, spatial scale, and model structure. Some models use a single term to represent a composite set of transmission mechanisms, while others include multiple timescales of transmission \cite{codeco2001,tien2010}. In addition to standard mass-action transmission \cite{tien2010, tuite2011}, many models use nonlinear transmission functions for waterborne transmission \cite{codeco2001,  Mukandavire2011} to reflect the dose response for cholera in the water. Models may also include an asymptomatic transmission pathway, a hyperinfectious state for the bacteria immediately after shedding, and other ecological and environmental factors in the environmental reservoir, such as effects of vibriophages, plankton, weather, and climate \cite{tien2010, king2008, codeco2001, hartley2006, shuai2012, pascual2006, koelle2005b, Eisenberg2013b, millerneilan2010, rinaldo2012, akman2015evolutionary}.

In addition to the variation in transmission mechanisms, loss of immunity to cholera is poorly understood and therefore, modeled with many different assumptions. Estimates of the length of immunity in the literature range widely from several months to three to ten years \cite{Levine1981, king2008, koelle2004disentangling}. Immunity to cholera is of particular interest given the recent and ongoing oral cholera vaccine campaigns worldwide, including in Haiti, Bangladesh, and Thailand \cite{tohme2015oral,o2015oral,deen2016scenario,phares2016mass}, which raise additional questions of how vaccine-derived immunity compares to immunity derived from infection.

As modeling gains prevalence among policy makers in public health \cite{Abrams2012, moyer2014screening, auchincloss2008new, grad2012cholera, lofgren2014opinion, lipsitch2011improving}, comparative or ensemble modeling approaches have been increasingly viewed as a way to ensure that the results of parameter estimation, forecasting efforts, and the evaluation of intervention strategies are conserved across the range of realistic model structures \cite{koopman2004, meza2014comparative}. Two related concepts are useful to consider in these efforts---\textit{model distinguishability} addresses whether candidate models can be distinguished by their fits to empirical data \cite{walter1984structural}, and \textit{inference robustness assessments} examine whether conclusions drawn from a particular model are robust to realistic variations in the model structure \cite{koopman2004}. Both of these concepts are important to evaluate in the model-building process when model results are used as the basis for decision making. The importance of parameter and model uncertainty for cholera has been highlighted by several recent studies, both in general \cite{fung2014cholera,akman2015evolutionary}, and in the context of the recent Haiti epidemic specifically \cite{fung2014cholera, rinaldo2012}. 

In this paper, we examine the effects of uncertainty in model structure on cholera disease dynamics and inference by considering five models with different transmission and loss of immunity mechanisms. The models under consideration share a common base, the SIWR model of Tien and Earn \cite{tien2010}. The SIWR model is an extension of the classic Susceptible-Infected-Recovered (SIR) \cite{kermack1927} with an added compartment for pathogen concentration in an aquatic reservoir (W) \cite{tien2010}. In addition to the person-to-person and water transmission pathways of the base SIWR model, we evaluate two additional transmission-related features: a nonlinear Hill-function dose response for waterborne transmission and an asymptomatic infection and transmission pathway \cite{dunworth2011, codeco2001, king2008, millerneilan2010}. We also consider three loss of immunity features not included in the SIWR model: exponential loss of immunity, stage-progression gamma-distributed loss of immunity, and a novel model that features progressively increasing susceptibility after recovery from infection. 

Using these five deterministic SIWR-based models, we first simulate data from each model with different types of added noise. In the frame of model distinguishability and inference robustness assessment, we estimate model parameters and fit model data for all five models to each simulated dataset and an empirical dataset from the 2006 cholera epidemic in Angola. We determine how well each model recaptures the underlying parameter and $\Ro$ values, as well as how each model fits to simulated data generated from different models. Finally, we forecast trajectories using parameters estimated from truncated simulated incidence data and compare model forecasts with that of the true simulated data. 

\section{Methods}

\subsection{Model descriptions} 

\begin{figure}
\centering
\includegraphics[width=0.32\textwidth]{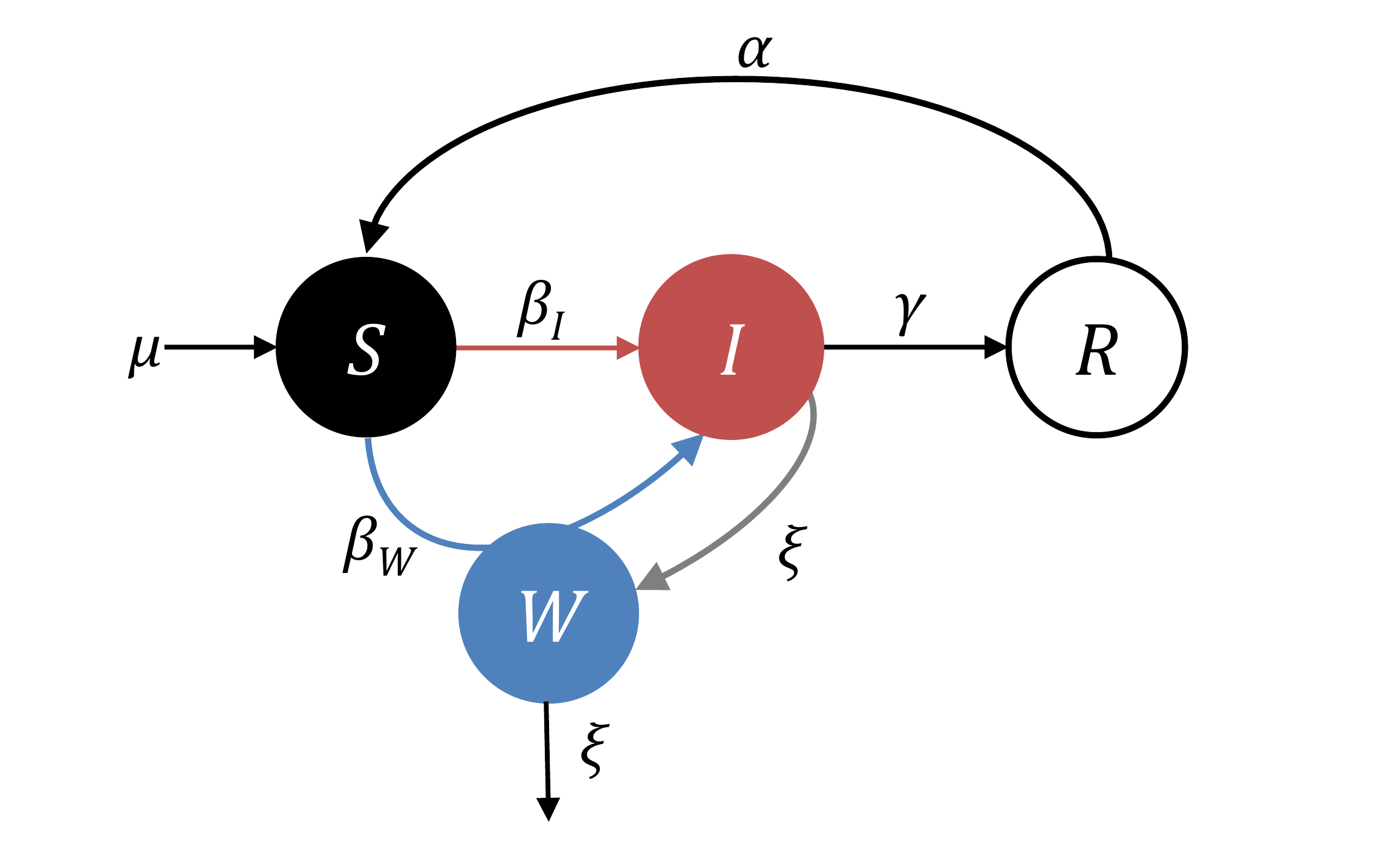}
\includegraphics[width=0.32\textwidth]{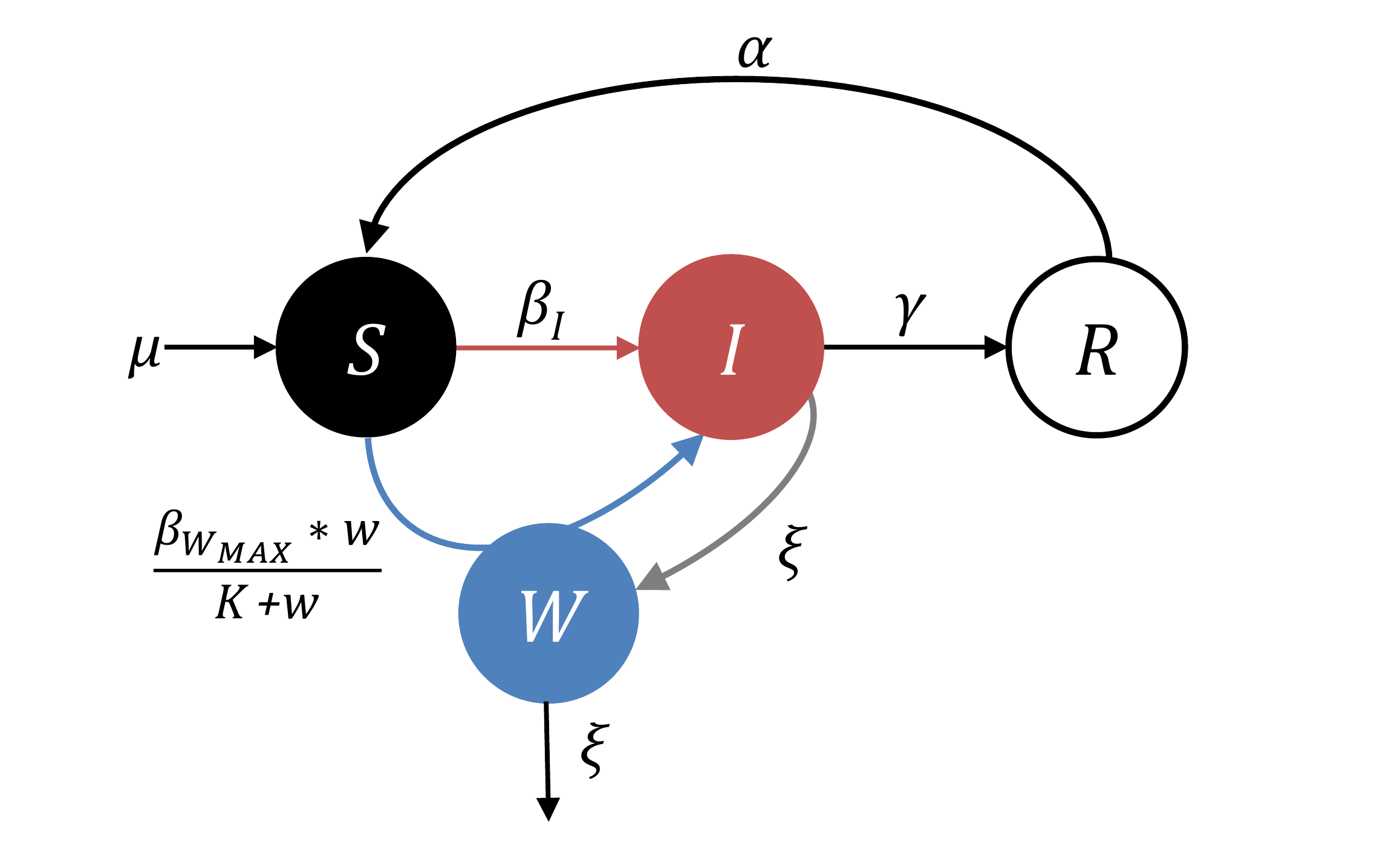}
\includegraphics[width=0.32\textwidth]{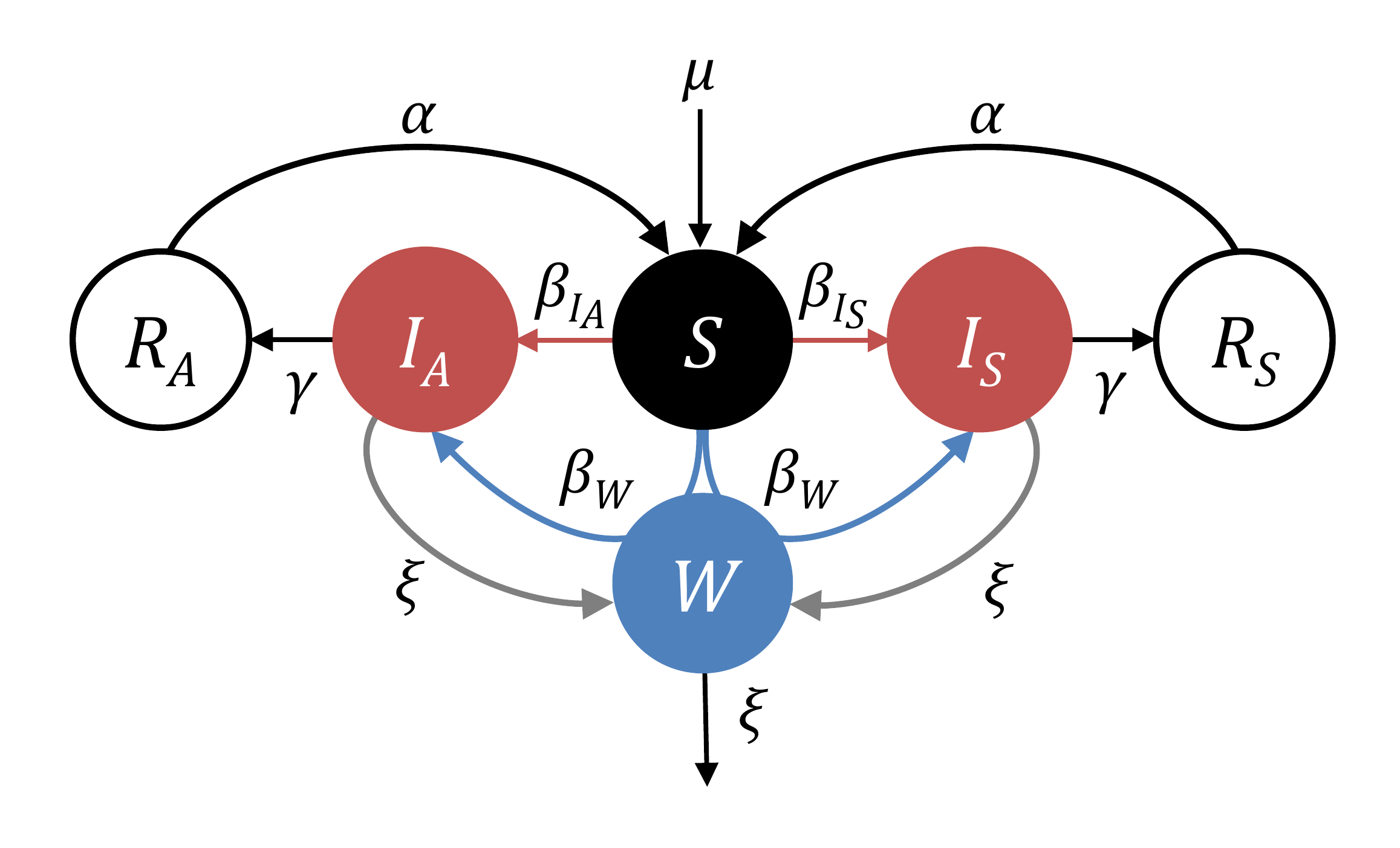}\\
\
\includegraphics[width=0.32\textwidth]{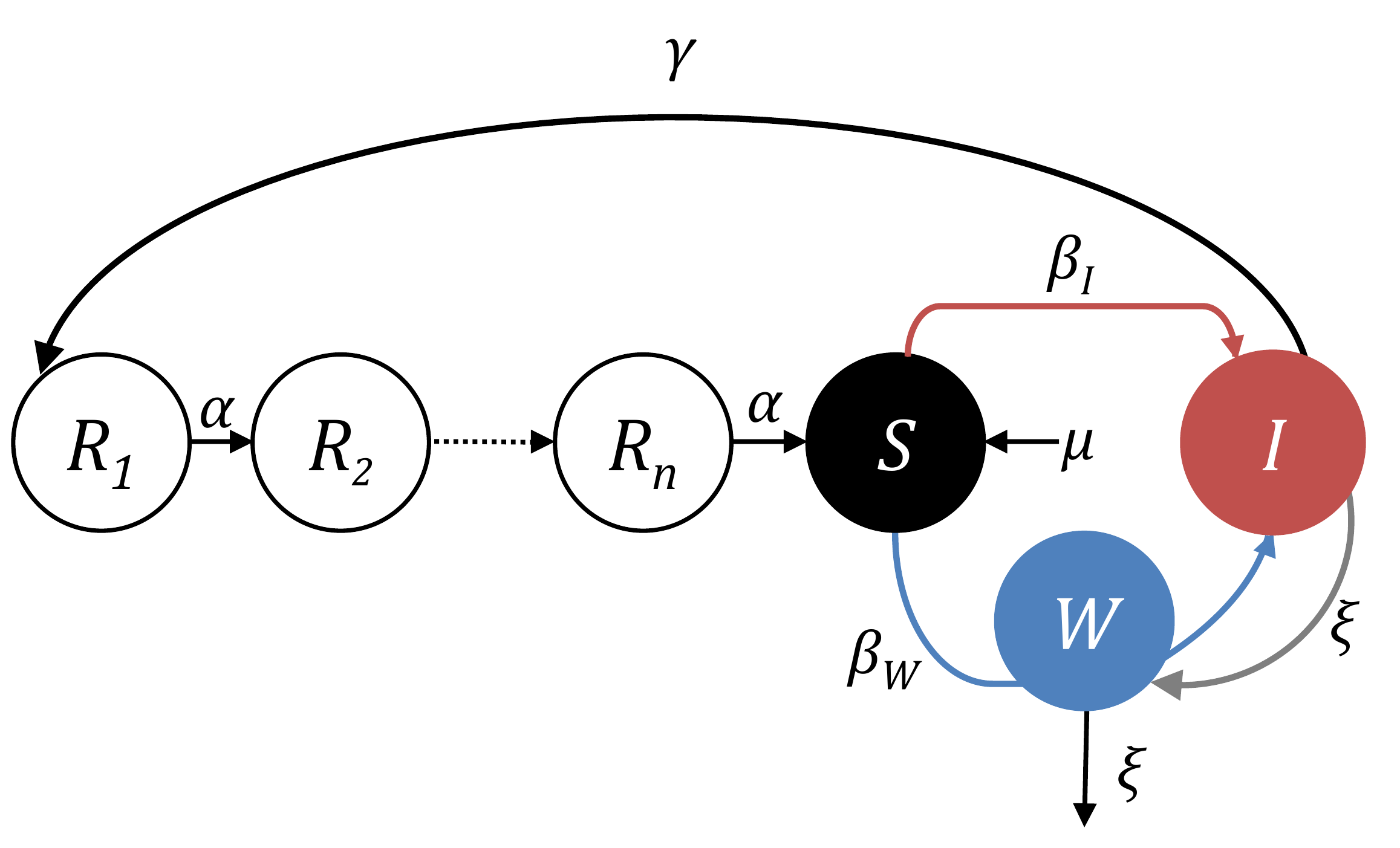} \hspace{1cm}
\includegraphics[width=0.32\textwidth]{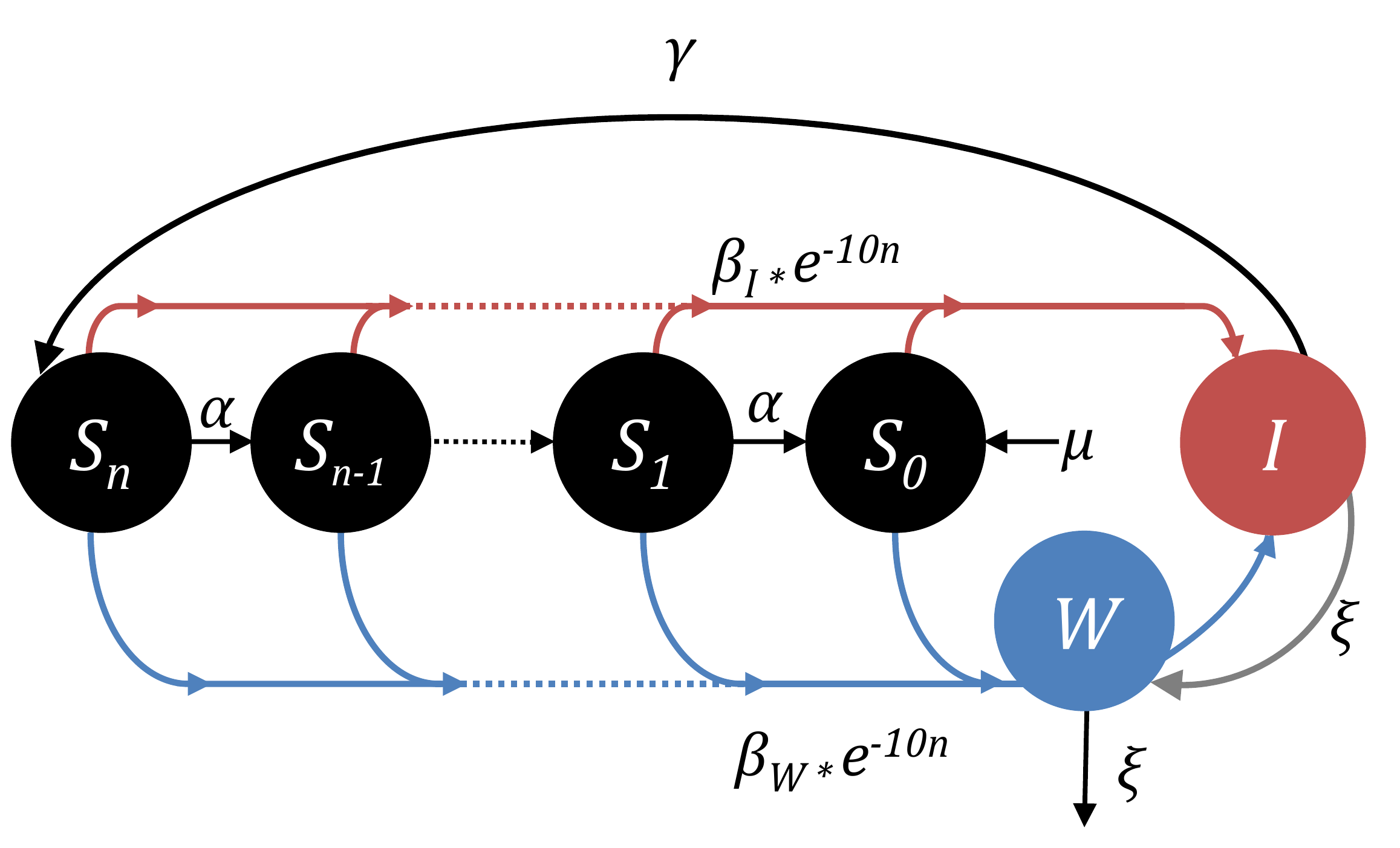}
\caption{\textbf{Diagrams of study models.} First row (left to right): Exponential model, Dose Response model, Asymptomatic model. Second row (left to right): Gamma model and Waning Immunity model. Red compartments represent the infected population and red arrows represent person-person transmission. Blue compartments represent pathogen concentration in water while blue arrows represent pathogen-person transmission. Black compartments are susceptible or partially susceptible, while white compartments are immune. Grey arrows indicate pathogen shedding.} \label{fig:AllModels}
\end{figure}

Here we introduce the five models of epidemic cholera disease dynamics that were used in our analyses. We base these models on the SIWR model of Tien and Earn \cite{tien2010}, which includes two timescales of transmission: a fast, direct route (represented by $\beta_I$) which incorporates direct person-person transmission, foodborne and household transmission, and other non-environmental transmission pathways; and a slow, indirect route ($\beta_W$) which represents long-term transmission mediated by bacteria in the water. The scaled non-dimensional model equations are given by \cite{tien2010}:
\begin{equation}
\begin{aligned}
s' &=  \mu -\beta_Isi-\beta_Wsw - \mu s\\
i' &=  \beta_W sw + \beta_I si - \gamma i - \mu i\\
w' &= \xi(i -w)\\
r' &= \gamma i - \mu r
\end{aligned}
\label{eq:SIWR}
\end{equation}
where $s$, $i$, and $r$ represent the fractions of the population that are susceptible, infectious, and recovered, respectively. The $w$ variable is proportional to the concentration of \textit{V.\ cholerae} in the environment, and $\xi$ represents the decay rate of the pathogen in the water (noting that the nondimensionalization in \cite{tien2010} rescales the water so that the parameter for pathogen shedding into the water is canceled and replaced with $\xi$, yielding the form of $w$ seen above---see Supplementary Information for nondimensionalization details). Parameter definitions and values are given in Table \ref{tab:params}. The basic reproduction number ($\Ro$) for the baseline model includes both transmission pathways and is given by
\begin{eqnarray}\label{BaselineR0}
\Ro = \frac{\beta_I + \beta_W}{\gamma + \mu},
\end{eqnarray}
where $\gamma$ is the recovery rate and $\mu$ is the natural death rate in the population. The five models in our study build upon this baseline SIWR model (Figure \ref{fig:AllModels}), with the aim of capturing a range of features commonly considered in loss of immunity and cholera dynamics. These features include: loss of immunity following exponential or gamma distributions, nonlinear dose response effects on transmission, and asymptomatic cases. 

To connect these models with data, we use the same measurement equation of $y = \kappa i$ (i.e. data are assumed to have mean $y$ with some measurement error), as has been used with a range of data sets using the SIWR model \cite{tuite2011, Eisenberg2013,Eisenberg2013b} (including the same epidemic dataset as considered here). The parameter $\kappa$ represents a combination of factors including the reporting rate, at-risk population size, potentially the infectious period (if used as an approximation for incidence data), and potentially the symptomatic fraction (depending on the model considered), among others. We define $k = 1/\kappa$ and fit $k$ rather than $\kappa$ as this constrains the parameter estimation to be within $(0,1)$ rather than $(1,\infty)$, making $k$ easier to estimate \cite{Eisenberg2013}.

\noindent \textbf{Exponential model.} The Exponential model builds on the baseline SIWR model by enabling recovered individuals to return to the susceptible state. We assume those in the recovered class ($r$) become susceptible ($s$) again after losing their immunity with rate parameter $\alpha$. The duration of immunity among recovered individuals follows an exponential distribution. $\Ro$ for this model has the same structure as that of the baseline model (Eq. \ref{BaselineR0}). The equations are given by:
\begin{equation}
\begin{aligned}
s' &=  \mu -\beta_Isi-\beta_Wsw - \mu s + \alpha r\\
i' &=  \beta_W sw + \beta_I si - \gamma i - \mu i\\
w' &= \xi(i -w)\\
r' &= \gamma i - \mu r - \alpha r.
\end{aligned}
\label{eq:Exponential}
\end{equation}

\noindent \textbf{Dose Response model.} The Dose Response model adds to the Exponential model by accounting for nonlinear dose response effects in the waterborne transmission pathway \cite{haas1999quantitative,codeco2001, dunworth2011,fung2014cholera}. Challenge studies in which volunteers ingested cholera and other pathogens at various doses have shown that the probability of transmission is a nonlinear function of dose size, with sigmoidal or Hill function-shaped curves that show low probabilities of infection at low concentrations and saturating probability of infection once the bacteria attain sufficiently high concentrations in the aquatic reservoir. Following previous studies, this model incorporated a Hill function dose response curve in the parameter for waterborne transmission \cite{codeco2001, dunworth2011}. We assume susceptible individuals become infected at a rate $\frac{\beta_{Wmax}w}{K + w}$, where $\beta_{Wmax}$ is the maximum force of infection from the water and $K$ is the concentration of \textit{V.\ cholerae} that gives a $50\%$ chance of water-transmitted cholera infection. As in the Exponential model, individuals lose immunity at a rate $\alpha$. The model equations are given by:
\begin{equation}
\begin{aligned}
s' &=  \mu -\beta_Isi-\frac{\beta_{Wmax}w}{K + w}s - \mu s + \alpha r\\
i' &=  \frac{\beta_{Wmax}w}{K + w}s + \beta_I si - \gamma i - \mu i\\
w' &= \xi(i -w)\\
r' &= \gamma i - \mu r - \alpha r
\end{aligned}
\label{eq:DoseResponse}
\end{equation}
and $\Ro$ for this model is given by: 
\[R_0 = \frac{\beta_I + \frac{\beta_{Wmax}}{K}}{\gamma+\mu}.\]

\noindent \textbf{Asymptomatic model.} This model expands upon the Exponential model to include an asymptomatic infection pathway. Estimates of the ratio of asymptomatic to symptomatic cholera infections range from 3 to 100 \cite{king2008}. We assumed 20\% of infections were symptomatic ($q = 0.2$) \cite{fung2014cholera}. We assumed that all pathogens decay at the same rate $\xi$, regardless of which class they are shed from. Shedding rates differ by symptomatic/asymptomatic status \cite{fung2014cholera}, with differential shedding rates $\sigma_S$ and $\sigma_A$. Scaling by these shedding rates and the other model parameters in the nondimensionalization (given in the Supplementary Information) results in the following model equations:
\begin{equation}
\begin{aligned}
s' &=  \mu -\beta_{IA}si_A -\beta_{IS}si_S -\beta_W sw - \mu s + \alpha_A r_A + \alpha_S r_S\\
i'_S &=  q(\beta_W sw + \beta_{IS}si_S + \beta_{IA}si_A) - \gamma i_S - \mu i_S\\
i'_A &=  (1-q)(\beta_W sw + \beta_{IS}si_S + \beta_{IA}si_A) - \gamma i_A - \mu i_A\\
w' &= \xi(i_A + i_S -w) \\
r'_S &= \gamma i_S - \mu r_S - \alpha_S r_S\\
r'_A &= \gamma i_A - \mu r_A - \alpha_A r_A
\end{aligned}
\label{eq:Asymptomatic}
\end{equation}
where $\Ro$ for this model is given by: 
\[R_0 = \frac{1}{\gamma+\mu}\left[q(\beta_{I_S}+\beta_W)+(1-q)(\beta_{I_A}+\beta_W)\right].\]

\noindent \textbf{Gamma model.} As opposed to exponential loss of immunity, the Gamma model represents waning immunity through a chain of $n$ recovered classes, which results in an immunity duration that has a gamma distribution (See \cite{wearing2005} and \cite{lloyd2001b} for examples). The number of classes depends on the assumed distribution on length of immunity; a greater number of compartments yields a duration that begins to approximate a time delay. In this model, individuals in the chain of recovered classes retain complete immunity until they return to the susceptible class, with equations given by:
\begin{equation}
\begin{aligned}
s' &=  \mu -\beta_Isi-\beta_Wsw - \mu s + n\alpha r_n\\
i' &=  \beta_W sw + \beta_I si - \gamma i - \mu i\\
w' &= \xi(i -w)\\
r'_1 &= \gamma i - \mu r_1 - n\alpha r_1\\
  & \vdots \\
r'_n &= n\alpha r_{n-1} - \mu r_n - n\alpha r_n
\end{aligned}
\label{eq:Gamma}
\end{equation}
and $\Ro$ for this model is same as Equation \ref{BaselineR0}. We note that the progression through each stage of immunity is at rate $n\alpha$, so that this model has the same overall duration of immunity as the previous models.

\noindent \textbf{Waning Immunity model.} Given the wide range of estimates for the length of immunity in cholera, we also developed a new model that allows for progressively increasing susceptbility after infection over time. This may more realistically reflect the dynamics of loss of immunity, which is likely to be progressive rather than switch-like. Similar models of progressive waning has been used in a range of other contexts \cite{hethcote1999simulations}, and a distribution of cholera antibody titers are often observed in cholera vaccine studies \cite{mosley1968serological}. The Waning Immunity model again represents loss of immunity through a chain of $n$ classes, but individuals are at least partially susceptible at all points in the chain. The transmission rates are given as functions of location in the chain (Figure \ref{fig:AllModels}), where more recently recovered individuals are less likely to become infected. As individuals progress through the chain, immunity to water-borne and person-person infection wanes, with susceptibility given by functions $f$ and $g$, respectively. The forms of $f$ and $g$ were chosen so that the probability of infection is close to zero for individuals in the first susceptible class ($s_n$) and equal to one for individuals in the last susceptible class ($s_0$). The model equations are:
\begin{equation}
\begin{aligned}
i' &=  \sum_{k=1}^n \beta_I si\cdot g(k) + \sum_{k=1}^n \beta_W sw\cdot f(k)  - \gamma i - \mu i\\
w' &= \xi(i -w)\\
s'_1 &=  \mu + n\alpha s_2 -\beta_I s_1 i  -  \beta_W s_1 w - \mu s \\
& \vdots\\
s'_k &= n\alpha s_{k+1} - n\alpha s_k - \beta_I g(k) s_k i - \beta_W f(k) s_k w - \mu s_k \\
& \vdots \\
s'_n &= \gamma i - n\alpha s_n - \beta_I g(n) s_n i - \beta_W f(n) s_n w - \mu s_n,
\end{aligned}
\label{eq:WaningImmunity}
\end{equation}
where we take $f(k) = e^{-u_1(k-1)/(n-1)}$ and $g(k) = e^{-u_2(k-1)/(n-1)}$. For simplicity, we assumed baseline values of $u_1 = u_2 = n$, i.e. equal rates of waning immunity for waterborne and direct transmission. This form was chosen to make $f$ and $g$ equal to 1 for $s_n$ and near zero for $s_1$. We also make the same adjustment to the waning immunity rates as in the Gamma model. Again, $\Ro$ for this model is given by Equation \ref{BaselineR0}.

\subsection{Data simulation and baseline parameters}\label{sec:methods_simdata}
Model parameters were taken from previous modeling and empirical studies of epidemic cholera, primarily from the estimates in \cite{Eisenberg2013}, which used a SIWR model fitted to the 2006 outbreak of cholera in Angola (Table \ref{tab:params}). However, some updates to the parameter values were made, based on literature estimates and known biology, as described in more detail in the Supplementary Information (Section \ref{SuppParameterSection}). For the epidemic scenarios, we simulated six sets of data from each model to represent several possible combinations of measurement error assumptions---noise-free, normal noise ($\mu$ = model output, $\sigma = 0.1 \times data$), and poisson noise ($\lambda$ = model output), and simulation durations---epidemic (100 days) and long-term (3 years). Throughout the paper, we use the term \textit{simulation model} to refer to the model that simulated the data for the analysis. These 30 simulated datasets (five models $\times$ three noise options $\times$ two simulation durations) were used for both parameter estimation/model fitting and forecasting. 

We used the average life expectancy of Angola, 55 years \cite{CIAFactbook}, to determine $\mu$. Length of immunity to cholera ($\alpha$) acquired after infection is not well understood, with estimates ranging from weeks to months to years \cite{Levine1981, king2008, koelle2004disentangling}. For our simulated data, we assumed immunity lasted one year. The complete set of values, descriptions, and sources for the baseline parameters used for data simulation is given in Table \ref{tab:params}.  

While most parameters were common across models, some models required additional, specialized parameters. For the Asymptomatic model, we chose $\beta_{IA}$ and $\beta_{IS}$ such that $\beta_{IS}$ was four times greater than $\beta_{IA}$ and the weighted average of $\beta_{IA}$ and $\beta_{IS}$ (where symptomatic infections represent $q = 20\%$ of the infected population) was equal to $\beta_I$ for all of the other models. To compare $\beta_I$ across models, in the Results for the Asymptomatic model we report this weighted average of the two transmission parameters. Reasoning that individuals with symptomatic infections will have longer lasting immunity than those with asymptomatic infections, we assumed that immunity from symptomatic infections ($\alpha_S$) would last two years and immunity from asymptomatic infections ($\alpha_A$) would last 0.5 years. Similarly, the reported estimates of $\alpha$ for this model are given as the weighted average of $\alpha_S$ and $\alpha_A$. Additionally, the $k$ value used for the Asymptomatic model was adjusted based on the fact that only symptomatic cases are observed (Table \ref{tab:params}), although the unadjusted form was reported to facilitate model comparison in the Results.

The Dose Response model parameter structure also deviated from the other models, with the waterborne transmission term expanded to form a Hill function. The half-saturation constant, $K$, was determined based on previous literature estimates \cite{dunworth2011nonlinear}. To determine the value of the maximum transmission level, $\beta_{Wmax}$, there are in general several natural ways that one might match up the Hill function with the single waterborne transmission parameter $\beta_W$ value used in the original SIWR model. For example, one might match this value to the linear approximation of the Hill function (i.e. let $\beta_W = \beta_{Wmax}/K$, which approximates the Hill function for low values of $w$), or to the maximum level of waterborne transmission (i.e. let $\beta_W = \beta_{Wmax}$), or to some intermediate value of transmission. The Dose Response model behaved quite similarly to the Exponential model when the linear approximation was used; we chose the saturated, maximum value of transmission to match the models, letting $\beta_{Wmax} = \beta_W$ for our baseline parameters, in order to accentuate the differences between models.

For the simulated data, we assumed simulation model initial conditions with 1\% of the population infected and the remaining population susceptible. For the Asymptomatic model, we assumed that 0.2\% of the population was $I_S$, 0.8\% of the population was $I_A$, and the remaining population was susceptible. All model simulations and parameter estimation were done in MATLAB. We also evaluated the baseline $\Ro$ for each model using the baseline parameter values---with these parameters, all models have an $\Ro$ of 3, except for the Dose Response model, which has an $\Ro$ of 6.

\begin{table}[h]
\centering
\caption{Model parameters. Additional details about the baseline values for each parameter are given in Section \ref{sec:methods_simdata} and the Supplementary Information (SI).}
 \small
 \begin{tabular}{| c | p{5cm} | c | p{1.25cm} | p{3.75cm} | p{2.25cm} |}
 \hline
Parameter		& Description		&	Value  & Units	& Source &Models	\\ \hline
$\mu$		& natural birth/death rate in Angola	& 	$\frac{1}{(55*365)}$ & $days^{-1}$ & \cite{ciafactbook_angola} &	All	\\ \hline
$\gamma$	& 1/(mean duration of infectiousness), or recovery rate 	& 	$\frac{1}{4}$ & $days^{-1}$ & \cite{Weil2009, factsheet,tuite2011, Eisenberg2013} & All \\ \hline
$\beta_I$		& person-person transmission rate & 	0.25 & $days^{-1}$ & \cite{Eisenberg2013, tien2010, tuite2011} & All except Asymptomatic \\ \hline
$\beta_{IA}$ 	& person-person transmission rate in asymptomatics &	$\frac{5}{32}$ & $days^{-1}$ & See Model Description, Section \ref{sec:methods_simdata} & Asymptomatic	 \\ \hline
$\beta_{IS}$ 	& person-person transmission rate in symptomatics &	$\frac{5}{8}$ & $days^{-1}$ &See Model Description, Section \ref{sec:methods_simdata} & Asymptomatic	 \\ \hline
$\beta_W$ 	& water-borne transmission rate &	0.50	& $days^{-1}$ &\cite{Eisenberg2013, tien2010, tuite2011} & All \\ \hline
$\alpha$		& loss of immunity rate	&	$\frac{1}{365}$	& $days^{-1}$ &\cite{Levine1981, king2008, koelle2004disentangling}, See Section \ref{sec:methods_simdata} & All except Asymptomatic \\ \hline
$\alpha_{IA}$	& loss of immunity rate in asymptomatics	& $\frac{1}{182.5}$  & $days^{-1}$ &See Section \ref{sec:methods_simdata} &	Asymptomatic \\ \hline
$\alpha_{IS}$	& loss of immunity rate in symptomatics	&	$\frac{1}{730}$ &  $days^{-1}$ &See Section \ref{sec:methods_simdata} &	Asymptomatic \\ \hline
$\xi$			& decay rate of cholera in water	& $\frac{1}{100}$ & $days^{-1}$ & \cite{Eisenberg2013, tuite2011} & All	\\ \hline
$k$		& 1/(population at risk $\cdot$ reporting rate)	&	$\frac{1}{50,000}$ & $people^{-1}$ & \cite{Eisenberg2013}, See Section \ref{sec:methods_simdata}, SI & All except Asymptomatic \\ \hline
$k$			& proportion symptomatic cases / (population at risk $\cdot$ reporting rate)	&	$\frac{0.2}{50,000}$ & $people^{-1}$ &\cite{Eisenberg2013}, See Section \ref{sec:methods_simdata}, SI & Asymptomatic \\ \hline
$K$			& bacteria concentration that gives $50\%$ chance of infection	&	0.4 & $unitless$ &\cite{dunworth2011nonlinear}, See Model Description, Section \ref{sec:methods_simdata} &	Dose Response \\ \hline
$u_1$			& modifies the susceptibility of individuals to direct infection	& 10 & $unitless$ &See Model Description &	Waning Immunity \\ \hline
$u_2$			& modifies the susceptibility of individuals to waterborne infection	& 10 & $unitless$ & See Model Description &	Waning Immunity \\ \hline
\end{tabular}
\label{tab:params}
\end{table}

\subsection{Identifiability of parameters} 

Identifiability and uncertainty quantification methods allow us to address whether, and with what degree of certainty, it is possible to uniquely recover the parameters of a model for a given dataset. In general, two broad classes of identifiability are often considered---structural and practical identifiability \cite{Cobelli1980}. Structural identifiability addresses whether it is possible to estimate the parameters based purely on the structure of the model and the measurements, typically assuming a best-case of perfect, noise-free data. Practical identifiability considers how the amount and quality of the data (e.g. level of measurement error, number of replicates) may affect estimation of the parameters. One commonly used method of evaluating identifiability is via the Fisher Information Matrix (FIM), a symmetric matrix that represents the amount of information about the parameters that is contained in the data \cite{Rothenberg1971, Jacquez1985}. 
The rank of the FIM corresponds to the number of (locally structurally) identifiable parameters, and taking the inverse of the FIM gives an approximation of the covariance matrix for the parameters using the Cramer-Rao bound \cite{Rothenberg1971,Komorowski2011,CintronArias2009, Cobelli1980}. We calculated the FIM for each model (assuming the same parameters and starting parameters as for the simulated data, with daily data for one year), and evaluated both the rank and covariance matrix. From the parameter variances, we calculated the percent coefficient of variation ($\%CV$) to evaluate parameter uncertainty. The $\%CV$ takes into account the size of the parameter value when evaluating uncertainty, with $\%CV = \frac{\sigma_p}{p} \times100$ (where $\sigma_p$ is the standard deviation and $p$ is the value of the parameter). We used a tolerance value \cite{EisenbergHayashi}, where $\%CV \leq 100$ indicates parameter identifiability. Typically, structurally unidentifiable parameters will have extremely large (e.g., numerically near-infinite, typically $>10^6$ \cite{EisenbergHayashi}) $\%CV$s, while practically unidentifiable parameters have finite but large $\%CV$s.

\subsection{Parameter estimation and model fitting}

Using maximum likelihood estimation and Nelder-Mead optimization in MATLAB, we fit each of the five model structures to the 30 simulated datasets. When fitting our models to data, we estimated all parameters except the birth/death rate, $\mu$, and the recovery rate, $\gamma$, which are known \cite{CIAFactbook,tuite2011, Weil2009}. Throughout the paper, we use the term \textit{fitting model} to refer to the model that was used to fit an epidemic dataset. To perform model fitting and parameter estimation from the simulated data, we used two sets of parameters to initialize the optimization: \textit{informed starting parameters}, which were the true parameters from the simulation model (i.e. the parameters used in generating the simulated data, given in Table \ref{tab:params}), and \textit{naive starting parameters}, which were randomly chosen at $\pm 20\%$ of the true parameters. To compare model fits to simulated data, we calculated and compared Akaike information criterion (AIC) values for the models \cite{akaike1987factor}; for ease of comparison, we report $\Delta AIC$ values, defined as the difference in AIC between the model of interest and the best fit model (i.e., model with the lowest AIC) for a given dataset. To compare parameter estimates among the fitting models, we examined the percent deviation from the true value, which was calculated as the difference between the estimate and the true value, divided by the true value. In this descriptive analysis, a deviation less than 20\% of the true value was considered `accurately recaptured' because the variation should be smaller than the potential deviation in starting parameters. Additional details on the parameter estimation methods are also given in the Supplementary Information.

For model initial conditions when fitting, we let $i(0) = k z(0)$, where $z(0)$ is the initial data value, and then assumed the remaining population to be susceptible. For the Asymptomatic model, we assumed that the observed infected were symptomatic, and generated an additional fraction of asymptomatically infected individuals based on the proportionality parameter $q$ (with the remainder of the population again assumed susceptible).

\subsection{Forecasts of epidemic data}

In addition to parameter estimation, we evaluated the ability for the study models to forecast epidemic trajectories when provided with truncated and noisy simulated epidemic data. Using 10, 30, and 50 days of simulated incidence data (out of 100-day simulated epidemic data) with normal noise from the five study models, we estimated parameters from each fitting model. These parameter estimates were then used to generate a trajectory forecast for the remainder of the 100-day epidemic. Throughout the study, we use \textit{forecasting model} to refer to the model used to estimate these parameters and generate the forward-looking trajectory. For the parameter estimates in this section, we used the informed starting values.

\subsection{Application to 2006 cholera outbreak in Angola}

We investigated the ability for the five study models to fit and forecast data from an empirical cholera epidemic. We applied the same methods of parameter estimation, model fitting, and forecasting to data from the recent 2006 cholera outbreak in Angola (data courtesy of the WHO Cholera Task Force), shown in Figure \ref{fig:AngolaFits}. The outbreak began in Luanda province (February 13, 2006), and eventually spread to 16 of 18 provinces, resulting in 82,204 cases and 3,092 deaths \cite{Sack2006, Eisenberg2013}. The `true' parameter values are unknown, so we considered both the `informed' and `naive' sets of starting parameter values used in the simulated data for fitting to real data as well. In this case, the `informed' starting parameter values of course do not represent the true parameter values, but are in some sense more `informed', as the baseline values given in Table \ref{tab:params} are based on the best fit parameters for the the original SIWR model to this data set in \cite{Eisenberg2013}. For fitting, we used weighted least squares assuming a weight of $\sigma^2 = data$ \cite{Eisenberg2013}. The same initial condition setup was used in these fits as for fitting to the simulated data.

\section{Results}

\subsection{Identifiability of parameters}
The FIMs for all models except for the Asymptomatic model were full rank. All FIM's were numerically invertible, with the percent coefficients of variation ($\%CV$s) shown in Table \ref{tab:id}. Apart from the aforementioned rank deficiency in the Asymptomatic model, all parameter $\%CV$s were small enough that the parameters would be considered structurally identifiable. 

\begin{table}
\centering
\caption{Estimated $\%CV$ for each parameter. A $\%CV>100$ means that the parameter is practically unidentifiable. A `-' means that the parameter is not applicable to the model.}
\label{tab:id}
\small
\begin{tabular}{| p{2cm} | c | c | c | c | c |}
\hline
Parameters & Exponential & Dose Response & Asymptomatic & Gamma & Waning Immunity \\ \hline
$\beta_I$  &  0.016  &  0.019  &  -  &  0.011  &  0.016\\
$\beta_{IS}$  &  -  &  -  &  $1.11 \times 10^{12}$  &  -  & - \\
$\beta_{IA}$ &  -  &  -  &  $1.11 \times 10^{12}$  &  -  & - \\
$\beta_W$  &  0.56  &  0.27  &  0.13  &  0.34  &  0.48\\
$\alpha$  &  0.34  &  0.094  &    &  0.11  &  0.23\\
$\alpha_S$  &  -  &  -  &  0.60  &  -  & - \\
$\alpha_A$  &  -  &   - &  0.072  &  -  & - \\
$\xi$  &  0.66  &  0.33  &  0.097  &  0.37  &  0.55\\
$k$  &  0.025  &  0.0081  &  0.028  &  0.042  &  0.044\\
$u_1$  &  -  &  -  &  -  &   - &  0.55\\
$u_2$  &  -  &  -  &  -  &  -  &  8.33\\ \hline
\end{tabular} 
\end{table}

\begin{figure}
        \centering
        \includegraphics[width=0.5\textwidth]{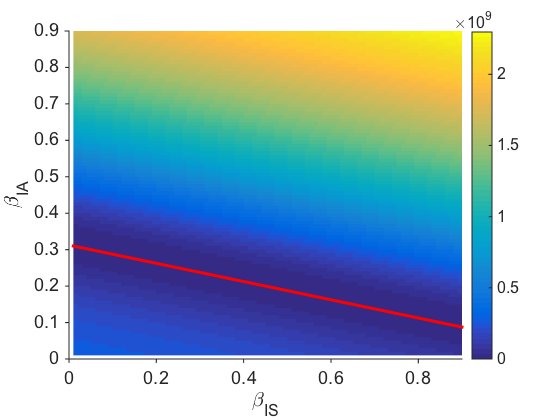}
       	\caption{Goodness-of-fit surface for $\beta_{IA}$ and $\beta_{IS}$ in the Asymptomatic model, highlighting the identifiable combination. The red line highlights the minimum sum of squares, which corresponds to an identifiable combination given by the line $q\beta_{I_S}+(1-q)\beta_{I_A}$.}\label{fig:asympcombo}
\end{figure}

The Asymptomatic model was rank deficient by one, although it was still possible to numerically invert the FIM, which showed that the direct transmission parameters for symptomatic vs. asymptomatic transmission form an identifiable combination. To determine the form of the identifiable combination, we simulated noise-free data using the Asymptomatic model as described in the Data Simulation section, and then plotted the sum-of-squares goodness-of-fit surface across a range of values of $\beta_{IS}$ and $\beta_{IA}$, shown in Figure \ref{fig:asympcombo}. We found a linear canyon which achieved the same minimum goodness-of-fit as the true parameters, corresponding to a structurally identifiable combination \cite{Raue2009, EisenbergHayashi}. The form of the identifiable combination was given by the sum of the two transmission parameters, weighted by the proportion of symptomatic cases, $q$: $\beta_I = q\beta_{I_S}+(1-q)\beta_{I_A}$. If we fixed either $\beta_{IA}$ or $\beta_{IS}$ using the identifiable combination as a constraint, and re-calculated the FIM, all $\%CV$s were $<1.5\%$. Because the dynamics for the two infected compartments ($i_S$ and $i_A$) are the same up to a constant scaling factor, it is impossible to distinguish the proportion of cases generated from one compartment vs. the other. Given these results, in our subsequent analyses for the Asymptomatic model we report only this identifiable combination $\beta_I$, rather than the individual unidentifiable parameters $\beta_{I_S}$ and $\beta_{I_A}$.

\subsection{Parameter estimation and model fitting to epidemic data}

\begin{figure}[H]
	\centering
	\includegraphics[width=0.85\textwidth]{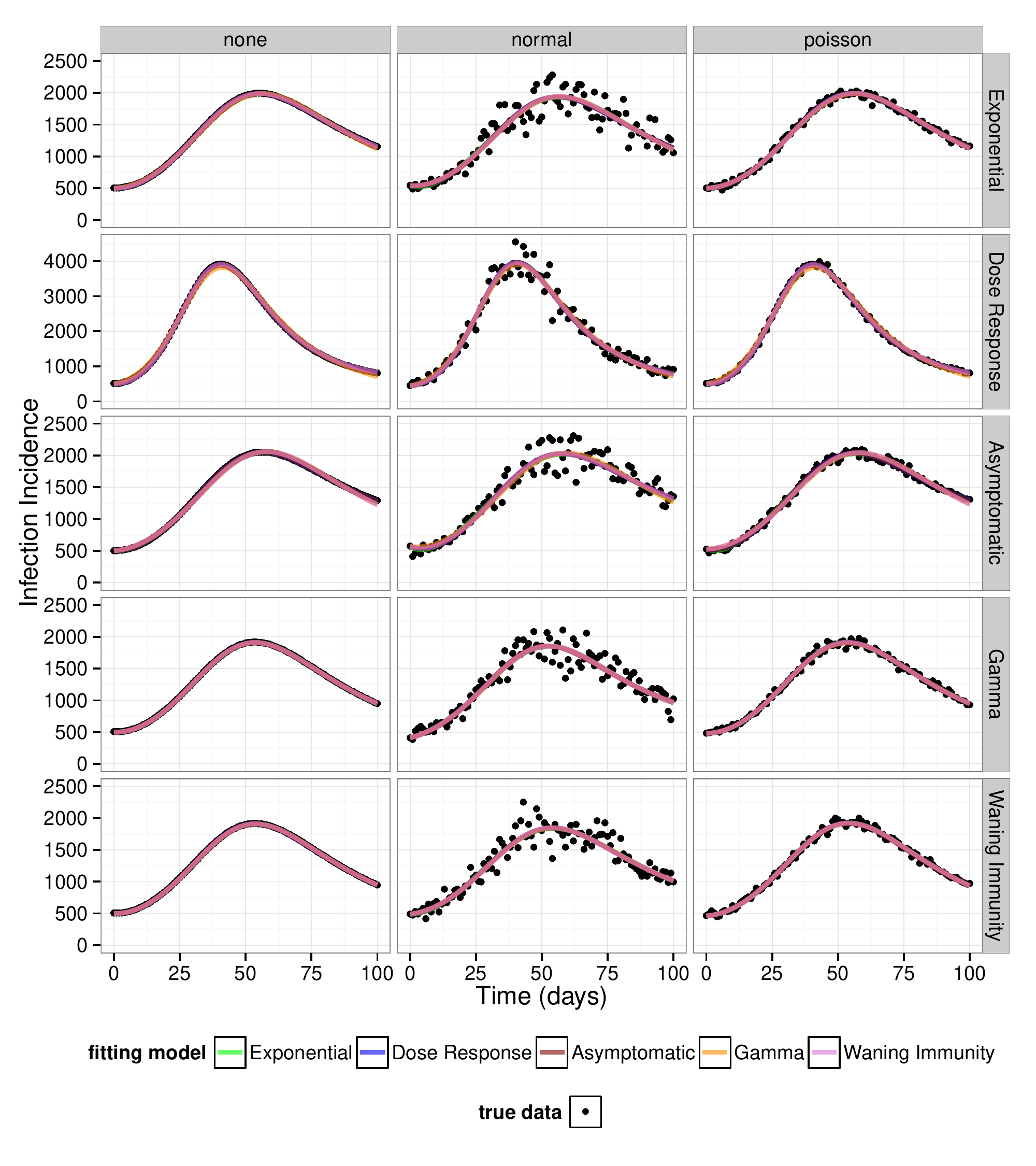}
    \caption{Fits to 100-day simulated model data (indicated by row) without noise (left column), with normal noise (middle column), and with poisson noise (right column), using naive starting parameters. Model fits are overlaid, thus obscuring some of the model fits in the figure. All fitting models were able to capture the mean epidemic data despite added noise and model misspecification.}\label{fig:naiveFits}
\end{figure}

For the 100-day simulation duration, across the three noise cases, all five models fit the simulated data well (Figure \ref{fig:naiveFits} and Supplementary Figure \ref{fig:informedFits}). Nearly all model fits matched the simulated noisy data closely throughout the trajectory, with few runs (strings of consecutive data points on one side of the fit) of obviously correlated residuals. Correspondingly, the AIC values were nearly indistinguishable across model fits, but the Exponential or Dose Response model fits consistently had the lowest (i.e., best) numerical values in 25 out of 30 cases, while the Asymptomatic and Waning Immunity model fits frequently matched or followed close behind (Supplementary Tables \ref{NNIP} - \ref{PNNP}).

The true parameter values of direct transmission ($\beta_I$) and measurement scaling ($k$) were recovered much more accurately than the other parameters that were common across all models (Figure \ref{fig:ParamPlots}), and neither of these parameters had identifiability issues in our study (Table \ref{tab:id}) or a previous cholera modeling study \cite{Eisenberg2013}. For $\beta_I$, the magnitude in parameter estimate deviations across four of five fitting models was under 20\%; the Asymptomatic model under-estimated $\beta_I$ by 33\% in one instance. For $k$, the magnitude in parameter estimate deviations across three of five fitting models was under 20\%; the Gamma model underestimated $k$ by 21-22\% in two instances and the Waning Immunity model both underestimated and overestimated $k$ with a deviation magnitude greater than 20\%.

\begin{figure}
        \centering
        \begin{subfigure}[b]{0.46\textwidth}
                \includegraphics[width=\textwidth]{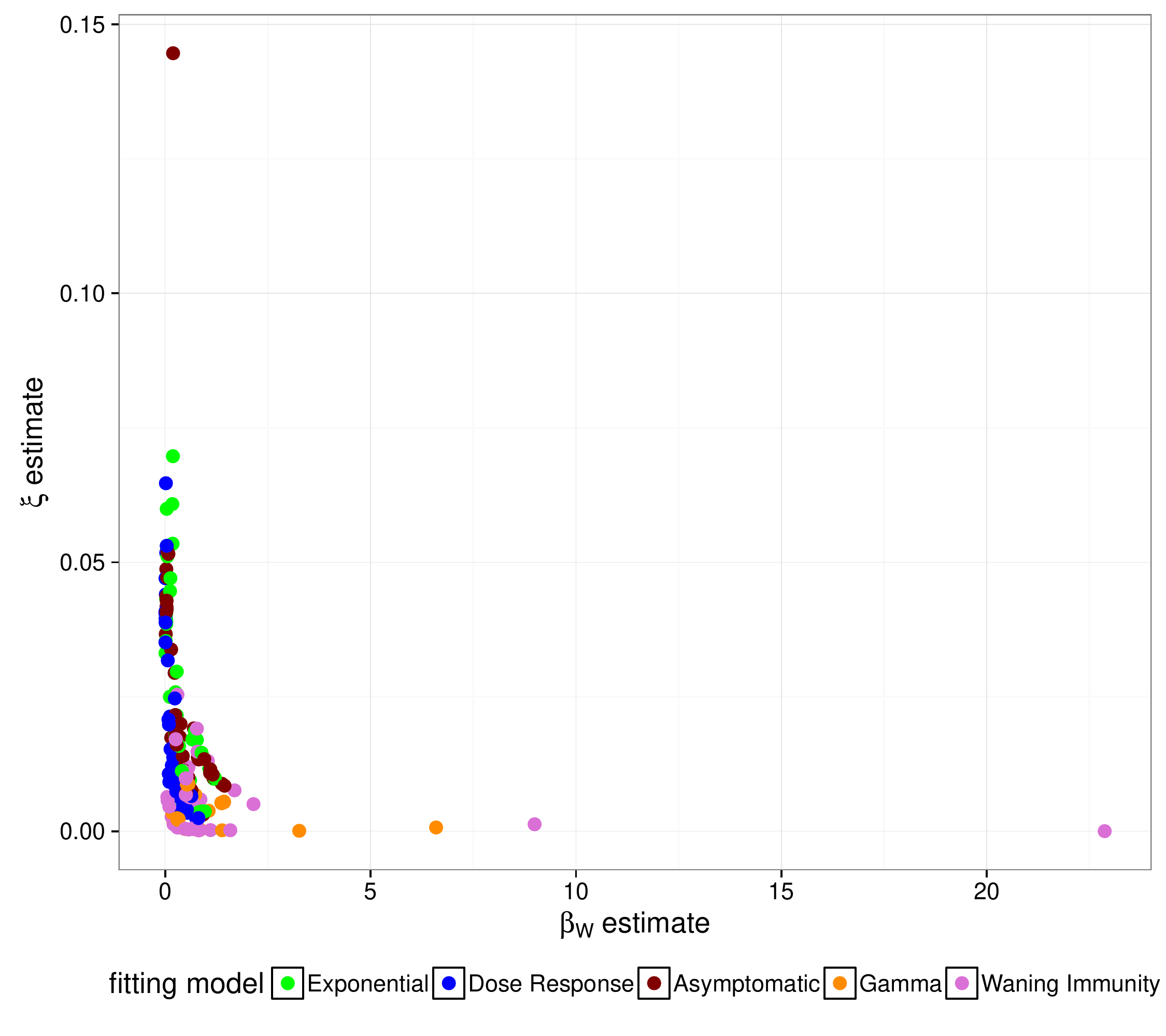}
        \end{subfigure}
        \begin{subfigure}[b]{0.46\textwidth}
        \includegraphics[width=\textwidth]{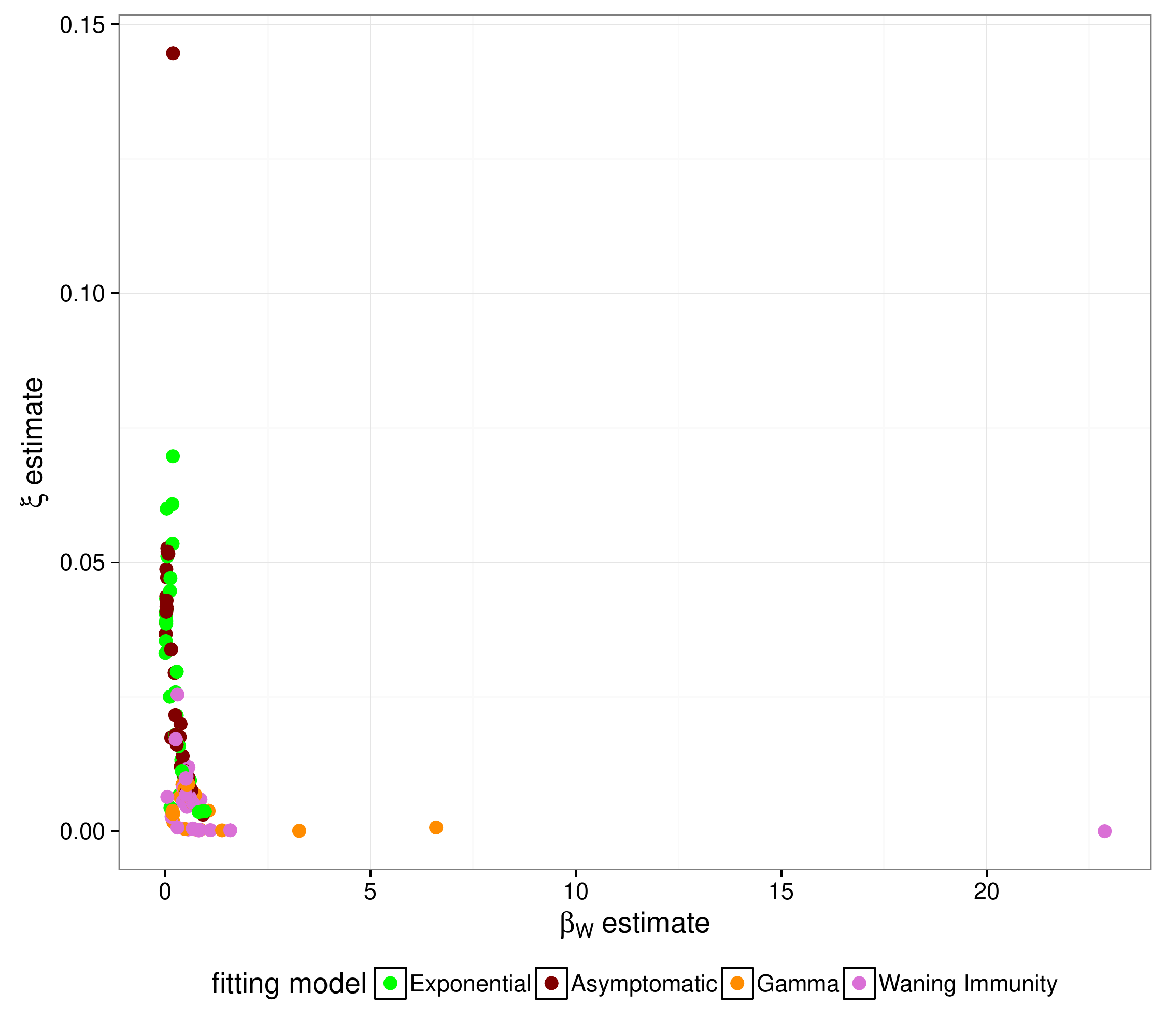}
        \end{subfigure}
        \caption{Scatterplot of $\beta_{W}$ and $\xi$ estimates by colored fitting model with all data points (left) and without Dose Response model estimates and fits (right).}\label{fig:betaW-xi-identif}
\end{figure}

In congruence with a previous study \cite{Eisenberg2013}, we found that $\beta_W$ and $\xi$ formed a practically unidentifiable combination in all models (Figure \ref{fig:ParamPlots} and Supplementary Figure \ref{fig:betaW-xi-identif}), which might explain their large deviations from the true parameter value. We also considered that some of the deviations in $\beta_W$ (and by extension $\xi$), may be due to the different choice of structure for $\beta_W$ in the Dose Response model. We evaluated the parameter variation without the Dose Response parameter estimates or simulated data, and found that while there was some attenuation of the more extreme outliers for the estimates of $\beta_W$ across all models, $\beta_W$ still showed a much broader range than $\beta_I$ or $\xi$, and the remaining parameter distributions remained largely identical (shown in Supplementary Figure \ref{fig:betaW-xi-identif}).

While most of the parameters were neither consistently underestimated nor overestimated by any of the models, the estimates of $\alpha$ tended to be underestimated for both the Gamma and Waning Immunity models when fitted to epidemic data (although other runs of estimation, e.g. using different optimizer settings did not necessarily yield this). 
Deviations were systematically smaller when models were fitting Exponential model data, and as expected, parameter estimate recapture accuracy improved as simulated data decreased in noise (Figure \ref{fig:ParamPlots2}). Parameter estimates for all combinations of simulation model data and fitting models are reported in Supplementary Tables \ref{NNIP} - \ref{PNNP}. 

\subsection{Parameter estimation and model fitting to long-term data}

For the 3-year data, all models were able to fit the qualitative dynamics of the simulated data, and with a few exceptions, visually all models matched all data sets well (Supplementary Figures \ref{fig:informedFits_3y} and \ref{fig:naiveFits_3y}). However, unlike the 100-day data, the lowest AIC values were most often achieved when the same model which generated the data was used for fitting (though often other models yielded very close AIC values, with differences $\leq 5$). AIC values were consistently lower when models where the waning immunity distributions matched those of the data simulation model (exponentially-distributed waning immunity: Exponential, Dose Response, and Asymptomatic models; gamma-distributed waning immunity: Gamma and Waning Immunity models) (Supplementary Tables \ref{NNIP3} - \ref{PNNP3}). Similarly, estimates from models where the waning immunity distributions matched those of the data simulation model were more accurate (Supplementary Figure \ref{fig:ParamPlots4}). We note that our ability to distinguish model fits through large differences in AIC was facilitated by the large volume of simulated data---daily data over three years. In practice, time series data would be much sparser, making it more difficult to distinguish the different models. 

\begin{figure}
        \centering
        \begin{subfigure}[b]{0.48\textwidth}
                \includegraphics[width=\textwidth]{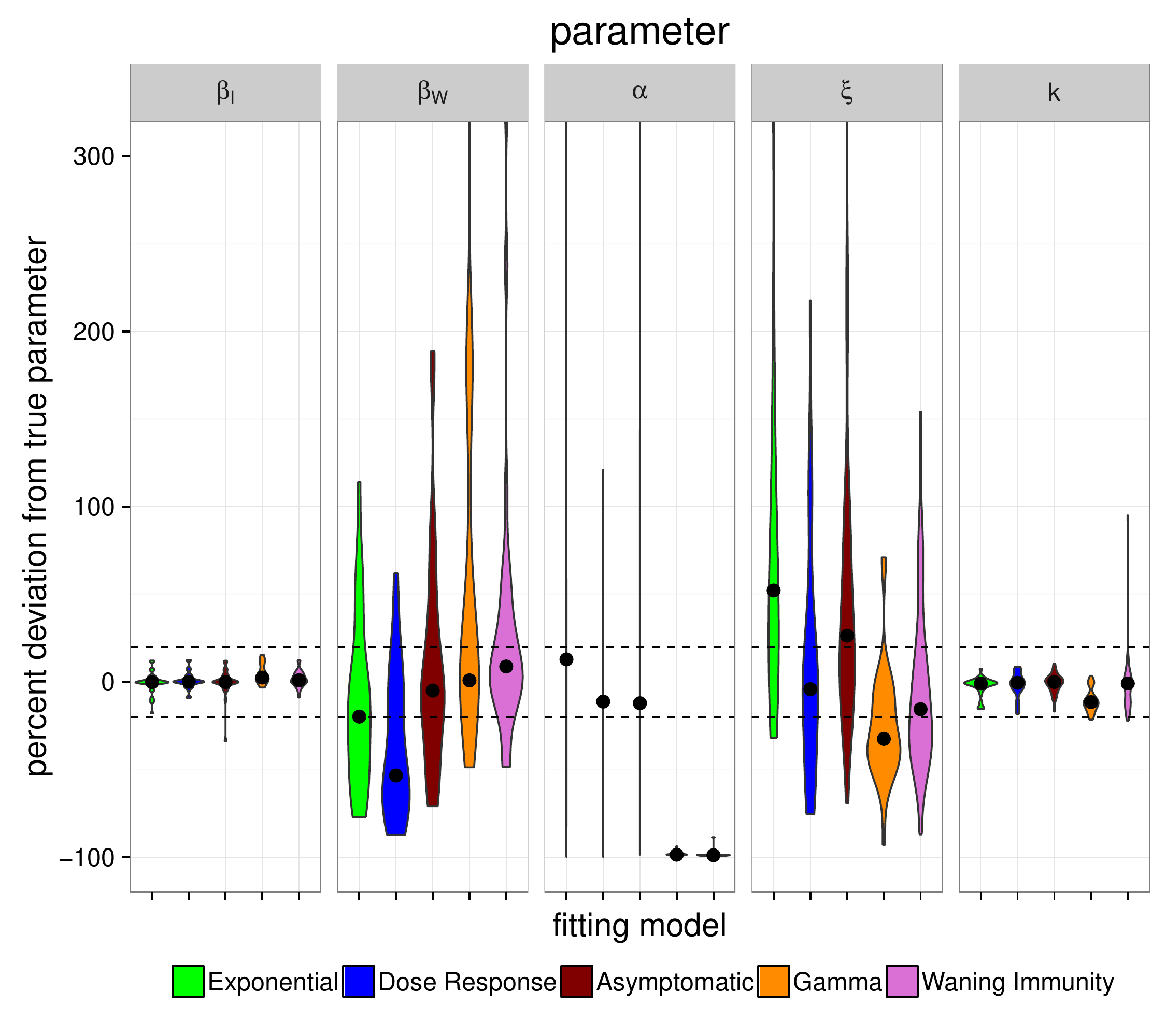}
        \end{subfigure}
        \begin{subfigure}[b]{0.48\textwidth}
                \includegraphics[width=\textwidth]{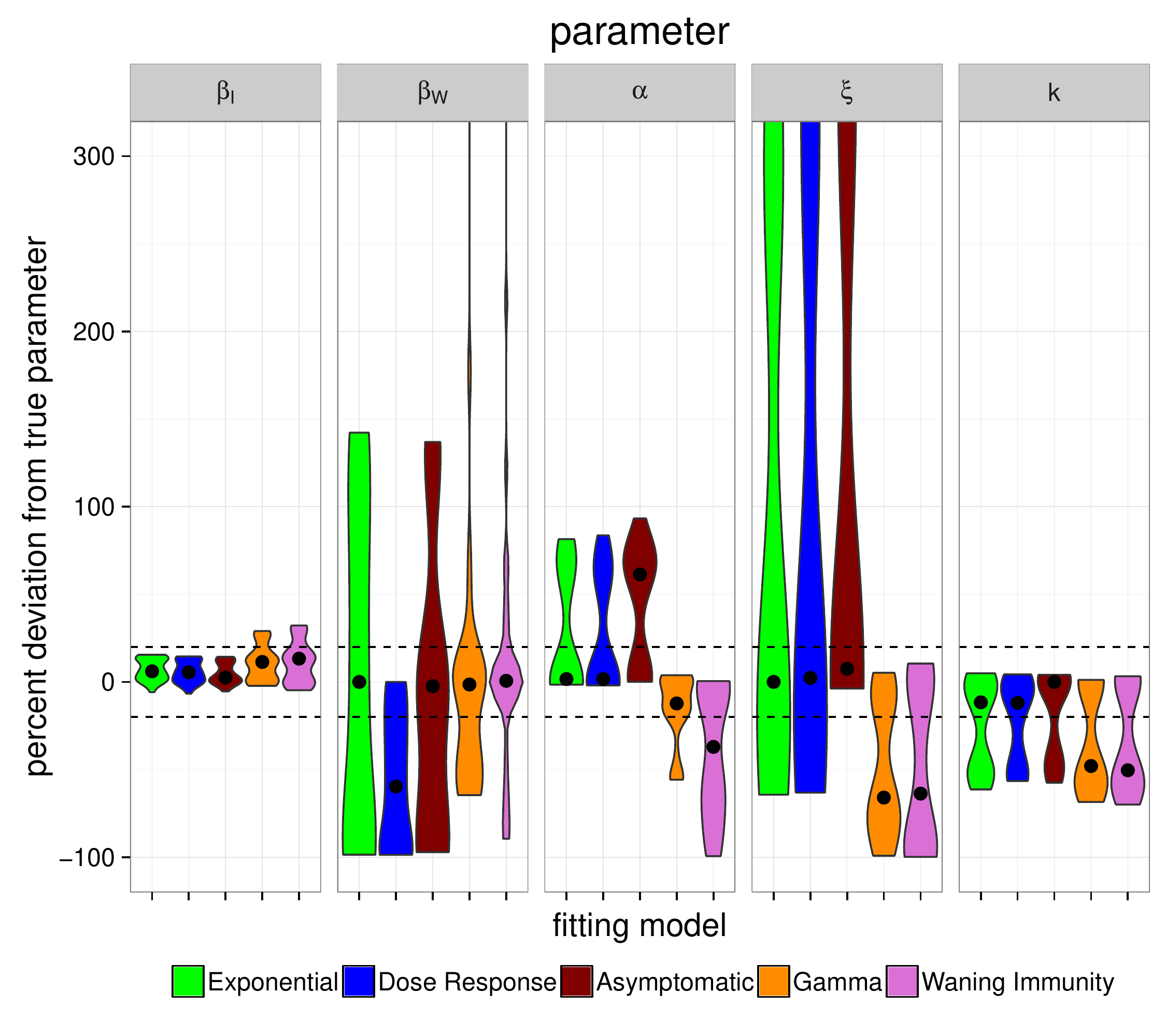}
        \end{subfigure}
        \begin{subfigure}[b]{0.48\textwidth}
                \includegraphics[width=\textwidth]{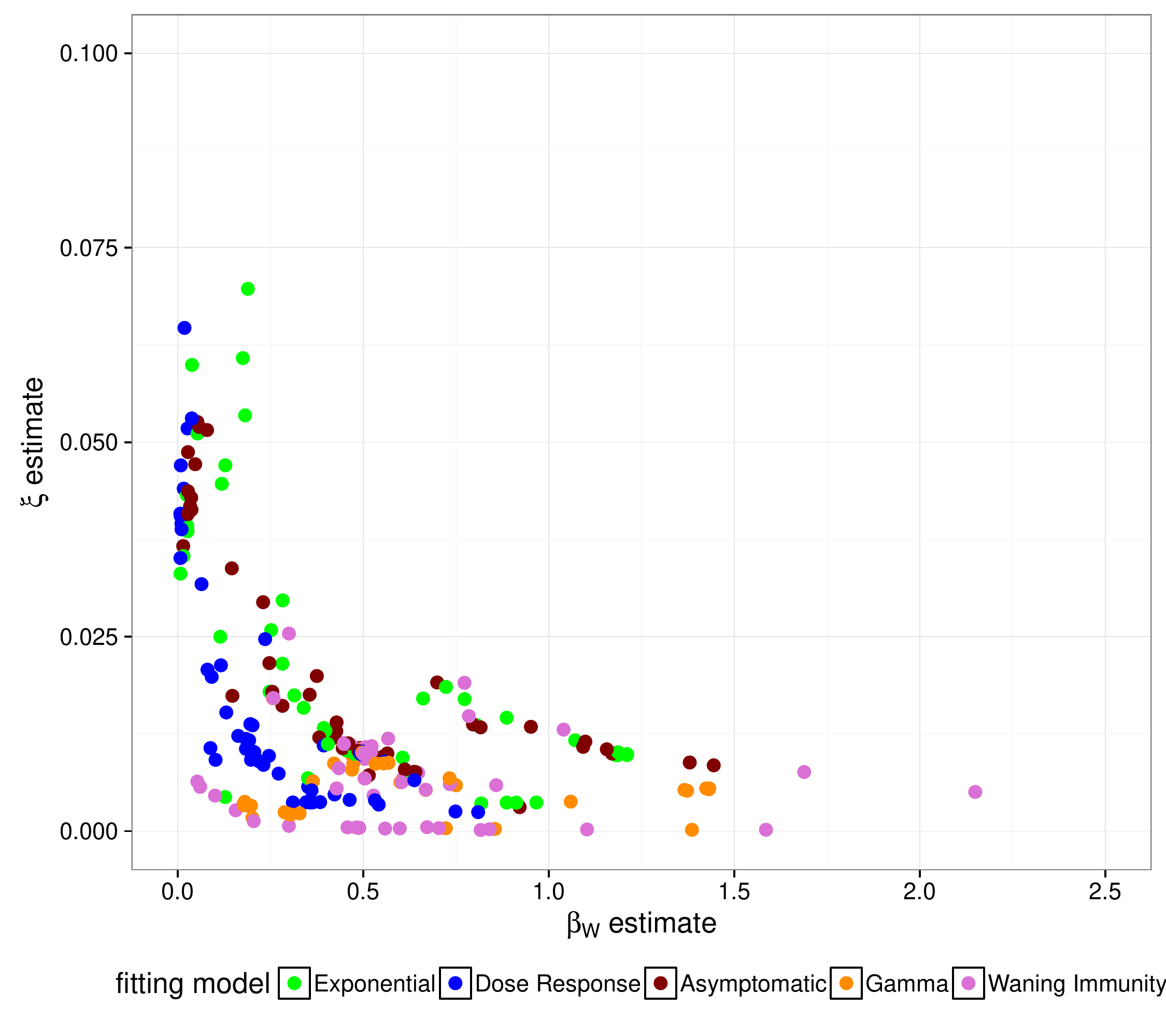}
        \end{subfigure}
        \caption{\emph{Top:} Percent deviation of estimates from true parameter values, grouped by parameter for the simulated 100-day data (left) and the simulated 3-year data (right). The model used to fit the data and estimate the parameter is indicated by color. The median across all estimates (i.e., across added noise type, simulation model, and data duration) is marked with a black point in the distribution and the black dashed line represents $\pm 20\%$ deviation. Distribution ranges for the $\beta_w$, $\alpha$, and $\xi$ subplots are trimmed for visibility. \emph{Bottom}: Scatterplot of $\beta_{W}$ and $\xi$ estimates by colored fitting model. Four data points in the distribution tails are truncated for ease of viewing. See Supplementary Figure \ref{fig:betaW-xi-identif} for the full plot.}\label{fig:ParamPlots}
\end{figure}

The parameter estimate accuracy did not appear to differ strongly as a function of the type of added noise (Supplementary Figure \ref{fig:ParamPlots4}). As with the 100-day data analyses, $\beta_I$ was the most accurate estimated parameter across all models, followed by $k$, although $k$ estimates appeared to be biased when the waning immunity distribution of the data simulation model did not match that of the fitting model  (Figure \ref{fig:ParamPlots}, Supplementary Tables \ref{NNIP3} - \ref{PNNP3}). In general, $\alpha$ was underestimated for gamma-distributed waning models fitted to exponentially-distributed waning models, and overestimated in the reverse case. The effect of the practically identifiable combination between $\beta_W$ and $\xi$ was also present in the 3-year data, with underestimates of $\beta_W$ tending to be compensated by overestimates of $\xi$ (Figure \ref{fig:ParamPlots}, Supplementary Tables \ref{NNIP3} - \ref{PNNP3}).

\subsection{Forecasts from simulated epidemic data} 

\begin{figure}
	\centering
	\includegraphics[width=0.9\textwidth]{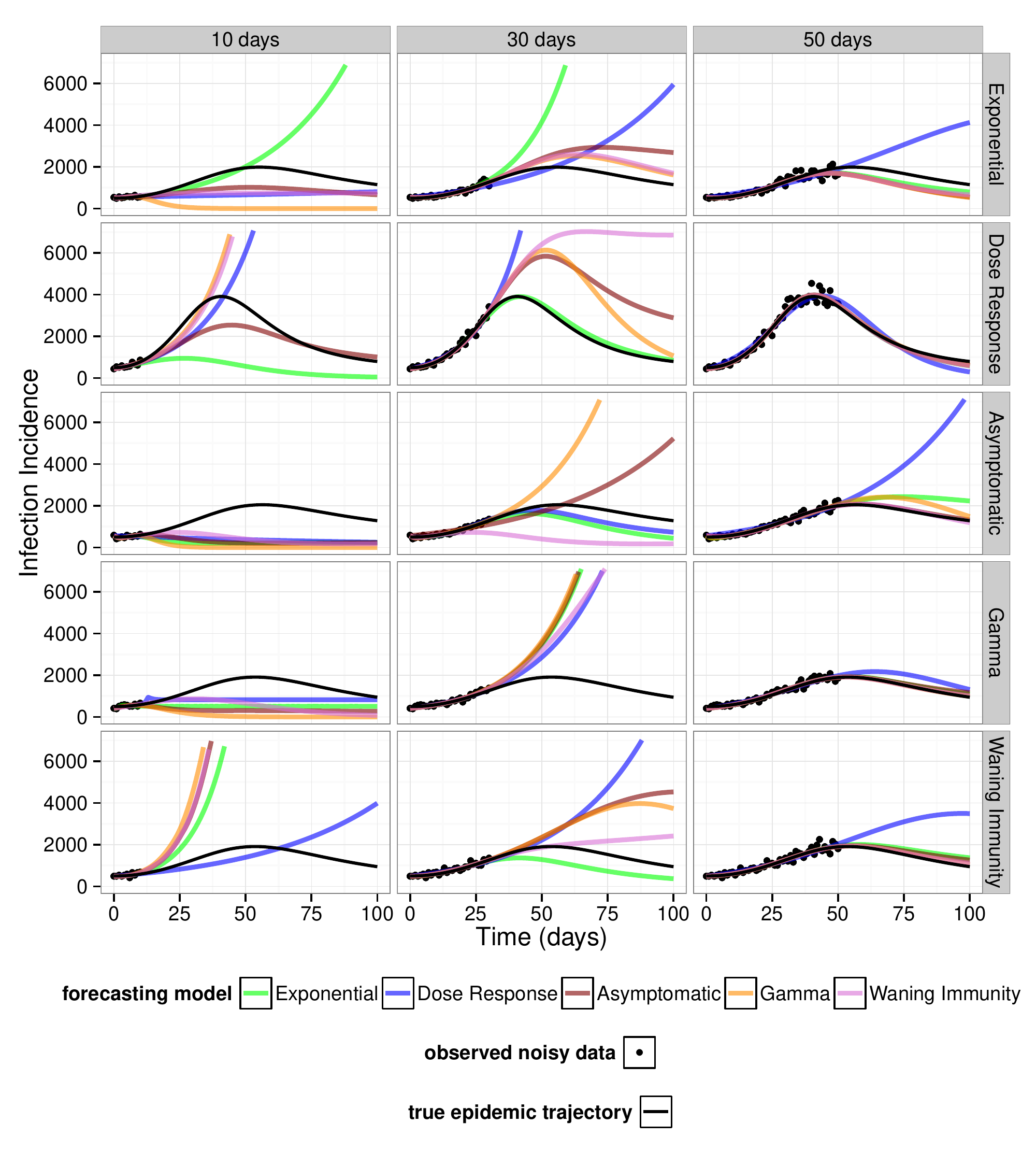}
	\caption{Forecasts to 100-day data (indicated by row) up though 10 days, (left column), 30 days (middle column), and 50 days (right column). Model fits are overlaid, thus obscuring some of the model fits in the figure.}
	\label{fig:forecasts}
\end{figure}

Epidemic trajectories were projected forward based on 10, 30, and 50 days of observed, simulated normal noise data for each study model. All forecasting models generated trajectories that were poorly representative of the true simulated data when only 10 or 30 data points were observed (Figure \ref{fig:forecasts}). In particular, even when forecasting models matched simulation models, these truncated epidemic datasets did not allow for accurate forecasts. While forecasts generated from 30 observations as compared to 10 observations were improved, performance remained poor overall, with only one out of 25 forecasts able to accurately capture both the timing and magnitude of the simulated epidemic peak (Exponential model forecasting from Dose Response simulated data). In fact, 11 of 25 forecasts continued to project exponential growth upwards past the 100-day simulation period.

Forecasts generated from 50 observed data points were significantly improved, with 16 of 25 forecasts roughly capturing the downward trajectory of the true data. Notably, most models underestimated the peak infection incidence and timing of the simulated Exponential model data, and the Dose Response model forecasts were least able to capture the correct disease dynamics of any model besides its own. As with the previous forecasts, a match between the simulation and forecasting model did not necessarily generate the most accurate forecast.

\subsection{Application to 2006 cholera outbreak in Angola} 
\begin{figure}
\centering
\includegraphics[width=0.4\textwidth]{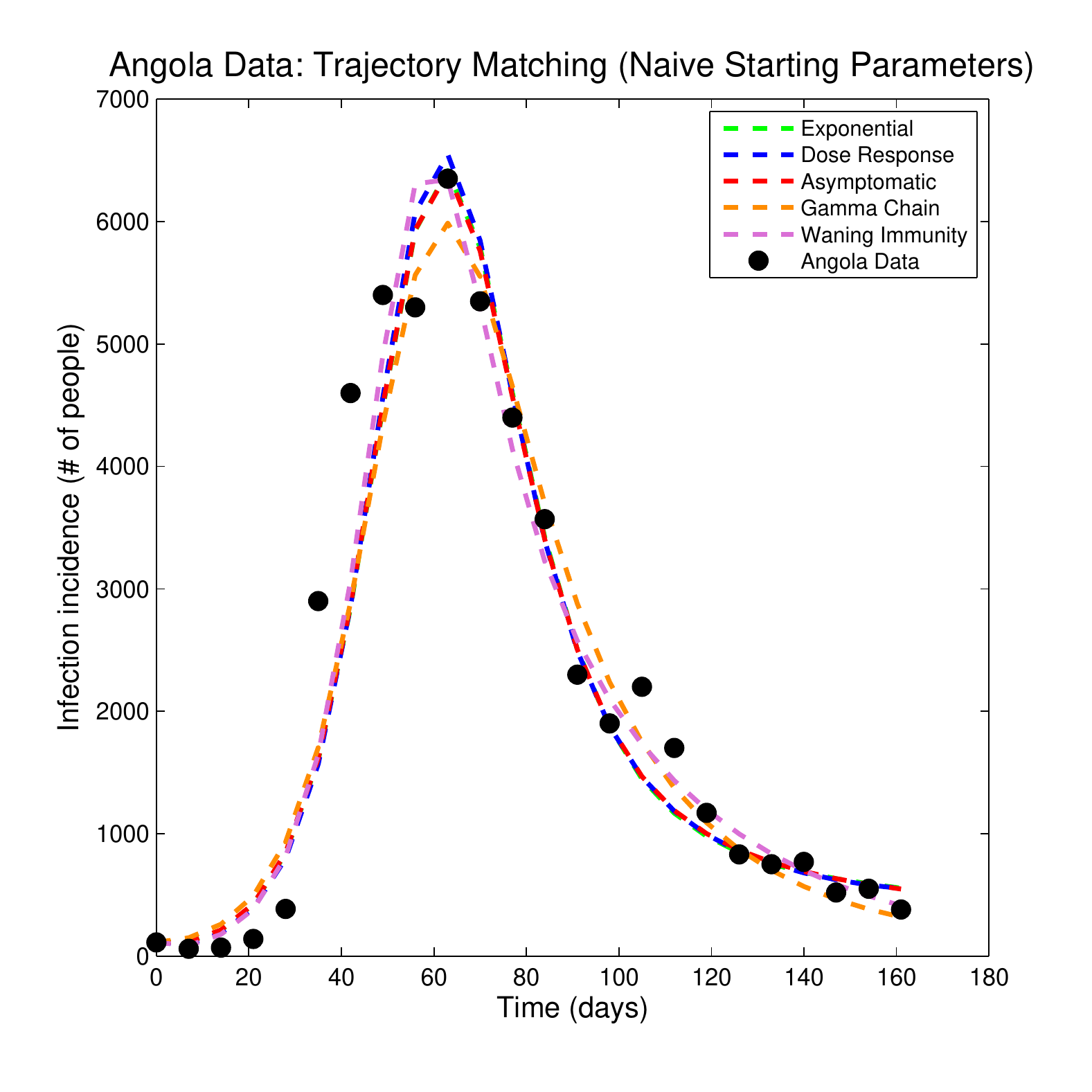}
\includegraphics[width=0.4\textwidth]{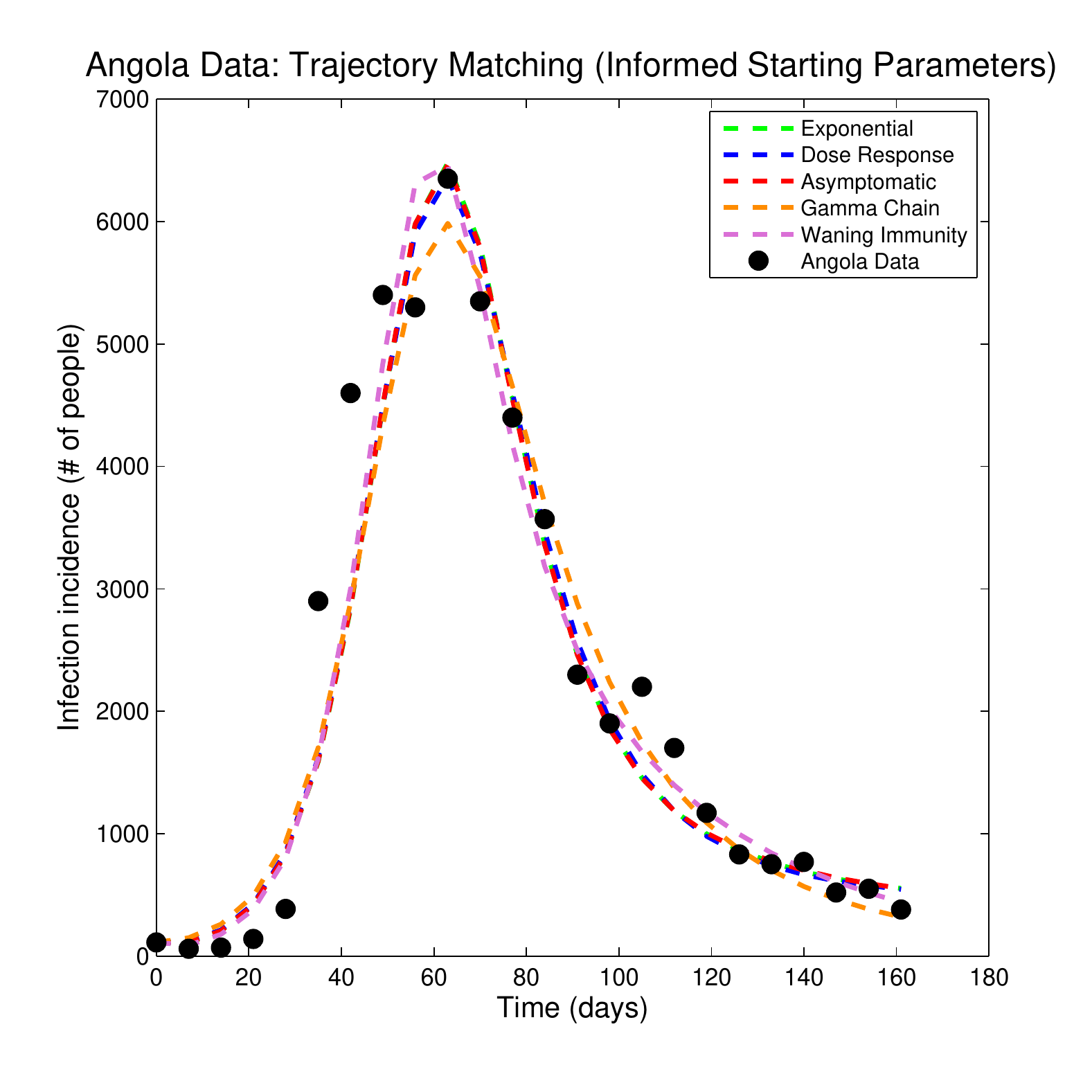}
\caption{\textit{``Naive''} (left) and \textit{``informed''} (right) starting parameter fits to 2006 Angola epidemic data. }
\label{fig:AngolaFits}
\end{figure}

We now turn our attention to data collected from a 2006 cholera outbreak in Angola. The fits of the models to the data are shown in Figure \ref{fig:AngolaFits}. The corresponding parameter estimates and AIC values for these fits, accounting for both informed and naive starting parameters, are given in Table \ref{fig:AP}. All models fit the data well, with similar trajectories that captured the overall shape, peak, and duration of the epidemic. The lowest $\%CV$s were found in the parameters $k$ and $\beta_W$, for both the `informed' and `naive' starting parameters (where `informed' here indicates the values chosen from the literature for data simulation, although these are not the true values as they were for the simulated data). The Waning Immunity model gave the lowest AIC for both naive and informed starting parameters (2940 and 2864, compared to AICs greater than 3400 for all other models), suggesting that the Waning Immunity model provides the best fit of the Angola data. Interestingly, this model showed very little direct transmission ($\beta_I$), suggesting that waterborne transmission may be enough to explain the outbreak.  

\begin{table}
\centering
\caption{Parameter estimates for the 2006 Angola epidemic data.}
\label{fig:AP}
\subcaption{\textnormal{\textit{Informed} starting parameters}\label{AInformed1}}{
    \tiny
   \centering
 \begin{tabular}{ | c | l | c | c | c | c | c | c | }
 \hline
Parameters/AIC &&	Exponential &	Dose Response & Asymptomatic	&	Gamma	&	Waning Immunity\\ \hline
$\beta_I$		&&	0.1460	&	0.1981	&	0.1600	&	0.2629	&	0.0358  \\ \hline
$\beta_W$ 	&&	0.6033	&	0.5055	&	0.6814	&	1.2485	&	1.0063 \\ \hline
$\alpha$		&&	0.0051	&	0.0020	&	0.0028	&	0.000026	&	0.000048 \\ \hline
$\xi$			&&	0.0544	&	0.0148	&	0.0413	&	0.0074	&	0.0530 \\ \hline
$k$			&&	1.37e-5	&	1.30e-5	&	1.37e-5	&	1.13e-5	&	1.95e-5 \\ \hline
$AIC$		&&	3667		&	3679		&	3657	 	&	4311 	&	2940 \\ \hline
$\Ro$		&&	3.00		&	5.85		&	3.03		&	6.04		&	4.17 \\ \hline
\end{tabular}} \\
\subcaption{\textnormal{\textit{Naive} starting parameters}\label{fig:ANaive1}}{
    \tiny
   \centering
       \begin{tabular}{ | c | l | c | c | c | c | c | c | }
    \hline
Parameters/AIC &&	Exponential	&	Dose Response	&	Asymptomatic	&	Gamma	&	Waning Immunity \\ \hline
$\beta_I$		&&	0.1865	&	0.0655	&	0.1827	&	0.2629	&	3.5e-8 \\ \hline
$\beta_W$ 	&&	0.8420	&	0.3101	&	0.7384	&	1.2479	&	0.9696 \\ \hline
$\alpha$		&&	0.0029	&	0.0050	&	0.0021	&	0.00017	&	0.000089 \\ \hline
$\xi$			&&	0.0257	&	0.0651	&	0.0312	&	0.0074	&	0.0668 \\ \hline
$k$			&&	1.33e-5	&	1.36e-5	&	1.33e-5	&	1.13e-5	&	1.96e-5 \\ \hline
$AIC$		&&	3731		&	3431		&	3696		&	4311		&	2864	 \\ \hline
$\Ro$		&&	4.11		&	3.36		&	3.68		&	6.04		&	3.88 \\ \hline
\end{tabular}} \\
\footnotesize{Note: The average estimates across starting conditions for $\alpha_S$ and $\alpha_A$ for the Asymptomatic model were 2.61e-9 and 0.005, respectively.}
\end{table}

\section{Discussion}

Mathematical modelers often need to make decisions on how to model transmission and loss immunity mechanisms, so it is critical to understand the robustness and sensitivity of results to realistic variations in model structures.
Toward that end, we have examined how model uncertainty and misspecification affect parameter estimates, model fits, and model forecasting ability. We considered five deterministic SIWR-based model structures, each including different hypothesized mechanisms of cholera transmission and loss of immunity, using both simulated data and data from the 2006 cholera epidemic in Angola. We found that goodness-of-fit criteria were unable to distinguish misspecified model fits from those of the true model for simulated epidemic data, and that forecasting from short-term data of one month or less was universally poor, even when the model that generated the data and the model that produced the forecast were the same. However, model fits for long-term data were found to be specific to the type of waning immunity that was implemented (i.e., exponential- vs. gamma-distributed waning immunity). Moreover, some parameters were consistently estimated across all models and datasets, suggesting that it may be possible to estimate some parameters even when the underlying mechanistic model is unknown. 

We showed that each model was able to capture the epidemic trajectories for all datasets, real and simulated, regardless of the source of the simulated data, type of noise, or starting parameters. Indeed, often the best fit to a given epidemic dataset did not come from the model which generated it. This suggests that it may not be possible to infer the true underlying mechanisms of loss of immunity and transmission solely from epidemic cholera incidence or prevalence data.  
This highlights the distinction between the issues of identifiability (parameter uncertainty) and distinguishability (model uncertainty)---all model parameters were structurally identifiable (excepting the parameter combination for the Asymptomatic model), but the model structure itself may not be possible to infer from the data.

When a longer, 3-year data set was considered---as noted above, in this case it was generally possible to distinguish exponentially-distributed waning models (Exponential, Dose Response, and Asymptomatic) from gamma-distributed waning models (Gamma and Waning Immunity). Indeed, the best-fit model to the simulated 3-year data was often the model which generated it, although typically the other models with the same waning distribution had very similar AIC values (differences $\leq 5$), such that the models may not be distinguishable in practice. Additionally, the simulations here assumed frequent data taken daily, while more realistic data over three years would likely be weekly or monthly, making distinguishabilty by goodness of fit even less likely. Nonetheless, these results do suggest that some degree of model selection, at least for loss of immunity mechanisms, may be possible with longer term data sets. Other additional data might add to the distinguishability of realistic models, such as data on pathogen shedding, bacterial concentration in water sources, or immunity/serology data such as antibody titers. 

Our study design also enabled the systematic assessment of parameter estimates under conditions of model misspecification. As noted above, all parameters were structurally identifiable (apart from the identifiable combination for $\beta_{IA}$ and $\beta_{IS}$), but unfortunately, identifiability analyses only inform the validity of parameter estimates when the true underlying model is known. When we consider the variation of the parameter estimates across different model misspecification scenarios, the picture is more complex. As $\beta_I$ was recovered quite well for all simulated data sets, this suggests that the ``fast'' transmission parameter and reporting rate/population factor may be accurately estimated from any of the common SIWR model variants, even if the true underlying disease mechanisms are unknown. Interestingly, $k$ was more accurately recovered from the misspecified models when the 100-day data sets were used rather than the 3-year data (Figure \ref{fig:ParamPlots}), suggesting that when model misspecification is present, more data may not always be better. It is likely that the bias in $k$ was being used to help correct for fitting issues that came from using an incorrect model structure, but this would be difficult to assess in practice with real data. 

Previous work noted that in the SIWR model, $\beta_W$ and $\xi$ became practically unidentifiable when noisy data was considered, and formed a practically identifiable combination \cite{Eisenberg2013}. It was conjectured there that this issue of practical identifiability may extend more broadly to many models that include a waterborne or environmental transmission pathway. Indeed, here we confirmed that $\beta_W$ and $\xi$ are practically unidentifiable and form a practically identifiable combination across multiple SIWR-based models. This underscores the need for improved data collection regarding waterborne cholera transmission in order to better inform modeling and public health intervention efforts. However, we note that while the two parameters were practically unidentifiable in all models, the curve of the identifiable combination traced in Figure \ref{fig:ParamPlots} was quite similar across all models and data sets. This suggests that while these two parameters may be practically unidentifiable in all models, their combination may be more tightly estimated across models and datasets, similarly to $\beta_I$ and $k$. In part due to this practical unidentifiability of $\beta_W$, estimates of $\Ro$ varied widely across all models, both for the epidemic data in Angola and for all types of simulated data.

For the epidemic data, the estimates of $\alpha$ were uniformly underestimated for both the Gamma and Waning Immunity models when fitted to epidemic data. This may be due to the fact that both of these models have a gamma-distributed type of waning process, meaning that significant waning would be unlikely to be observed over the 100-day epidemic period. This may have resulted in highly insensitive $\alpha$ estimates for these two models, so that the optimization algorithm wandered to very low values for $\alpha$. In general, in the 3-year datasets, $\alpha$ was underestimated for gamma-distributed waning models fitted to exponential waning models, and underestimated in the reverse case. This may be because the gamma-distributed waning models used a slower loss of immunity to generate the flat trajectories after the initial peak, while the exponential-distributed waning models may have used a more rapid $\alpha$ to generate the subsequent peaks in the disease trajectory. 

The new Waning Immunity model presented here had the best absolute fit to the Angola epidemic data, but as previously noted, this does not necessarily indicate that the waning immunity mechanism was the true driver of these epidemic dynamics. Longitudinal data on sequential infections of cholera or on antibody titers over time are needed to validate the idea that recently infected individuals regain susceptibility progressively after recovery. If waning immunity does indeed best represent epidemic cholera dynamics, the parameter estimates in our study suggest the hypothesis that the Angola epidemic was spread primarily by waterborne transmission; the Waning Immunity model estimates for $\beta_I$ were at least two orders of magnitude smaller than those for $\beta_W$ (and much smaller than in any other model). We note that this somewhat contradicts a previous study using the base SIWR model, which showed stronger waterborne transmission but still significant direct transmission \cite{Eisenberg2013}. 

Further work is needed to examine the ability for SIWR-based models to project epidemic trajectories based on limited and noisy data early in an outbreak. Only projections that took place after the epidemic peak appeared to capture any semblance of disease prevalence; projections were also poor when the forecasting model matched the simulation model, perhaps suggesting that it may not be possible to forecast accurately with deterministic SIWR models until the latter half of the epidemic. This is to be expected in the very early part of the epidemic, during the exponential growth phase, as during this phase the epidemic is largely linear on a log-scale and so can be explained by only two parameters. This idea is borne out in other real-time forecasting efforts---for example, efforts to forecast the trajectory of the 2014 Ebola epidemic in West Africa met with difficulty \cite{Butler2014, Rivers2014Nature, weitz2015modeling, Eisenberg2016}. We note that our results illustrate how agreement across a range of models may not guarantee accuracy; however once the epidemic peak was observed, all models tended to converge on similar and more accurate forecasts. This suggests that once sufficient data is obtained, forecasting is possible even if the model structures do not match the underlying mechanisms. Future work may examine whether early outbreak forecasting, discrimination of model structures, and inference from forecasting comparisons may be aided by additional data on weather and climate variables such as rainfall or sea surface temperature \cite{pascual2000, pascual2008, Reiner2012, rinaldo2012,Eisenberg2013b}.  

There are several limitations and future directions for this work, to consider different model structures, frameworks, and empirical datasets. Alternative cholera transmission mechanisms (such as hyperinfectivity \cite{alam2005, hartley2006, pascual2006, shuai2012,millerneilan2010}), spatial structure \cite{bertuzzo2010, tuite2011, bertuzzo2011, Eisenberg2013c}, and climate drivers \cite{pascual2000, pascual2008, Reiner2012, Eisenberg2013b, rinaldo2012} were also realistic features that could have been added to the suite of model structures we considered. Model comparison with hyperinfectivity and two potential environmental reservoirs were considered in study by Rinaldo et al. \cite{rinaldo2012}, using a larger, detailed spatial model of the 2010 Haiti cholera epidemic (including human and pathogen movement, river networks, climate drivers, etc.). We also did not evaluate our conclusions outside of the deterministic ODE framework, although partial differential equation (PDE), agent-based, and stochastic models are also frequently used in similar contexts \cite{Righetto2011, Alexanderian2011, Reiner2012, king2008}. In addition, most of the parameter values were motivated from a single outbreak dataset (Angola, 2006), and may not reflect the breadth of parameter space for cholera epidemics. Additional work to further examine the identifiability, uncertainty of these models, and to test alternative parameter estimation methods (e.g. Bayesian and/or global approaches) would be warranted as well. It would also be useful to consider alternate data assumptions (e.g. weekly or monthly data frequency, additional noise and error assumptions), and even to consider the generalizability of these results to other infectious disease models. Further, here we only considered the effects of model structure on parameter estimation and forecasting. Additional work could use model comparison to evaluate the effects of model and parameter uncertainty on the optimal control of interventions, similar to that proposed by Akman and Schaefer \cite{akman2015evolutionary}. Their focus on the optimal balance of vaccination and sanitation strategies may be particularly interesting, as the lack of distinguishability between models may be less problematic if multiple models yield consistent control strategies. 

In conclusion, our study presents a systematic framework for the comparison of model mechanisms and inference that can be applied broadly in infectious disease epidemiology. Modeling requires simplifying assumptions to be an effective tool, but we show here that the choice of simplifications can have significant effects on the parameter estimates, and that it may be difficult or impossible to use fit to data as a criteria to distinguish between plausible simplified models. Of course, in real-world contexts, there is unlikely to be a `true' model in the sense that we considered in our simulated data---instead, each of the models represents a simplification, albeit with realistic elements. We note that our results also show that the most realistic model---a `super-model' containing all of the elements considered individually across the five models here---would almost certainly be unidentifiable from case data alone. Public health decision makers desire tools that demonstrate the sensitivity of results to model assumptions (on the analysis side) and policy decisions (on the implementation side). Attempts to address the divide between decision maker needs and modelers' ability to present useful, actionable information are on the rise, with the development of adaptive management, cost-benefit, and game theoretic frameworks in public health \cite{galvani2007long, merl2009statistical, yaesoubi2011dynamic, fenichel2011adaptive, shea2014adaptive, moore2015optimal, hayashi2016effects}. In demonstrating that goodness-of-fit is not sufficient to infer transmission and immunity mechanisms in a disease system, and that some parameter estimation may be valid even when mechanisms remain elusive, we similarly seek to inform modelers and decision makers about the relative utility of different analyses for specific types of policy questions. Our research raises questions about the appropriateness and equality of model structures and types for epidemiological inference, and future work should seek to further characterize the sensitivity of public health decision making in the context of model choice.

\section{Acknowledgments} 

This paper began as a group project at the joint NIMBioS-MBI-CAMBAM Graduate Summer School in 2013 (\url{http://www.nimbios.org/education/WS_grad2013})---many thanks to Suzanne Lenhart, Paul Hurtado, Ariel Cintron-Arias, and the other co-organizers and presenters. The summer school was hosted by the National Institute for Mathematical and Biological Synthesis (NIMBioS), an Institute sponsored by the National Science Foundation through NSF Award DBI-1300426, with additional support from The University of Tennessee, Knoxville. NIMBioS hosted the summer school jointly with Mathematical Biosciences Institute (MBI) at The Ohio State University (National Science Foundation grant DMS 68 0931642) and the Centre for Applied Mathematics in Bioscience and Medicine (CAMBAM) at McGill University. We thank the World Health Organization Cholera Task Force for providing us with data from the 2006 cholera outbreak in Angola. The NIMBioS summer school provided support to ECL, MRK, BO, SMA, KM, and MCE, and SMA acknowledges partial travel support from PIMS IGTC, Canada for summer school attendance. This work was supported by the National Science Foundation grant OCE-1115881 (to JHT and MCE) and the National Institute of General Medical Sciences of the National Institutes of Health under Award Number U01GM110712 (to MCE). The content is solely the responsibility of the authors and does not necessarily represent the official views of the National Institutes of Health.

{\small
\bibliographystyle{ieeetr}
\bibliography{reflib.bib}
}

%
%
%
%
%
%
%
%
%
%
%
%
%
%
%
%
%

\pagebreak
\beginsupplement
\section{Supplementary Information}

\subsection{Dimensional and nondimensional forms of each model}
As in \cite{tien2010, Eisenberg2013}, we use the nondimensional forms for each model throughout in order to reduce issues of model unidentifiability and simplify their presentation. Here we briefly introduce the dimensional and nondimensional forms for each model. We begin with the original SIWR model of Tien and Earn \cite{tien2010}. The dimensional form is given as:
\begin{equation}
\begin{aligned}
S' &= \mu N - b_W W S - b_I S I - \mu S\\
I' &= b_W W S + b_I S I - \gamma I - \mu I\\ 
W' &= \sigma I - \xi W\\
R' &= \gamma I - \mu R
\end{aligned}
\end{equation}
with constant population $N = S+I+R$.  Here $b_W$ represents the rate at which susceptibles become infected due to contact with pathogen in the water ($W$), $b_I$ is the transmission parameter for direct transmission, $\sigma$ is the rate that infected individuals shed pathogen into the water, $\xi$  the rate of decay of pathogen in the water, $\gamma$ the recovery rate, and $\mu$ the population turnover rate. We nondimensionalize the model by letting $s = S/N$, $i = I/N$, $r = R/N$, $w = \frac{\xi}{\sigma N} W$, $\beta_I = b_I N$, and $\beta_W = b_W N\sigma/\xi$:
\begin{equation}
\begin{aligned}
s' &= \mu - \beta_Wsw - \beta_Isi - \mu s\\
i' &= \beta_Wsw + \beta_I si -\gamma i - \mu i\\
w' &= \xi \left(i-w\right)\\
r' &= \gamma i - \mu r 
\end{aligned}
\end{equation}
The Exponential model can be nondimensionalized using the same scaling as given for the SIWR model (since it is the same model but with an added loss of immunity term). Similarly, an equivalent scaling can be used for the Gamma and Waning Immunity models, with $r_j = R_j$ and $s_j = S_j$ for $j=1,\dots,n$, for each model respectively. 

For the Dose Response model, the dimensional form of the model is given by:
\begin{equation}
\begin{aligned}
S' &=  \mu N -b_ISI-\frac{b_{Wmax}W}{\tilde{K} + W}S - \mu S + \alpha R\\
I' &=  b_ISI + \frac{b_{Wmax}W}{\tilde{K} + W}S - \gamma I - \mu I\\
W' &= \sigma I - \xi W\\
R' &= \gamma I - \mu R - \alpha R
\end{aligned}
\end{equation}
To nondimensionalize, we take $s = S/N$, $i = I/N$, $r = R/N$, $w = \frac{\xi}{\sigma N} W$, and $\beta_I = b_I N$ as before, but now we take $K =  \frac{\xi}{\sigma N}\tilde{K}$ and simply let $\beta_{Wmax} = b_{Wmax}$, yielding the final form given in \eqref{eq:DoseResponse}. (We note that there are other options for nondimensionalization, e.g. letting $w = W/\tilde{K}$, but for consistency with the other models we opted for this.) 

Lastly, for the Asymptomatic model, the dimensional form of the model is given by:
\begin{equation}
\begin{aligned}
S' &= \mu N - b_W W S - b_{IS} S I_S  - b_{IA} S I_A - \mu S\\
I_S' &= q(b_W W S + b_{IS} S I_S + b_{IA} S I_A) - (\gamma + \mu) I_S\\ 
I_A' &= (1-q)(b_W W S + b_{IS} S I_S + b_{IA} S I_A) - (\gamma + \mu) I_A\\ 
W' &= \sigma_S I_S  + \sigma_A I_A - \xi W\\
R_S' &= \gamma I_S - \mu R_S\\
R_A' &= \gamma I_A - \mu R_A
\end{aligned}
\end{equation}
The nondimensionalization is similar to the previous, except that we note that $I_S$ and $I_A$ are at a fixed ratio to one another, with the fraction $q$ of all infections being symptomatic and $1-q$ being asymptomatic. This is because the infected initial conditions and the new cases are both at this ratio, and both equations have the same loss rates ($\gamma + \mu$). Thus we take $I = I_S + I_A$, and write $W' = (q \sigma_S + (1-q) \sigma_A) I - \xi W$. Then let $\sigma = (q \sigma_S + (1-q) \sigma_A)$, and we can take an analogous rescaling as for the SIWR model above, letting $s = S/N$, $i_S = I_S/N$, $i_A = I_A/N$, $r_S = R_S/N$, $r_A = R_A/N$, $w = \frac{\xi}{\sigma N} W$, $\beta_{IS}= b_{IS} N$, $\beta_{IA}= b_{IA} N$, and $\beta_W = b_W N\sigma/\xi$. This rescaling yields the final model given in \eqref{eq:Asymptomatic}.

\subsection{Baseline parameter values and estimation details} \label{SuppParameterSection}
As noted in the Methods section, we made a few modifications of the original parameters for the SIWR model fitted to the Angola epidemic, based other studies in the literature. We increased the value of the pathogen decay rate $\xi$ to 1/100 days (pathogen lifetime approximately three months), to better match the faster decay rates measured in other studies \cite{tuite2011, Feachem1983, Hood1981, codeco2001}.
As $\xi$ and $\beta_W$ form a practically identifiable combination (Figure \ref{fig:betaW-xi-identif}) \cite{Eisenberg2013}, we also reduced $\beta_W$ from ~1 \cite{Eisenberg2013} to 0.5, which resulted in approximately the same epidemic shape and gave similar values for both waterborne and direct transmission parameters. 
The values of $\beta_I$ and $k$ were based on previous Angola data estimates\cite{Eisenberg2013}, with $k$ adjusted after our changes to $\beta_W$ and $\xi$ to make roughly the same size epidemic as in \cite{Eisenberg2013}. 

Additionally, for parameter estimation, we found that in the normal noise case, maximum likelihood (i.e. weighted least squares) often yielded poor performance when the simulation model did not match the fitting model (e.g. Gamma fitted to Asymptomatic data). As the true underlying error distribution would not be known a priori, we instead fitted using weighted least squares with a weight of $\sigma^2 = data$, as suggested in \cite{Eisenberg2013} and used for fitting the Angola data here. This gives a balance between weighted (where $\sigma^2$ would be proportional to $(data)^2$) and unweighted least squares (where the weights are proportional to $(data)^0$, i.e. constant weighting). We also note that for the Angola data, the prevalence measurement equation used is only an approximation for the weekly incidence data (similar to previous studies \cite{Eisenberg2013, Eisenberg2013b}). However the approximation was quite close, with a total (cumulative) error of $1\%$ over the entire epidemic (simulating weekly incidence vs. using the prevalence approximation using the default parameters).

\subsection{Parameter estimation results}

\subsubsection{100-day epidemic data}
\begin{figure}
        \centering
        \begin{subfigure}[b]{0.46\textwidth}
                \includegraphics[width=\textwidth]{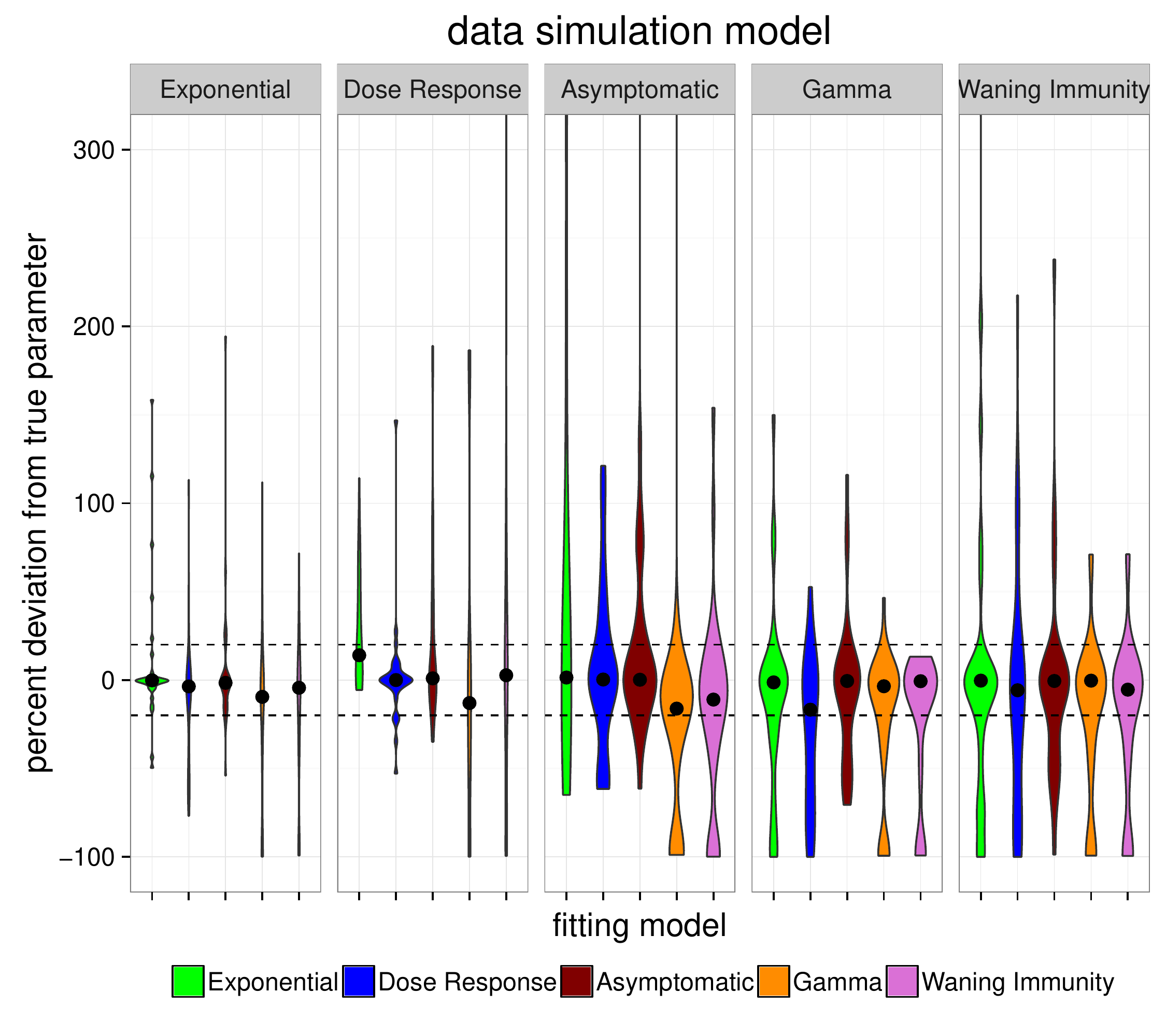}
        \end{subfigure}
        \begin{subfigure}[b]{0.46\textwidth}
                \includegraphics[width=\textwidth]{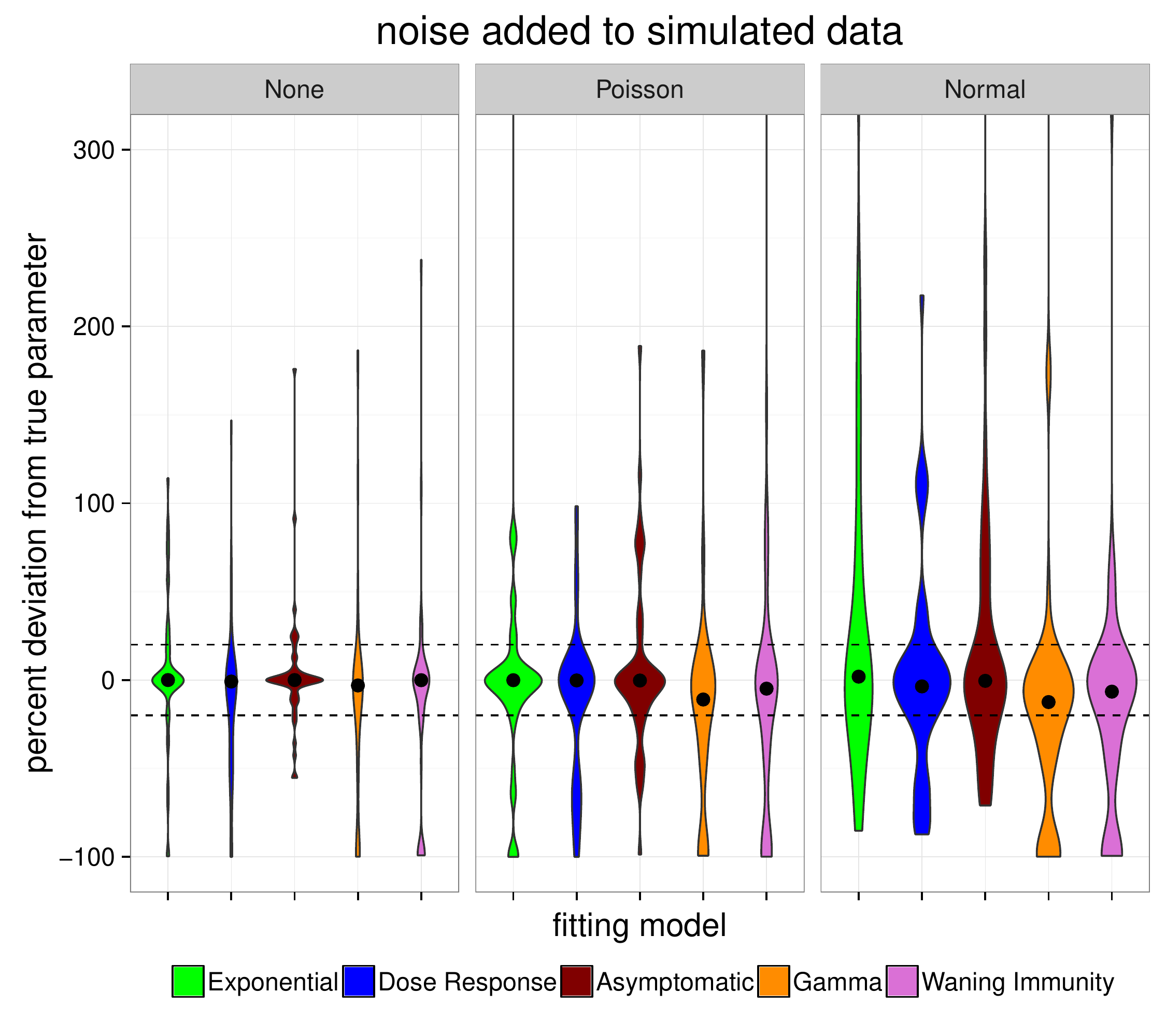}
        \end{subfigure}
        \caption{Percent deviation of parameter estimates from true values, grouped by model used to simulate the initial data (left) and type of noise added to data (right) for the simulated 100-day data. The model used to fit the data and estimate the parameter is indicated by color. The median across all estimates (i.e., across added noise type and simulation data) is marked with a black point in the distribution and the black dashed line represents $\pm 20\%$ deviation. Distribution ranges are truncated for visibility.}\label{fig:ParamPlots2}
\end{figure}

\begin{figure}
        \centering
        \begin{subfigure}[b]{0.46\textwidth}
                \includegraphics[width=\textwidth]{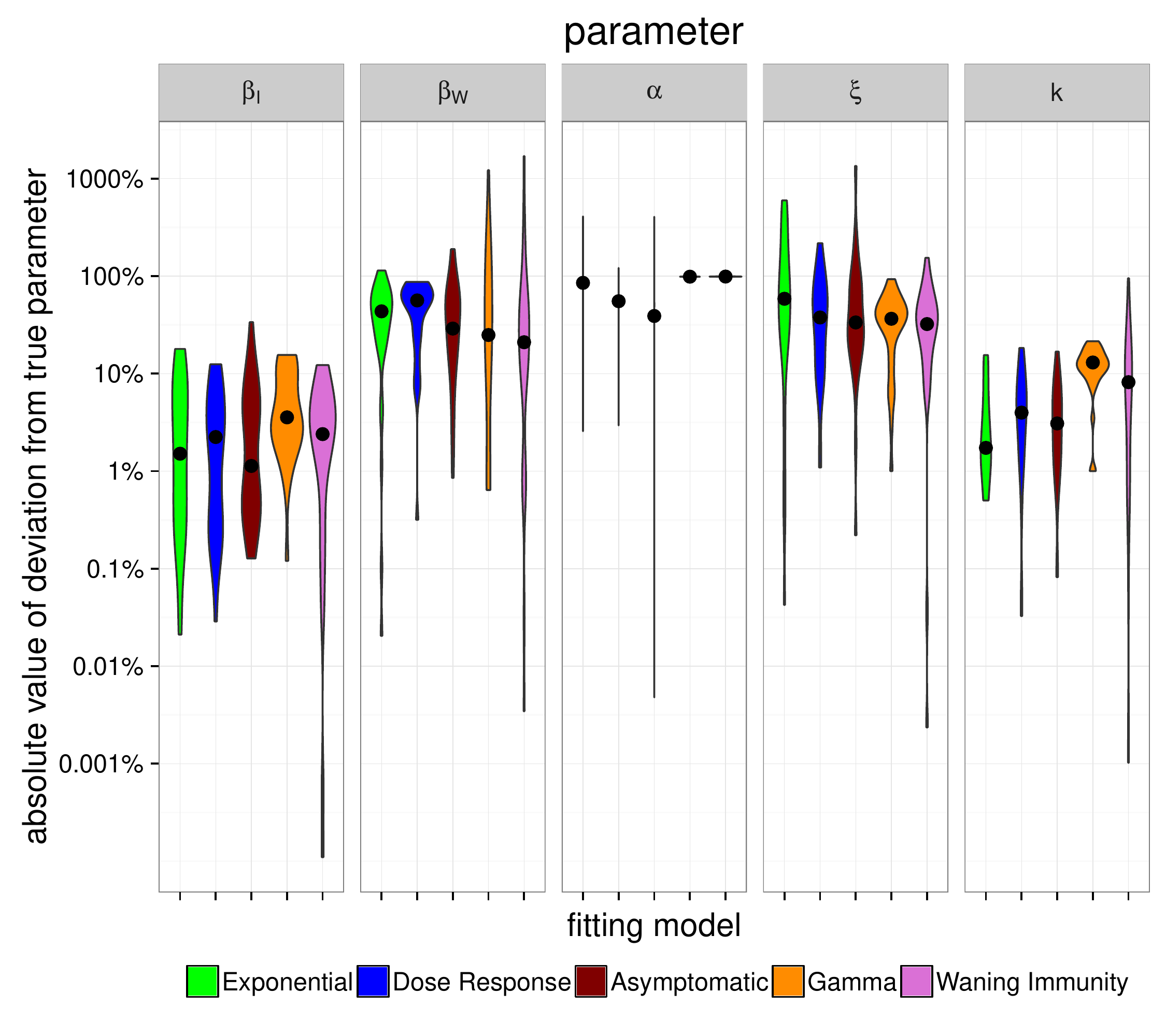}
        \end{subfigure}
        \caption{Absolute percent deviation of parameter estimates on a log scale, grouped by parameter. The model used to fit the data and estimate the parameter is indicated by color.}\label{fig:ParamPlots3}
\end{figure}

In addition to pooling parameter estimate deviations across parameters (Figure \ref{fig:ParamPlots}), we also grouped them by the model used to generate the data and the type of noise added to the simulated data (Figure \ref{fig:ParamPlots2}). No data simulation model enables better recapture of parameter estimates, but the Exponential and Dose Response models had notably smaller deviations from the true parameter when fitting their own simulated data. In addition, the Gamma and Waning Immunity models recaptured true parameter values from the Gamma simulated data well (less than 10\% deviation). Median deviations from the true parameter remained relatively constant across added noise categories, but deviations tended to be smaller with less noise, as evidenced by the longer lower tails of the distributions in the `None' and `Poisson' panels. As model complexity in the dimension of loss of immunity increased, the absolute deviation for the loss of immunity parameter ($\alpha$) decreased (Figure \ref{fig:ParamPlots3}).

\subsubsection{3-year long-term data}
\begin{figure}
        \centering
        \begin{subfigure}[b]{0.46\textwidth}
                \includegraphics[width=\textwidth]{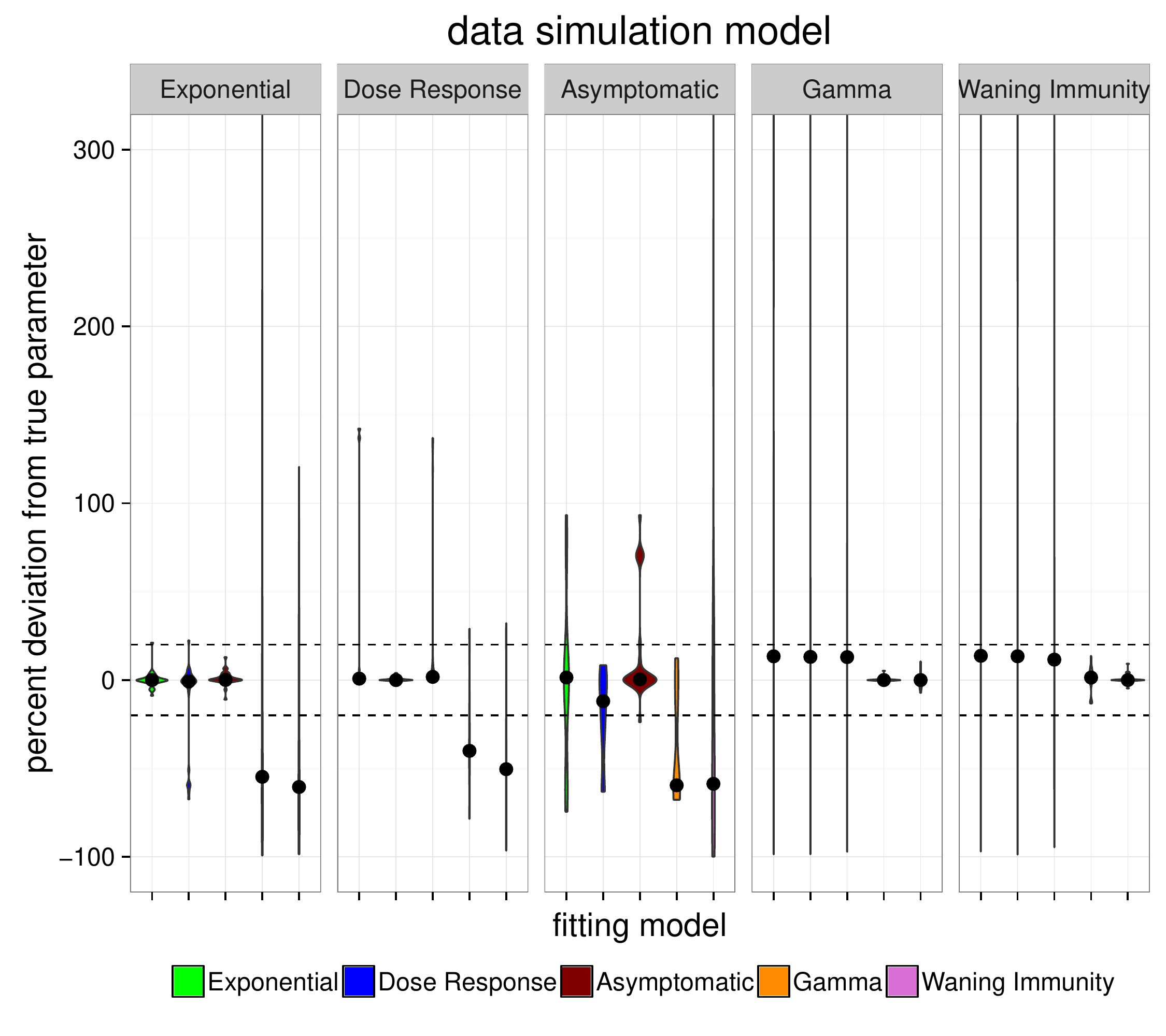}
        \end{subfigure}
        \begin{subfigure}[b]{0.46\textwidth}
                \includegraphics[width=\textwidth]{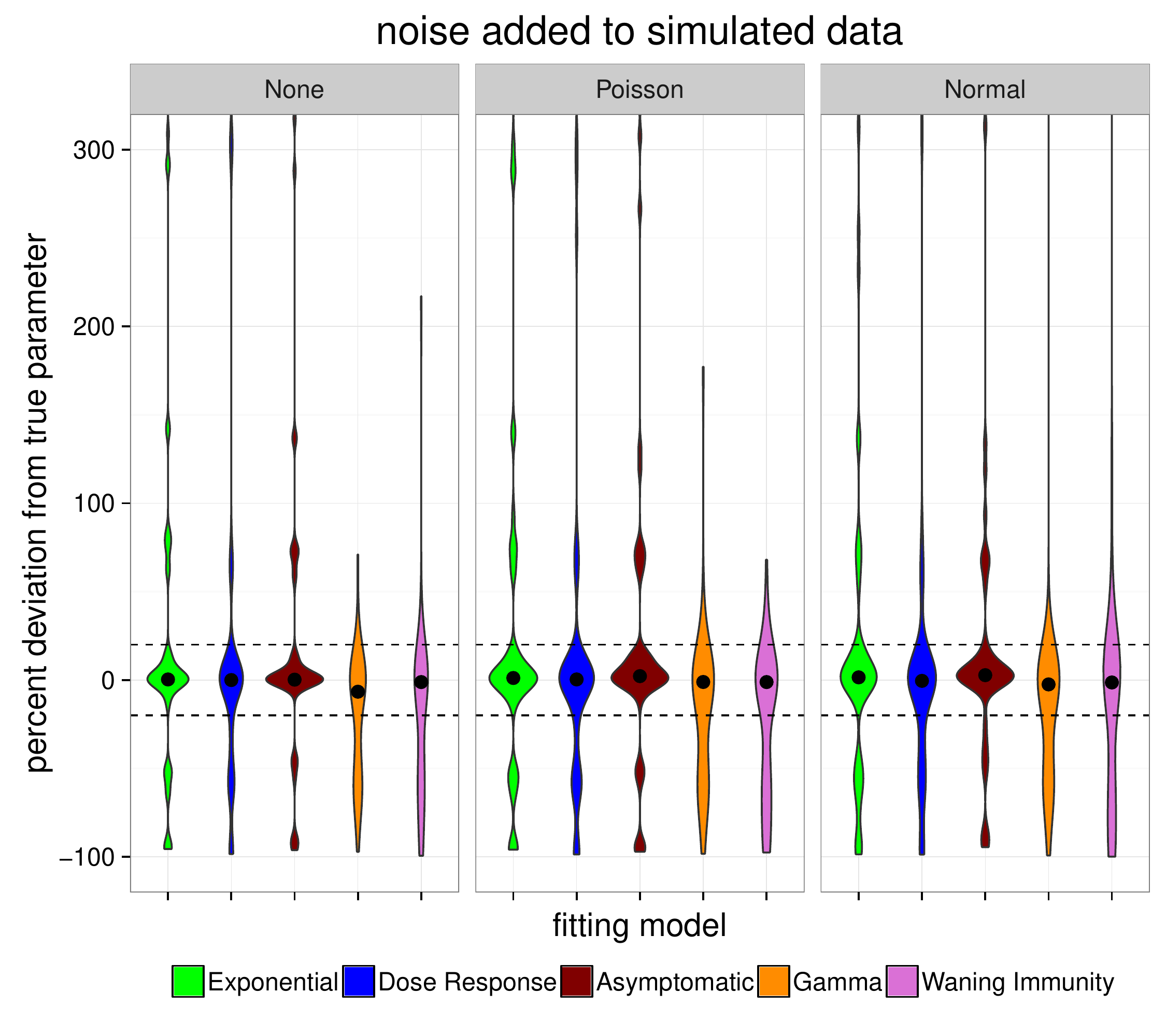}
        \end{subfigure}
        \caption{Percent deviation of parameter estimates from true values, grouped by model used to simulate the initial data (left) and type of noise added to data (right) for the simulated 3-year data. The model used to fit the data and estimate the parameter is indicated by color. The median across all estimates (i.e., across added noise type and simulation data) is marked with a black point in the distribution and the black dashed line represents $\pm 20\%$ deviation. Distribution ranges are truncated for visibility.}\label{fig:ParamPlots4}
\end{figure}

In examining the pooled parameter estimates for the simulated 3-year data, we found that parameter estimates were more accurate when models fit to data generated from models with the same distribution of waning immunity (exponentially-distributed waning immunity: Exponential, Dose Response, and Asymptomatic models; gamma-distributed waning immunity: Gamma and Waning Immunity models) (Figure \ref{fig:ParamPlots4}). There were no discernible patterns in parameter estimate accuracy among added noise types (Figure \ref{fig:ParamPlots4}). We also note that for the case of the Gamma Model fitted to data from the Exponential model, the convergence criterion for optimization had to be increased to ensure convergence.

\clearpage

\subsection{Model fits to simulated epidemic (100-day) data}
\begin{figure}[H]
	\centering
	\includegraphics[width=0.9\textwidth]{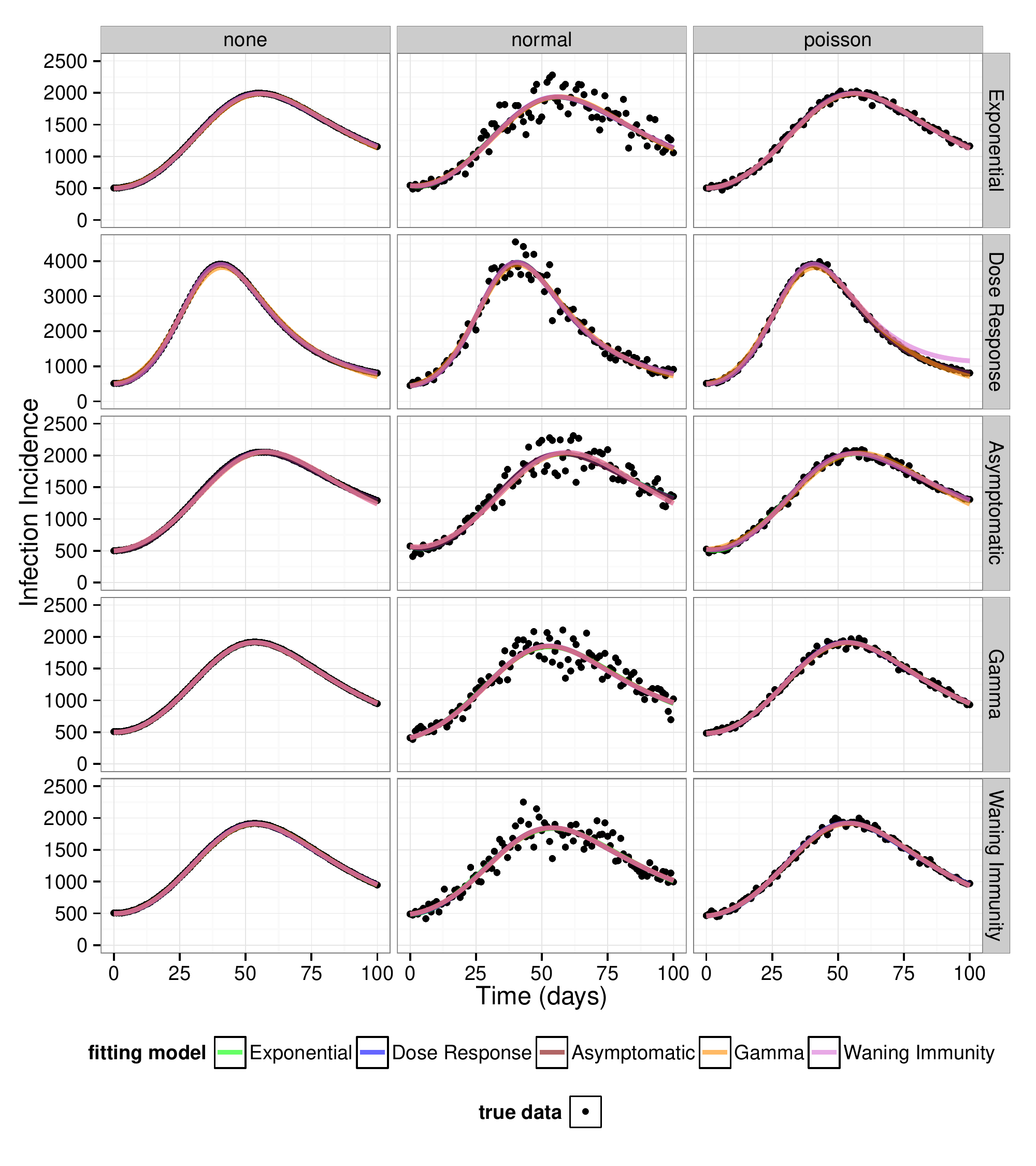}
	\caption{Fits to simulating model data (indicated by row) without noise (left column), with normal noise (middle column), and with poisson noise (right column), using informed starting parameters. Model fits are overlaid, thus obscuring some of the model fits in the figure.}
	\label{fig:informedFits}
\end{figure}

Shown below are the model fits for all simulating model datasets with informed starting parameters (Figure \ref{fig:informedFits}).

---------------------------------------------

\pagebreak

\subsection{Model fits to simulated long-term (3-year) data}
\begin{figure}[H]
	\centering
	\includegraphics[width=0.9\textwidth]{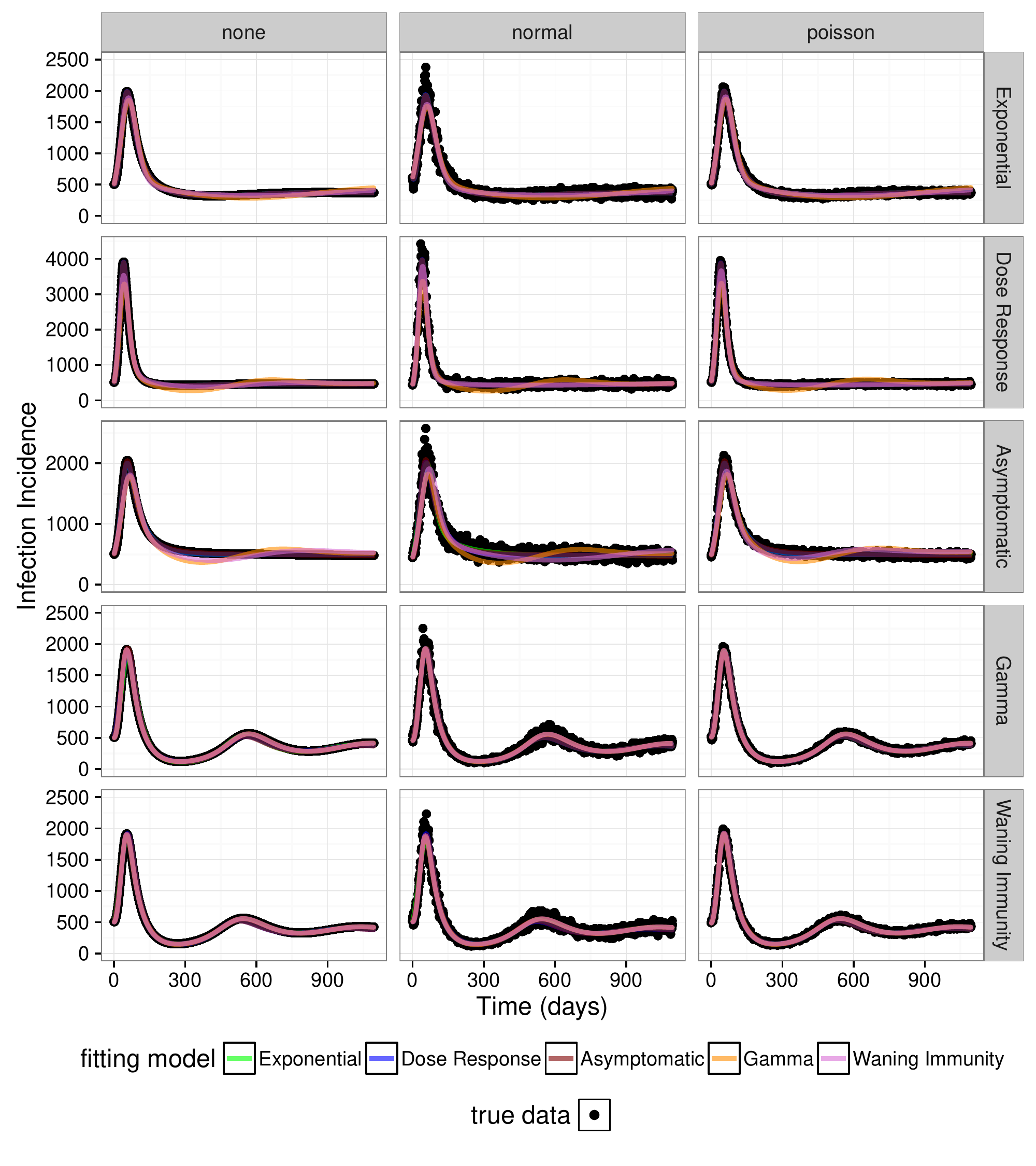}
	\caption{Fits to simulating model data (indicated by row) without noise (left column), with normal noise (middle column), and with poisson noise (right column), using informed starting parameters. Model fits are overlaid, thus obscuring some of the model fits in the figure.}
	\label{fig:informedFits_3y}
\end{figure}

\begin{figure}[H]
	\centering
	\includegraphics[width=0.9\textwidth]{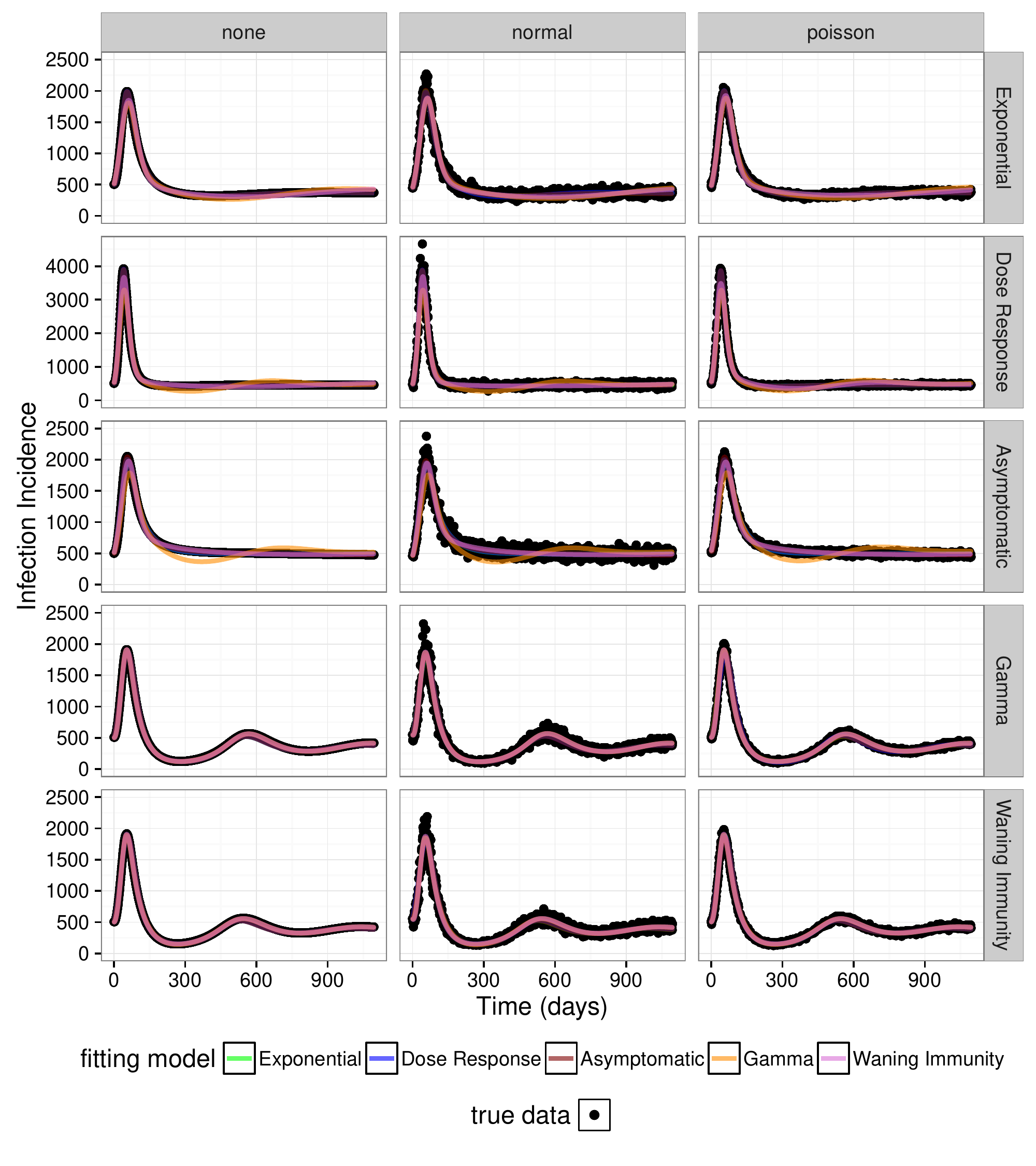}
	\caption{Fits to simulating model data (indicated by row) without noise (left column), with normal noise (middle column), and with poisson noise (right column), using naive starting parameters. Model fits are overlaid, thus obscuring some of the model fits in the figure.}
	\label{fig:naiveFits_3y}
\end{figure}

Shown below are the model fits for all simulating model datasets with informed starting parameters (Figure \ref{fig:informedFits_3y}) and naive starting parameters (Figure \ref{fig:naiveFits_3y}). 

---------------------------------------------

---------------------------------------------

\pagebreak

\subsection{Parameter estimate tables to simulated epidemic data}

We report the estimates for common model parameters to all 100-day simulated data fits in Tables \ref{NNIP} through \ref{PNNP}. Across all fits with the Asymptomatic model, the mean and standard deviation of $\alpha_S$ and $\alpha_A$ were 0.002 (SD = 0.001) and 0.004 (SD = 0.005), respectively. The reported estimates of $\alpha$ for this model are given as the weighted average of $\alpha_S$ and $\alpha_A$ (as described in the methods).

\begin{center}
\begin{table}[H]
\centering
\caption{Noise-free with Informed Starting Parameters}
\label{NNIP}
\subcaption{\textnormal{Exponential Data}\label{NNInformed1}}{
    \tiny
   \centering
 \begin{tabular}{ | c | l | c | c | c | c | c | c | }
 \hline
Parameters/AIC &&	Exponential	&	Dose Response	&	Asymptomatic	&	Gamma	&	Waning Immunity \\ \hline
$\beta_I$		&&	0.2494	&	0.2517	&	0.2494	&	0.2560	&	0.2505\\ \hline
$\beta_W$ 	&&	0.1809	&	0.5414	&	0.4131	&	1.0600	&	0.6476\\ \hline
$\alpha$ &&	0.0030&	0.0017&	0.0027&2.47e-06&4.69e-05\\ \hline
$\xi$			&&	0.0113	&	0.0034	&	0.0124	&	0.0038	&	0.0075\\ \hline
$k$			&&	1.96e-05	&	2.03e-5	&	1.99e-5	&	1.83e-5	&	2.02e-5\\ \hline
$\Delta AIC$ &&0&0&4&15&4\\ \hline
$\Ro$		&&	1.72		&	6.42		&	2.65		&	5.26		&	3.59 \\ \hline
\end{tabular}} 
\subcaption{\textnormal{Dose Response Data}\label{NNInformed2}}{
    \tiny
   \centering
       \begin{tabular}{ | c | l | c | c | c | c | c | c | }
    \hline
Parameters/AIC &&	Exponential	&	Dose Response	&	Asymptomatic	&	Gamma	&	Waning Immunity \\ \hline
$\beta_I$		&&	0.2506	&	0.2500	&	0.2480	&	0.2780		&	0.2449\\ \hline
$\beta_W$	&&	1.0709	&	0.5000	&	0.6989	&	1.4200		&	1.0399\\ \hline
$\alpha$ &&	0.0030&	0.0027&	0.0026&4.94e-06&3.46e-05\\ \hline
$\xi$			&&	0.0117	&	0.0100	&	0.0191	&	0.0055		&	0.0131\\ \hline
$k$			&&	2.02e-5 	&	2.00e-5 	&	2.00e-5	&	1.74e-5		&	2.10e-5\\ \hline
$\Delta AIC$ &&0&0&4&220&7\\ \hline
$\Ro$		&&	5.28		&	6.00		&	3.79		&	6.79			&	5.14 \\ \hline
\end{tabular}} \\
\subcaption{\textnormal{Asymptomatic Data}\label{NNInformed3}}{
    \tiny
   \centering
    \begin{tabular}{ | c | l | c | c | c | c | c | c | }
    \hline
Parameters/AIC &&	Exponential	&	Dose Response	&	Asymptomatic	&	Gamma	&	Waning Immunity \\ \hline
$\beta_I$		&&	0.2496	&	0.2503	&	0.2500	&	0.2590	&	0.2594\\ \hline
$\beta_W$ 	&&	0.3979	&	0.2308	&	0.5000	&	0.6680	&	0.6680\\ \hline
$\alpha$ &&	0.0050&	0.0043&	0.0034&2.84e-05&2.96e-05\\ \hline
$\xi$			&&	0.0129	&	0.0085	&	0.0100	&	0.0053	&	0.0053\\ \hline
$k$			&&	1.97e-5	&	1.97e-5	&	2.00e-5	&	1.62e-5	&	1.62e-5\\ \hline
$\Delta AIC$ &&0&0&3&40&38\\ \hline
$\Ro$		&&	2.59		&	3.31		&	3.00		&	3.71		&	3.71 \\ \hline
\end{tabular}} \\
\subcaption{\textnormal{Gamma Data}\label{NNInformed4}}{
    \tiny
   \centering
    \begin{tabular}{ | c | l | c | c | c | c | c | c | }
    \hline
Parameters/AIC &&	Exponential	&	Dose Response	&	Asymptomatic	&	Gamma	&	Waning Immunity\\ \hline
$\beta_I$		&&	0.2480	&	0.2494	&	0.2489	&	0.2500	&	0.2500 \\ \hline
$\beta_W$ 	&&	0.3139	&	0.3510	&	0.4596	&	0.5000	&	0.4998 \\ \hline
$\alpha$ &&	0.0011&	1.54e-9&	0.0012&3.09e-05&3.21e-05\\ \hline
$\xi$			&&	0.0175	&	0.0057	&	0.0113	&	0.0100	&	0.0100\\ \hline
$k$			&&	1.96e-5	&	2.08e-5	&	2.03e-5	&	2.00e-5	&	2.00e-5\\ \hline
$\Delta AIC$ &&0&1&4&5&4\\ \hline
$\Ro$		&&	2.25		&	4.51		&	2.83		&	3.00		&	3.00 \\ \hline
\end{tabular}} \\
\subcaption{\textnormal{Waning Immunity Data}\label{NNInformed5}}{
    \tiny
   \centering
       \begin{tabular}{ | c | l | c | c | c | c | c | c | }
    \hline
Parameters/AIC &&	Exponential	&	Dose Response	&	Asymptomatic	&	Gamma	&	Waning Immunity \\ \hline
$\beta_I$		&&	0.2484	&	0.2495	&	0.2489	&	0.2500	&	0.2500\\ \hline
$\beta_W$ 	&&	0.3390	&	0.2716	&	0.4617	&	0.5000	&	0.5000\\ \hline
$\alpha$ &&	0.0008&	3.37e-9&	0.0013&3.09e-05&3.58e-05\\ \hline
$\xi$			&&	0.0159	&	0.0074	&	0.0112	&	0.0100	&	0.0100\\ \hline
$k$			&&	1.97e-5	&	2.04e-5	&	2.03e-5	&	2.00e-5	&	2.00e-5\\ \hline
$\Delta AIC$ &&0&0&4&5&4\\ \hline
$\Ro$		&&	2.35		&	3.71		&	2.84		&	3.00		&	3.00 \\ \hline

\end{tabular}} \\
\end{table}
\end{center}
\begin{center}
\begin{table}[H]
\centering
\caption{Noise-free with Naive Starting Parameters	}
\label{NNNP}
\subcaption{\textnormal{Exponential Data}\label{NNNaive1}}{
    \tiny
   \centering
 \begin{tabular}{ | c | l | c | c | c | c | c | c | }
 \hline
Parameters/AIC &&	Exponential	&	Dose Response	&	Asymptomatic	&	Gamma	&	Waning Immunity \\ \hline
$\beta_I$	&&	0.2494	&	0.2503	&	0.2506	&	0.2560	&	0.2502 \\ \hline
$\beta_W$&&	0.4159	&	0.2186	&	0.6352	&	1.0800	&	0.5607 \\ \hline
$\alpha$ &&	0.0031&	0.0025&	0.0024&6.54e-05&2.35e-05\\ \hline
$\xi$	&&	0.0124	&	0.0090	&	0.0076	&	0.0037	&	0.0088 \\ \hline
$k$	&&	1.99e-5	&	1.96e-5	&	2.01e-5	&	1.83e-5	&	2.00e-5 \\ \hline
$\Delta AIC$ &&0&0&4&15&4\\ \hline
$\Ro$		&&	2.66	&	3.19	&	3.54	&	5.34	&	3.24 \\ \hline
\end{tabular}} \\
\subcaption{\textnormal{Dose Response Data}\label{NNNaive2}}{
    \tiny
   \centering
       \begin{tabular}{ | c | l | c | c | c | c | c | c | }
    \hline
Parameters/AIC &&	Exponential	&	Dose Response	&	Asymptomatic	&	Gamma	&	Waning Immunity\\ \hline
$\beta_I$	&&	0.2488	&	0.2438	&	0.2514	&	0.2780	&	0.2477 \\ \hline
$\beta_W$	&&	0.7729	&	0.2353	&	1.3799	&	1.4300	&	1.6883 \\ \hline
$\alpha$ &&	0.0033&	0.0033&	0.0024&5.06e-05&4.94e-05\\ \hline
$\xi$	&&	0.0170	&	0.0247	&	0.0088	&	0.0055	&	0.0076 \\ \hline
$k$	&&	2.00e-5	&	1.92e-5	&	2.04e-5	&	1.74e-5	&	2.27e-5 \\ \hline
$\Delta AIC$ &&0&0&4&220&5\\ \hline
$\Ro$		&&	4.09	&	3.33	&	6.52	&	6.83	&	7.74 \\ \hline
\end{tabular}} \\
\subcaption{\textnormal{Asymptomatic Data}\label{NNNaive3}}{
    \tiny
   \centering
    \begin{tabular}{ | c | l | c | c | c | c | c | c | }
    \hline
Parameters/AIC &&	Exponential	&	Dose Response	&	Asymptomatic	&	Gamma	&	Waning Immunity \\ \hline
$\beta_I$	&&	0.2491	&	0.2512	&	0.2507	&	0.2600	&	0.2601 \\ \hline
$\beta_W$	&&	0.3934	&	0.3603	&	0.6118	&	0.4370	&	0.4339 \\ \hline
$\alpha$ &&	0.0052&	0.0037&	0.0032&3.7e-05&3.33e-05\\ \hline
$\xi$	&&	0.0133	&	0.0052	&	0.0079	&	0.0080	&	0.0081 \\ \hline
$k$	&&	1.99e-5	&	2.00e-5	&	2.00e-5	&	1.56e-5	&	1.56e-5 \\ \hline
$\Delta AIC$ &&0&0&3&51&49\\ \hline
$\Ro$		&&	2.57	&	4.61	&	3.45	&	2.79	&	2.78\\ \hline
\end{tabular}} \\
\subcaption{\textnormal{Gamma Data}\label{NNNaive4}}{
    \tiny
   \centering
    \begin{tabular}{ | c | l | c | c | c | c | c | c | }
    \hline
Parameters/AIC &&	Exponential	&	Dose Response	&	Asymptomatic	&	Gamma	&	Waning Immunity \\ \hline
$\beta_I$	&&	0.2499	&	0.2494	&	0.2484	&	0.2500	&	0.2497 \\ \hline
$\beta_W$	&&	0.4943	&	0.4224	&	0.4910	&	0.5780	&	0.4483 \\ \hline
$\alpha$ &&	2.74e-5&	2.30e-8&	0.0018&2.22e-05&2.47e-05\\ \hline
$\xi$	&&	0.0102	&	0.0047	&	0.0107	&	0.0086	&	0.0113 \\ \hline
$k$	&&	2.00e-5	&	2.11e-5	&	2.06e-5	&	2.03e-5	&	1.99e-5 \\ \hline
$\Delta AIC$ &&9&11&0&16&14\\ \hline
$\Ro$		&&	2.98	&	5.22	&	2.96	&	3.31	&	2.79 \\ \hline
\end{tabular}} \\
\subcaption{\textnormal{Waning Immunity Data}\label{NNNaive5}}{
    \tiny
   \centering
       \begin{tabular}{ | c | l | c | c | c | c | c | c | }
    \hline
Parameters/AIC &&	Exponential	&	Dose Response	&	Asymptomatic	&	Gamma	&	Waning Immunity \\ \hline
$\beta_I$	&&	0.2499	&	0.2494	&	0.2485	&	0.2500	&	0.2501 \\ \hline
$\beta_W$ 	&&	0.5005	&	0.8094	&	0.5234	&	0.5680	&	0.4476 \\ \hline
$\alpha$ &&	9.66e-6&	1.78e-8&	0.0016&3.58e-05&3.83e-05\\ \hline
$\xi$	&&	0.0100	&	0.0024	&	0.0100	&	0.0088	&	0.0112 \\ \hline
$k$	&&	2.00e-5	&	2.17e-5	&	2.07e-5	&	2.03e-5	&	1.98e-5 \\ \hline
$\Delta AIC$ &&0&3&5&7&5\\ \hline
$\Ro$		&&	3.00	&	9.09	&	3.09	&	3.27	&	2.79\\ \hline
\end{tabular}} \\
\end{table}
\end{center}
\begin{center}
\begin{table}[H]
\centering
\caption{Normal Noise with Informed Starting Parameters	}
\label{NormNPI}
\subcaption{\textnormal{Exponential Data}\label{NormNInformed1}}{
    \tiny
   \centering
 \begin{tabular}{ | c | l | c | c | c | c | c | c | }
 \hline
Parameters/AIC &&	Exponential	&	Dose Response	&	Asymptomatic	&	Gamma	&	Waning Immunity \\ \hline
$\beta_I$	&&	0.2395	&	0.2384	&	0.2364	&	0.2480	&	0.2436 \\ \hline
$\beta_W$	&&	0.2816	&	0.1160	&	0.2298	&	0.7500	&	0.8582 \\ \hline
$\alpha$ &&	0.0040&	0.0040&	0.0044&1.23e-05&4.44e-05\\ \hline
$\xi$	&&	0.0215	&	0.0213	&	0.0294	&	0.0060	&	0.0059 \\ \hline
$k$	&&	1.96e-5	&	1.92e-5	&	1.96e-5	&	1.86e-5	&	2.07e-5 \\ \hline
$\Delta AIC$ &&0&0&0&23&11\\ \hline
$\Ro$		&&	2.08	&	2.11	&	1.86	&	3.99	&	4.41\\ \hline
\end{tabular}} \\
\subcaption{\textnormal{Dose Response Data}\label{NormNInformed2}}{
    \tiny
   \centering
       \begin{tabular}{ | c | l | c | c | c | c | c | c | }
    \hline
Parameters/AIC &&	Exponential	&	Dose Response	&	Asymptomatic	&	Gamma	&	Waning Immunity \\ \hline
$\beta_I$	&&	0.2677	&	0.2677	&	0.2679	&	0.2890	&	0.2660 \\ \hline
$\beta_W$ 	&&	0.8064	&	0.3931	&	0.8161	&	1.3700	&	2.1494 \\ \hline
$\alpha$ &&	0.0026&	0.0022&	0.0018&2.47e-06&3.09e-05\\ \hline
$\xi$	&&	0.0136	&	0.0110	&	0.0133	&	0.0053	&	0.0050 \\ \hline
$k$	&&	1.93e-5	&	1.90e-5	&	1.93e-5	&	1.71e-5	&	2.01e-5 \\ \hline
$\Delta AIC$ &&0&1&5&109&16\\ \hline
$\Ro$		&&	4.30	&	5.00	&	4.34	&	6.63	&	9.66\\ \hline
\end{tabular}} \\
\subcaption{\textnormal{Asymptomatic Data}\label{NormNInformed3}}{
    \tiny
   \centering
    \begin{tabular}{ | c | l | c | c | c | c | c | c | }
    \hline
Parameters/AIC &&	Exponential	&	Dose Response	&	Asymptomatic	&	Gamma	&	Waning Immunity\\ \hline
$\beta_I$	&&	0.2212	&	0.2274	&	0.1658	&	0.2420	&	0.2420 \\ \hline
$\beta_W$	&&	0.2825	&	0.2013	&	0.1927	&	0.5050	&	0.5047 \\ \hline
$\alpha$ &&	0.0091&	0.0061&	0.0138&3.58e-05&3.46e-05\\ \hline
$\xi$	&&	0.0297	&	0.0136	&	0.1446	&	0.0093	&	0.0093 \\ \hline
$k$	&&	2.15e-5	&	2.15e-5	&	1.98e-5	&	1.67e-5	&	1.67e-5 \\ \hline
$\Delta AIC$ &&42&58&0&124&126\\ \hline
$\Ro$		&&	2.01	&	2.92	&	1.43	&	2.99	&	2.99\\ \hline
\end{tabular}} \\
\subcaption{\textnormal{Gamma Data}\label{NormNInformed4}}{
    \tiny
   \centering
    \begin{tabular}{ | c | l | c | c | c | c | c | c | }
    \hline
Parameters/AIC &&	Exponential	&	Dose Response	&	Asymptomatic	&	Gamma	&	Waning Immunity \\ \hline
$\beta_I$	&&	0.2792	&	0.2811	&	0.2803	&	0.2810	&	0.2806 \\ \hline
$\beta_W$ 	&&	0.1144	&	0.1017	&	0.1470	&	0.4290	&	0.4279 \\ \hline
$\alpha$ &&	0.0051&	0.0011&	0.0027&5.19e-05&2.22e-05\\ \hline
$\xi$	&&	0.0250	&	0.0092	&	0.0174	&	0.0055	&	0.0055 \\ \hline
$k$	&&	1.70e-5	&	1.64e-5	&	1.66e-5	&	1.68e-5	&	1.68e-5 \\ \hline
$\Delta AIC$ &&0&9&8&14&12\\ \hline
$\Ro$		&&	1.57	&	2.14	&	1.71	&	2.84	&	2.83\\ \hline
\end{tabular}} \\
\subcaption{\textnormal{Waning Immunity Data}\label{NormNInformed5}}{
    \tiny
   \centering
       \begin{tabular}{ | c | l | c | c | c | c | c | c | }
    \hline
Parameters/AIC &&	Exponential	&	Dose Response	&	Asymptomatic	&	Gamma	&	Waning Immunity\\ \hline
$\beta_I$	&&	0.2568	&	0.2562	&	0.2623	&	0.2640	&	0.2642 \\ \hline
$\beta_W$ 	&&	0.1184	&	0.0638	&	0.5149	&	0.5060	&	0.5034 \\ \hline
$\alpha$ &&	0.0067&	0.0054&	0.0014&4.32e-05&1.98e-05\\ \hline
$\xi$	&&	0.0446	&	0.0317	&	0.0072	&	0.0068	&	0.0068 \\ \hline
$k$	&&	1.72e-5	&	1.79e-5	&	1.90e-5	&	1.80e-5	&	1.80e-5 \\ \hline
$\Delta AIC$ &&0&2&11&17&12\\ \hline
$\Ro$		&&	1.50	&	1.66	&	3.11	&	3.08	&	3.07\\ \hline
\end{tabular}} \\
\end{table}
\end{center}
\begin{center}
\begin{table}[H]
\centering
\caption{Normal Noise with Naive Starting Parameters}
\label{NormNNP}
\subcaption{\textnormal{Exponential Data}\label{NormNNaive1}}{
    \tiny
   \centering
 \begin{tabular}{ | c | l | c | c | c | c | c | c | }
 \hline
Parameters/AIC &&	Exponential	&	Dose Response	&	Asymptomatic	&	Gamma	&	Waning Immunity \\ \hline
$\beta_I$		&&	0.2373	&	0.2418	&	0.2417	&	0.2480	&	0.2483\\ \hline
$\beta_W$	&&	0.2515	&	0.1836	&	0.4272	&	0.4700	&	0.7331\\ \hline
$\alpha$ &&	0.0048&	0.0026&	0.0023&2.72e-05&2.22e-05\\ \hline
$\xi$			&&	0.0258	&	0.0119	&	0.0129	&	0.0097	&	0.0060\\ \hline
$k$			&&	1.97e-5	&	1.98e-5	&	2.02e-5	&	1.79e-5	&	1.88e-5\\ \hline
$\Delta AIC$ &&0&7&10&25&22\\ \hline
$\Ro$		&&	1.95		&	2.80	&	2.68	&	2.87	&	3.92\\ \hline
\end{tabular}} \\
\subcaption{\textnormal{Dose Response Data}\label{NormNNaive2}}{
    \tiny
   \centering
       \begin{tabular}{ | c | l | c | c | c | c | c | c | }
    \hline
Parameters/AIC &&	Exponential	&	Dose Response	&	Asymptomatic	&	Gamma	&	Waning Immunity \\ \hline
$\beta_I$		&&	0.2670	&	0.2689	&	0.2679	&	0.2890	&	0.2647\\ \hline
$\beta_W$	&&	0.6610	&	0.6375	&	0.7955	&	1.3700	&	0.7846\\ \hline
$\alpha$ &&	0.0028&	0.0021&	0.0018&6.54e-05&1.6e-05\\ \hline
$\xi$			&&	0.0171	&	0.0065	&	0.0137	&	0.0052	&	0.0148\\ \hline
$k$			&&	1.91e-5	&	1.94e-5	&	1.92e-5	&	1.71e-5	&	1.98e-5\\ \hline
$\Delta AIC$ &&0&2&5&109&23\\ \hline
$\Ro$		&&	3.71	&	7.45	&	4.25	&	6.63	&	4.20	\\ \hline
\end{tabular}} \\
\subcaption{\textnormal{Asymptomatic Data}\label{NormNNaive3}}{
    \tiny
   \centering
    \begin{tabular}{ | c | l | c | c | c | c | c | c | }
    \hline
Parameters/AIC &&	Exponential	&	Dose Response	&	Asymptomatic	&	Gamma	&	Waning Immunity \\ \hline
$\beta_I$		&&	0.2052	&	0.2282	&	0.2248	&	0.2430	&	0.2278\\ \hline
$\beta_W$	&&	0.1888	&	0.1954	&	0.3747	&	6.7300	&	0.5668\\ \hline
$\alpha$ &&	0.0132&	0.0058&	0.0064&3.7e-05&2.35e-05\\ \hline
$\xi$			&&	0.0697	&	0.0138	&	0.0199	&	0.0007	&	0.0119\\ \hline
$k$			&&	1.99e-5	&	2.11e-5	&	2.21e-5	&	1.84e-5	&	2.24e-5\\ \hline
$\Delta AIC$ &&0&45&42&101&44\\ \hline
$\Ro$		&&	1.58	&	2.87	&	2.40	&	27.89	&	3.18\\ \hline
\end{tabular}} \\
\subcaption{\textnormal{Gamma Data}\label{NormNNaive4}}{
    \tiny
   \centering
    \begin{tabular}{ | c | l | c | c | c | c | c | c | }
    \hline
Parameters/AIC &&	Exponential	&	Dose Response	&	Asymptomatic	&	Gamma	&	Waning Immunity \\ \hline
$\beta_I$		&&	0.2805	&	0.2811	&	0.2772	&	0.2810	&	0.2801\\ \hline
$\beta_W$	&&	0.3507	&	0.0878	&	0.9213	&	0.3640	&	0.5275\\ \hline
$\alpha$ &&	0.0004&	0.0015&	0.0013&4.07e-05&3.09e-05\\ \hline
$\xi$			&&	0.0068	&	0.0107	&	0.0031	&	0.0064	&	0.0046\\ \hline
$k$			&&	1.69e-5	&	1.63e-5	&	1.89e-5	&	1.66e-5	&	1.72e-5\\ \hline
$\Delta AIC$ &&0&0&9&6&4\\ \hline
$\Ro$		&&	2.52	&	2.00	&	4.79	&	2.58	&	3.23\\ \hline
\end{tabular}} \\
\subcaption{\textnormal{Waning Immunity Data}\label{NormNNaive5}}{
    \tiny
   \centering
       \begin{tabular}{ | c | l | c | c | c | c | c | c | }
    \hline
Parameters/AIC &&	Exponential	&	Dose Response	&	Asymptomatic	&	Gamma	&	Waning Immunity \\ \hline
$\beta_I$	&&	0.2532	&	0.2608	&	0.2578	&	0.2640	&	0.2642\\ \hline
$\beta_W$	&&	0.1279	&	0.0798	&	0.1455	&	0.5050	&	0.5038\\ \hline
$\alpha$ &&	0.0083&	0.0030&	0.0054&3.33e-05&2.47e-05\\ \hline
$\xi$	&&	0.0470	&	0.0208	&	0.0338	&	0.0068	&	0.0068\\ \hline
$k$	&&	1.86e-5	&	1.76e-5	&	1.81e-5	&	1.80e-5	&	1.80e-5\\ \hline
$\Delta AIC$ &&0&7&6&19&14\\ \hline
$\Ro$		&&	1.52		&1.84	&	1.61	&	3.08	&	3.07\\ \hline
\end{tabular}} \\
\end{table}
\end{center}
\begin{center}
\begin{table}[H]
\centering
\caption{Poisson Noise with Informed Starting Parameters}
\label{PNIP}
\subcaption{\textnormal{Exponential Data}\label{PNInformed1}}{
    \tiny
   \centering
 \begin{tabular}{ | c | l | c | c | c | c | c | c | }
 \hline
Parameters/AIC &&	Exponential	&	Dose Response	&	Asymptomatic	&	Gamma	&	Waning Immunity \\ \hline
$\beta_I$		&&	0.2494	&	0.2494	&	0.2497	&	0.2550	&	0.2554 \\ \hline
$\beta_W$	&&	0.4999	&	0.1967	&	0.5043	&	0.6140	&	0.6139 \\ \hline
$\alpha$ &&	0.0025&	0.0024&	0.0022&3.7e-05&2.47e-05\\ \hline
$\xi$			&&	0.0100	&	0.0101	&	0.0098	&	0.0067	&	0.0067 \\ \hline
$k$			&&	1.98e-5	&	1.94e-5	&	1.97e-5	&	1.78e-5	&	1.78e-5 \\ \hline
$\Delta AIC$ &&0&0&4&141&9\\ \hline
$\Ro$		&&	3.00		&	2.96		&	3.02		&	3.48		&	4.48 \\ \hline
\end{tabular}} \\
\subcaption{\textnormal{Dose Response Data}\label{PNInformed2}}{
    \tiny
   \centering
       \begin{tabular}{ | c | l | c | c | c | c | c | c | }
    \hline
Parameters/AIC &&	Exponential	&	Dose Response	&	Asymptomatic	&	Gamma	&	Waning Immunity \\ \hline
$\beta_I$		&&	0.2479	&	0.2496	&	0.2496	&	0.2810	&	0.2508 \\ \hline
$\beta_W$	&&	0.7228	&	0.4984	&	0.9517	&	0.8190	&	8.9934 \\ \hline
$\alpha$ &&	0.0035&	0.0028&	0.0024&2.59e-05&0.000312\\ \hline
$\xi$			&&	0.0185	&	0.0101	&	0.0134	&	0.0093	&	0.0013 \\ \hline
$k$			&&	2.01e-5	&	2.01e-5	&	2.03e-5	&	1.69e-5	&	2.11e-5 \\ \hline
$\Delta AIC$ &&0&0&4&2504&4\\ \hline
$\Ro$		&&	3.88		&	5.98		&	4.80		&	4.40		&	36.97 \\ \hline
\end{tabular}} \\
\subcaption{\textnormal{Asymptomatic Data}\label{PNInformed3}}{
    \tiny
   \centering
    \begin{tabular}{ | c | l | c | c | c | c | c | c | }
    \hline
Parameters/AIC &&	Exponential	&	Dose Response	&	Asymptomatic	&	Gamma	&	Waning Immunity \\ \hline
$\beta_I$		&&	0.2238	&	0.2433	&	0.2412	&	0.2550	&	0.2556 \\ \hline
$\beta_W$	&&	0.1752	&	0.1922	&	0.4279	&	0.5990	&	0.6030 \\ \hline
$\alpha$ &&	0.0140&	0.0052&	0.0049&3.58e-05&4.57e-05\\ \hline
$\xi$			&&	0.0608	&	0.0117	&	0.0140	&	0.0063	&	0.0063 \\ \hline
$k$			&&	2.06e-5	&	2.04e-5	&	2.14e-5	&	1.63e-5	&	1.63e-5 \\ \hline
$\Delta AIC$ &&0&7&9&411&40\\ \hline
$\Ro$		&&	1.60		&	2.89		&	2.68		&	3.42		&	3.43	\\ \hline
\end{tabular}} \\
\subcaption{\textnormal{Gamma Data}\label{PNInformed4}}{
    \tiny
   \centering
    \begin{tabular}{ | c | l | c | c | c | c | c | c | }
    \hline
Parameters/AIC &&	Exponential	&	Dose Response	&	Asymptomatic	&	Gamma	&	Waning Immunity \\ \hline
$\beta_I$		&&	0.2541	&	0.2531	&	0.2528	&	0.2540	&	0.2541 \\ \hline
$\beta_W$	&&	0.4763	&	0.7484	&	0.6401	&	0.5030	&	0.5028 \\ \hline
$\alpha$ &&	3.00e-8&	1.00e-8&	0.0011&3.09e-05&2.96e-05\\ \hline
$\xi$			&&	0.0099	&	0.0025	&	0.0076	&	0.0094	&	0.0094 \\ \hline
$k$			&&	1.97e-5	&	2.14e-5	&	2.08e-5	&	1.98e-5	&	1.99e-5 \\ \hline
$\Delta AIC$ &&0&2&5&108&4\\ \hline
$\Ro$		&&	2.92		&	8.49		&	3.57		&	3.03		&	3.03 \\ \hline
\end{tabular}} \\
\subcaption{\textnormal{Waning Immunity Data}\label{PNInformed5}}{
    \tiny
   \centering
       \begin{tabular}{ | c | l | c | c | c | c | c | c | }
    \hline
Parameters/AIC &&	Exponential	&	Dose Response	&	Asymptomatic	&	Gamma	&	Waning Immunity \\ \hline
$\beta_I$		&&	0.2576	&	0.2573	&	0.2582	&	0.2590	&	0.2592 \\ \hline
$\beta_W$	&&	0.4043	&	0.1973	&	0.2819	&	0.2570	&	0.2568 \\ \hline
$\alpha$ &&	3.00e-8&	4.74e-9&	0.0011&1.48e-05&3.46e-05\\ \hline
$\xi$			&&	0.0112	&	0.0092	&	0.0161	&	0.0171	&	0.0171 \\ \hline
$k$			&&	1.92e-5	&	1.93e-5	&	1.84e-5	&	1.77e-5	&	1.77e-5 \\ \hline
$\Delta AIC$ &&22&25&24&0&24\\ \hline
$\Ro$		&&	2.65		&	3.00		&	2.16		&	2.06		&	2.06 \\ \hline
\end{tabular}} \\
\end{table}
\end{center}

\begin{center}
\begin{table}[H]
\centering
\caption{Poisson Noise with Naive Starting Parameters}
\label{PNNP}
\subcaption{\textnormal{Exponential Data}\label{PNNaive1}}{
    \tiny
   \centering
 \begin{tabular}{ | c | l | c | c | c | c | c | c | }
 \hline
Parameters/AIC &&	Exponential	&	Dose Response	&	Asymptomatic	&	Gamma	&	Waning Immunity \\ \hline
$\beta_I$		&&	0.2497	&	0.2509	&	0.2491	&	0.2550	&	0.2552\\ \hline
$\beta_W$	&&	0.5188	&	0.4631	&	0.4233	&	0.6130	&	0.6067\\ \hline
$\alpha$ &&	0.0023&	0.0015&	0.0025&2.72e-05&3.21e-05\\ \hline
$\xi$			&&	0.0096	&	0.0040	&	0.0120	&	0.0067	&	0.0068\\ \hline
$k$			&&	1.98e-5	&	2.00e-5	&	1.96e-5	&	1.78e-5	&	1.78e-5\\ \hline
$\Delta AIC$ &&0&0&4&141&9\\ \hline
$\Ro$		&& 	3.07		&	5.63		&	2.69		&	3.47		&	3.45 \\ \hline
\end{tabular}} \\
\subcaption{\textnormal{Dose Response Data}\label{PNNaive2}}{
    \tiny
   \centering
       \begin{tabular}{ | c | l | c | c | c | c | c | c | }
    \hline
Parameters/AIC &&	Exponential	&	Dose Response	&	Asymptomatic	&	Gamma	&	Waning Immunity \\ \hline
$\beta_I$		&&	0.2493	&	0.2501	&	0.2511	&	0.2790	&	0.2409\\ \hline
$\beta_W$	&&	0.8864	&	0.5567	&	1.4445	&	1.4300	&	0.7726\\ \hline
$\alpha$ &&	0.0033&	0.0028&	0.0027&0.000159&4.07e-05\\ \hline
$\xi$			&&	0.0146	&	0.0089	&	0.0085	&	0.0054	&	0.0191\\ \hline
$k$			&&	2.02e-5	&	2.02e-5	&	2.05e-5	&	1.74e-5	&	2.14e-5\\ \hline
$\Delta AIC$ &&0&0&4&2504&9\\ \hline
$\Ro$		&&	4.54		&	6.57		&	6.78		&	6.83		&	4.05 \\ \hline
\end{tabular}} \\
\subcaption{\textnormal{Asymptomatic Data}\label{PNNaive3}}{
    \tiny
   \centering
    \begin{tabular}{ | c | l | c | c | c | c | c | c | }
    \hline
Parameters/AIC &&	Exponential	&	Dose Response	&	Asymptomatic	&	Gamma	&	Waning Immunity \\ \hline
$\beta_I$		&&	0.2274	&	0.2444	&	0.2402	&	0.2550	&	0.2348\\ \hline
$\beta_W$	&&	0.1813	&	0.5314	&	0.3553	&	0.5950	&	0.2995\\ \hline
$\alpha$ &&	0.0130&	0.0046&	0.0048&0.000138&2.47e-06\\ \hline
$\xi$			&&	0.0535	&	0.0040	&	0.0175	&	0.0064	&	0.0254\\ \hline
$k$			&&	2.06e-5	&	2.17e-5	&	2.13e-5	&	1.63e-5	&	3.90e-5\\ \hline
$\Delta AIC$ &&0&7&8&411&7\\ \hline
$\Ro$		&&	1.63		&	6.29		&	2.38		&	3.40		&	2.14 \\ \hline
\end{tabular}} \\
\subcaption{\textnormal{Gamma Data}\label{PNNaive4}}{
    \tiny
   \centering
    \begin{tabular}{ | c | l | c | c | c | c | c | c | }
    \hline
Parameters/AIC &&	Exponential	&	Dose Response	&	Asymptomatic	&	Gamma	&	Waning Immunity \\ \hline
$\beta_I$		&&	0.2538	&	0.2531	&	0.2519	&	0.2540	&	0.2540\\ \hline
$\beta_W$	&&	0.6268	&	0.1305	&	0.2467	&	0.5030	&	0.5058\\ \hline
$\alpha$ &&	8.97e-8&	0.0006&	0.0015&1.73e-05&2.96e-05\\ \hline
$\xi$			&&	0.0076	&	0.0153	&	0.0216	&	0.0094	&	0.0093\\ \hline
$k$			&&	2.03e-5	&	1.88e-5	&	1.91e-5	&	1.98e-5	&	1.99e-5\\ \hline
$\Delta AIC$ &&0&1&5&108&4\\ \hline
$\Ro$		&&	3.52		&	2.32		&	1.99		&	3.03		&	3.04 \\ \hline
\end{tabular}} \\
\subcaption{\textnormal{Waning Immunity Data}\label{PNNaive5}}{
    \tiny
   \centering
       \begin{tabular}{ | c | l | c | c | c | c | c | c | }
    \hline
Parameters/AIC &&	Exponential	&	Dose Response	&	Asymptomatic	&	Gamma	&	Waning Immunity \\ \hline
$\beta_I$		&&	0.2592	&	0.2587	&	0.2582	&	0.2590	&	0.2591\\ \hline
$\beta_W$	&&	0.2472	&	0.0913	&	0.2548	&	0.2560	&	0.2582\\ \hline
$\alpha$ &&	0.0001&	0.0003&	3.58e-5&2.47e-05&1.48e-05\\ \hline
$\xi$			&&	0.0179	&	0.0198	&	0.0179	&	0.0171	&	0.0171\\ \hline
$k$			&&	1.77e-5	&	1.71e-5	&	1.81e-5	&	1.77e-5	&	1.78e-5\\ \hline
$\Delta AIC$ &&20&20&24&0&24\\ \hline
$\Ro$		&&	2.03		&	1.95		&	2.05		&	2.06		&	2.07 \\ \hline
\end{tabular}} \\
\end{table}
\end{center}

\pagebreak

\subsection{Parameter estimate tables to simulated long-term data}
We report the estimates for common model parameters to all simulated 3-year data fits in Tables \ref{NNIP3} through \ref{PNNP3}. The reported estimates of $\alpha$ for this model are given as the weighted average of $\alpha_S$ and $\alpha_A$ (as described in the methods).

\begin{center}
\begin{table}[H]
\centering
\caption{Noise-free with Informed Starting Parameters}
\label{NNIP3}
\subcaption{\textnormal{Exponential Data}\label{NNInformed1}}{
    \tiny
   \centering
 \begin{tabular}{ | c | l | c | c | c | c | c | c | }
 \hline
Parameters/AIC & & Exponential & Dose Response & Asymptomatic & Gamma & Waning \\ \hline
$\beta_i$ & & 0.25 & 0.2485 & 0.25 & 0.2843 & 0.2862 \\ \hline
$\beta_w$ & & 0.5 & 0.2019 & 0.4997 & 0.855 & 0.5586 \\ \hline
$\xi$ & & 0.01 & 0.0102 & 0.01 & 0.0003 & 0.0003 \\ \hline
$\alpha$ & & 0.0027 & 0.0027 & 0.0027 & 0.0013 & 0.001 \\ \hline
$k$ & & 2e-05 & 1.9838e-05 & 2e-05 & 7.091e-06 & 7.045e-06 \\ \hline
$\Delta AIC$ & & 0 & 2 & 2 & 5621 & 2663 \\ \hline
$\Ro$ & & 4.7041 & 4.686 & 2.9994 & 1.8204 & 1.7454 \\ \hline

\end{tabular}} 
\subcaption{\textnormal{Dose Response Data}\label{NNInformed2}}{
    \tiny
   \centering
       \begin{tabular}{ | c | l | c | c | c | c | c | c | }
    \hline
Parameters/AIC & & Exponential & Dose Response & Asymptomatic & Gamma & Waning \\ \hline
$\beta_i$ & & 0.2541 & 0.25 & 0.2536 & 0.3174 & 0.3289 \\ \hline
$\beta_w$ & & 1.2106 & 0.5 & 1.1845 & 0.2998 & 0.0602 \\ \hline
$\xi$ & & 0.0098 & 0.01 & 0.0101 & 0.0022 & 0.0057 \\ \hline
$\alpha$ & & 0.0028 & 0.0027 & 0.0028 & 0.0022 & 0.0017 \\ \hline
$k$ & & 2.0052e-05 & 2e-05 & 2.0075e-05 & 1.0406e-05 & 1.3733e-05 \\ \hline
$\Delta AIC$ & & 3 & 0 & 5 & 19378 & 4312 \\ \hline
$\Ro$ & & 1.2408 & 1.2184 & 1.2986 & 2.9994 & 2.9852 \\ \hline
\end{tabular}} \\
\subcaption{\textnormal{Asymptomatic Data}\label{NNInformed3}}{
    \tiny
   \centering
    \begin{tabular}{ | c | l | c | c | c | c | c | c | }
    \hline
Parameters/AIC & & Exponential & Dose Response & Asymptomatic & Gamma & Waning \\ \hline
$\beta_i$ & & 0.2641 & 0.2638 & 0.25 & 0.2781 & 0.2806 \\ \hline
$\beta_w$ & & 0.9121 & 0.3632 & 0.5 & 0.1771 & 0.1558 \\ \hline
$\xi$ & & 0.0037 & 0.0037 & 0.01 & 0.0035 & 0.0027 \\ \hline
$\alpha$ & & 0.0028 & 0.0028 & 0.0047 & 0.0024 & 0.0018 \\ \hline
$k$ & & 1.7712e-05 & 1.7622e-05 & 2e-05 & 8.083e-06 & 9.176e-06 \\ \hline
$\Delta AIC$ & & 435 & 417 & 0 & 16535 & 12318 \\ \hline
$\Ro$ & & 2.9994 & 3.0124 & 2.9982 & 4.5562 & 3.3787 \\ \hline
\end{tabular}} \\
\subcaption{\textnormal{Gamma Data}\label{NNInformed4}}{
    \tiny
   \centering
    \begin{tabular}{ | c | l | c | c | c | c | c | c | }
    \hline
Parameters/AIC & & Exponential & Dose Response & Asymptomatic & Gamma & Waning \\ \hline
$\beta_i$ & & 0.2721 & 0.2845 & 0.272 & 0.25 & 0.2496 \\ \hline
$\beta_w$ & & 0.0381 & 0.0081 & 0.0527 & 0.5 & 0.4968 \\ \hline
$\xi$ & & 0.0599 & 0.0406 & 0.0526 & 0.01 & 0.0101 \\ \hline
$\alpha$ & & 0.0045 & 0.0044 & 0.0043 & 0.0027 & 0.0026 \\ \hline
$k$ & & 9.928e-06 & 8.969e-06 & 1.1052e-05 & 2e-05 & 2.0025e-05 \\ \hline
$\Delta AIC$ & & 3165 & 2363 & 2965 & 0 & 6 \\ \hline
$\Ro$ & & 5.8577 & 5.9988 & 5.751 & 2.4685 & 1.5559 \\ \hline
\end{tabular}} \\
\subcaption{\textnormal{Waning Immunity Data}\label{NNInformed5}}{
    \tiny
   \centering
       \begin{tabular}{ | c | l | c | c | c | c | c | c | }
    \hline
Parameters/AIC & & Exponential & Dose Response & Asymptomatic & Gamma & Waning \\ \hline
$\beta_i$ & & 0.2841 & 0.2683 & 0.2762 & 0.2535 & 0.25 \\ \hline
$\beta_w$ & & 0.0253 & 0.0261 & 0.0468 & 0.5338 & 0.5 \\ \hline
$\xi$ & & 0.0393 & 0.0518 & 0.0472 & 0.0087 & 0.01 \\ \hline
$\alpha$ & & 0.0048 & 0.0045 & 0.0048 & 0.0028 & 0.0027 \\ \hline
$k$ & & 9.698e-06 & 1.186e-05 & 1.1116e-05 & 1.984e-05 & 2e-05 \\ \hline
$\Delta AIC$ & & 2016 & 2319 & 2467 & 53 & 0 \\ \hline
$\Ro$ & & 1.2374 & 1.3334 & 1.2919 & 3.1485 & 2.9994 \\ \hline

\end{tabular}} \\
\end{table}
\end{center}

\begin{center}
\begin{table}[H]
\centering
\caption{Noise-free with Naive Starting Parameters	}
\label{NNNP3}
\subcaption{\textnormal{Exponential Data}\label{NNNaive1}}{
    \tiny
   \centering
 \begin{tabular}{ | c | l | c | c | c | c | c | c | }
 \hline
Parameters/AIC & & Exponential & Dose Response & Asymptomatic & Gamma & Waning \\ \hline
$\beta_i$ & & 0.25 & 0.2485 & 0.25 & 0.284 & 0.2859 \\ \hline
$\beta_w$ & & 0.5 & 0.2019 & 0.4998 & 0.2011 & 0.4573 \\ \hline
$\xi$ & & 0.01 & 0.0102 & 0.01 & 0.0017 & 0.0005 \\ \hline
$\alpha$ & & 0.0027 & 0.0027 & 0.0027 & 0.0016 & 0.0011 \\ \hline
$k$ & & 2.0001e-05 & 1.9839e-05 & 2.0001e-05 & 7.813e-06 & 9.256e-06 \\ \hline
$\Delta AIC$ & & 0 & 2 & 2 & 6335 & 3268 \\ \hline
$\Ro$ & & 4.704 & 4.6854 & 2.945 & 1.8204 & 7.485 \\ \hline
\end{tabular}} \\
\subcaption{\textnormal{Dose Response Data}\label{NNNaive2}}{
    \tiny
   \centering
       \begin{tabular}{ | c | l | c | c | c | c | c | c | }
    \hline
Parameters/AIC & & Exponential & Dose Response & Asymptomatic & Gamma & Waning \\ \hline
$\beta_i$ & & 0.2541 & 0.25 & 0.2536 & 0.3174 & 0.3257 \\ \hline
$\beta_w$ & & 1.2107 & 0.5 & 1.1845 & 0.2998 & 0.2044 \\ \hline
$\xi$ & & 0.0098 & 0.01 & 0.0101 & 0.0022 & 0.0013 \\ \hline
$\alpha$ & & 0.0028 & 0.0027 & 0.0028 & 0.0022 & 0.0011 \\ \hline
$k$ & & 2.0051e-05 & 2e-05 & 2.0074e-05 & 1.0405e-05 & 9.781e-06 \\ \hline
$\Delta AIC$ & & 3 & 0 & 5 & 19378 & 4687 \\ \hline
$\Ro$ & & 1.2251 & 1.2184 & 1.2165 & 2.9995 & 2.9614 \\ \hline
\end{tabular}} \\
\subcaption{\textnormal{Asymptomatic Data}\label{NNNaive3}}{
    \tiny
   \centering
    \begin{tabular}{ | c | l | c | c | c | c | c | c | }
    \hline
Parameters/AIC & & Exponential & Dose Response & Asymptomatic & Gamma & Waning \\ \hline
$\beta_i$ & & 0.2641 & 0.2638 & 0.2499 & 0.2781 & 0.2863 \\ \hline
$\beta_w$ & & 0.9121 & 0.3631 & 0.4865 & 0.1771 & 1.5854 \\ \hline
$\xi$ & & 0.0037 & 0.0037 & 0.0103 & 0.0035 & 0.0002 \\ \hline
$\alpha$ & & 0.0028 & 0.0028 & 0.0047 & 0.0024 & 0 \\ \hline
$k$ & & 1.7712e-05 & 1.7622e-05 & 2.0002e-05 & 8.083e-06 & 7.517e-06 \\ \hline
$\Delta AIC$ & & 435 & 417 & 0 & 16534 & 1088 \\ \hline
$\Ro$ & & 2.9994 & 3.0124 & 2.9985 & 1.94 & 2.9719 \\ \hline
\end{tabular}} \\
\subcaption{\textnormal{Gamma Data}\label{NNNaive4}}{
    \tiny
   \centering
    \begin{tabular}{ | c | l | c | c | c | c | c | c | }
    \hline
Parameters/AIC & & Exponential & Dose Response & Asymptomatic & Gamma & Waning \\ \hline
$\beta_i$ & & 0.284 & 0.2845 & 0.2852 & 0.25 & 0.2481 \\ \hline
$\beta_w$ & & 0.0224 & 0.0081 & 0.019 & 0.5 & 0.4924 \\ \hline
$\xi$ & & 0.0409 & 0.0406 & 0.0388 & 0.01 & 0.0106 \\ \hline
$\alpha$ & & 0.0044 & 0.0044 & 0.0044 & 0.0027 & 0.0025 \\ \hline
$k$ & & 9.227e-06 & 8.968e-06 & 8.983e-06 & 2e-05 & 2.0306e-05 \\ \hline
$\Delta AIC$ & & 2348 & 2363 & 2374 & 0 & 15 \\ \hline
$\Ro$ & & 5.8579 & 5.9991 & 5.751 & 2.4685 & 2.12 \\ \hline
\end{tabular}} \\
\subcaption{\textnormal{Waning Immunity Data}\label{NNNaive5}}{
    \tiny
   \centering
       \begin{tabular}{ | c | l | c | c | c | c | c | c | }
    \hline
Parameters/AIC & & Exponential & Dose Response & Asymptomatic & Gamma & Waning \\ \hline
$\beta_i$ & & 0.2843 & 0.2843 & 0.2818 & 0.2535 & 0.25 \\ \hline
$\beta_w$ & & 0.0246 & 0.0096 & 0.0329 & 0.5338 & 0.4999 \\ \hline
$\xi$ & & 0.039 & 0.0395 & 0.0418 & 0.0087 & 0.01 \\ \hline
$\alpha$ & & 0.0048 & 0.0048 & 0.0048 & 0.0028 & 0.0027 \\ \hline
$k$ & & 9.64e-06 & 9.495e-06 & 1.0268e-05 & 1.9839e-05 & 1.9997e-05 \\ \hline
$\Delta AIC$ & & 2015 & 2029 & 2036 & 53 & 0 \\ \hline
$\Ro$ & & 1.2353 & 1.2332 & 1.2586 & 3.1486 & 2.9993 \\ \hline
\end{tabular}} \\
\end{table}
\end{center}
\begin{center}

\begin{table}[H]
\centering
\caption{Normal Noise with Informed Starting Parameters	}
\label{NormNPI3}
\subcaption{\textnormal{Exponential Data}\label{NormNInformed1}}{
    \tiny
   \centering
 \begin{tabular}{ | c | l | c | c | c | c | c | c | }
 \hline
Parameters/AIC & & Exponential & Dose Response & Asymptomatic & Gamma & Waning \\ \hline
$\beta_i$ & & 0.2353 & 0.2333 & 0.2366 & 0.2797 & 0.2809 \\ \hline
$\beta_w$ & & 0.6058 & 0.2456 & 0.5649 & 0.4866 & 0.8163 \\ \hline
$\xi$ & & 0.0095 & 0.0097 & 0.01 & 0.0004 & 0.0002 \\ \hline
$\alpha$ & & 0.0027 & 0.0027 & 0.0027 & 0.0013 & 0.0009 \\ \hline
$k$ & & 2.0988e-05 & 2.0851e-05 & 2.0818e-05 & 6.297e-06 & 6.007e-06 \\ \hline
$\Delta AIC$ & & 19 & 0 & 35 & 5766 & 3140 \\ \hline
$\Ro$ & & 1.6652 & 4.5328 & 2.9955 & 1.9128 & 92.5967 \\ \hline
\end{tabular}} \\
\subcaption{\textnormal{Dose Response Data}\label{NormNInformed2}}{
    \tiny
   \centering
       \begin{tabular}{ | c | l | c | c | c | c | c | c | }
    \hline
Parameters/AIC & & Exponential & Dose Response & Asymptomatic & Gamma & Waning \\ \hline
$\beta_i$ & & 0.2643 & 0.2599 & 0.2628 & 0.3225 & 0.3304 \\ \hline
$\beta_w$ & & 1.18 & 0.492 & 1.0925 & 0.328 & 0.4811 \\ \hline
$\xi$ & & 0.0098 & 0.0099 & 0.0108 & 0.0023 & 0.0005 \\ \hline
$\alpha$ & & 0.0027 & 0.0027 & 0.0029 & 0.0023 & 0.0002 \\ \hline
$k$ & & 2.0178e-05 & 2.014e-05 & 2.0241e-05 & 1.1207e-05 & 1.0059e-05 \\ \hline
$\Delta AIC$ & & 13 & 0 & 0 & 20009 & 2408 \\ \hline
$\Ro$ & & 1.2759 & 1.2825 & 1.2786 & 2.9856 & 2.8721 \\ \hline
\end{tabular}} \\
\subcaption{\textnormal{Asymptomatic Data}\label{NormNInformed3}}{
    \tiny
   \centering
    \begin{tabular}{ | c | l | c | c | c | c | c | c | }
    \hline
Parameters/AIC & & Exponential & Dose Response & Asymptomatic & Gamma & Waning \\ \hline
$\beta_i$ & & 0.2888 & 0.2691 & 0.2545 & 0.2809 & 0.2827 \\ \hline
$\beta_w$ & & 0.1276 & 0.3458 & 0.4945 & 0.1974 & 22.871 \\ \hline
$\xi$ & & 0.0044 & 0.0037 & 0.01 & 0.0033 & 0 \\ \hline
$\alpha$ & & 0.0028 & 0.0028 & 0.0046 & 0.0024 & 0.0011 \\ \hline
$k$ & & 9.445e-06 & 1.7675e-05 & 2.0264e-05 & 8.613e-06 & 7.707e-06 \\ \hline
$\Delta AIC$ & & 374 & 356 & 0 & 17276 & 10728 \\ \hline
$\Ro$ & & 3.3636 & 3.388 & 3.2053 & 3.0646 & 4.388 \\ \hline
\end{tabular}} \\
\subcaption{\textnormal{Gamma Data}\label{NormNInformed4}}{
    \tiny
   \centering
    \begin{tabular}{ | c | l | c | c | c | c | c | c | }
    \hline
Parameters/AIC & & Exponential & Dose Response & Asymptomatic & Gamma & Waning \\ \hline
$\beta_i$ & & 0.2832 & 0.2812 & 0.283 & 0.2547 & 0.2517 \\ \hline
$\beta_w$ & & 0.0359 & 0.0158 & 0.0367 & 0.4918 & 0.4665 \\ \hline
$\xi$ & & 0.0413 & 0.044 & 0.0413 & 0.0099 & 0.0111 \\ \hline
$\alpha$ & & 0.0042 & 0.0042 & 0.0042 & 0.0027 & 0.0026 \\ \hline
$k$ & & 1.0557e-05 & 1.0611e-05 & 1.0614e-05 & 2.0103e-05 & 2.0242e-05 \\ \hline
$\Delta AIC$ & & 2742 & 2815 & 2755 & 0 & 45 \\ \hline
$\Ro$ & & 5.7759 & 5.9579 & 5.4204 & 2.6017 & 3.2457 \\ \hline
\end{tabular}} \\
\subcaption{\textnormal{Waning Immunity Data}\label{NormNInformed5}}{
    \tiny
   \centering
       \begin{tabular}{ | c | l | c | c | c | c | c | c | }
    \hline
Parameters/AIC & & Exponential & Dose Response & Asymptomatic & Gamma & Waning \\ \hline
$\beta_i$ & & 0.2859 & 0.2569 & 0.2626 & 0.25 & 0.245 \\ \hline
$\beta_w$ & & 0.0151 & 0.0376 & 0.0786 & 0.5541 & 0.521 \\ \hline
$\xi$ & & 0.0354 & 0.0531 & 0.0516 & 0.0087 & 0.0102 \\ \hline
$\alpha$ & & 0.005 & 0.0044 & 0.0046 & 0.0028 & 0.0027 \\ \hline
$k$ & & 8.886e-06 & 1.3397e-05 & 1.3031e-05 & 2.021e-05 & 2.0479e-05 \\ \hline
$\Delta AIC$ & & 2370 & 2707 & 2550 & 82 & 0 \\ \hline
$\Ro$ & & 1.204 & 1.4036 & 1.3645 & 3.2159 & 3.0635 \\ \hline
\end{tabular}} \\
\end{table}
\end{center}

\begin{center}
\begin{table}[H]
\centering
\caption{Normal Noise with Naive Starting Parameters}
\label{NormNNP3}
\subcaption{\textnormal{Exponential Data}\label{NormNNaive1}}{
    \tiny
   \centering
 \begin{tabular}{ | c | l | c | c | c | c | c | c | }
 \hline
Parameters/AIC & & Exponential & Dose Response & Asymptomatic & Gamma & Waning \\ \hline
$\beta_i$ & & 0.2534 & 0.2533 & 0.2519 & 0.2867 & 0.2882 \\ \hline
$\beta_w$ & & 0.5183 & 0.1628 & 0.5081 & 3.2638 & 1.103 \\ \hline
$\xi$ & & 0.0095 & 0.0122 & 0.0101 & 0.0001 & 0.0002 \\ \hline
$\alpha$ & & 0.0027 & 0.0029 & 0.0029 & 0.0012 & 0.0011 \\ \hline
$k$ & & 2.0344e-05 & 1.9617e-05 & 2.053e-05 & 7.543e-06 & 9.246e-06 \\ \hline
$\Delta AIC$ & & 0 & 290 & 85 & 5447 & 3114 \\ \hline
$\Ro$ & & 4.3526 & 4.1834 & 2.5527 & 1.8362 & 3.8392 \\ \hline
\end{tabular}} \\
\subcaption{\textnormal{Dose Response Data}\label{NormNNaive2}}{
    \tiny
   \centering
       \begin{tabular}{ | c | l | c | c | c | c | c | c | }
    \hline
Parameters/AIC & & Exponential & Dose Response & Asymptomatic & Gamma & Waning \\ \hline
$\beta_i$ & & 0.2584 & 0.2541 & 0.2577 & 0.3192 & 0.3259 \\ \hline
$\beta_w$ & & 1.1856 & 0.4961 & 1.1703 & 0.2877 & 0.5982 \\ \hline
$\xi$ & & 0.0098 & 0.0099 & 0.01 & 0.0024 & 0.0004 \\ \hline
$\alpha$ & & 0.0027 & 0.0027 & 0.0028 & 0.0023 & 0.0003 \\ \hline
$k$ & & 2.0184e-05 & 2.0149e-05 & 2.0215e-05 & 1.0794e-05 & 9.537e-06 \\ \hline
$\Delta AIC$ & & 0 & 29 & 3 & 18824 & 1699 \\ \hline
$\Ro$ & & 1.1697 & 1.2044 & 1.2238 & 2.9509 & 2.9878 \\ \hline
\end{tabular}} \\
\subcaption{\textnormal{Asymptomatic Data}\label{NormNNaive3}}{
    \tiny
   \centering
    \begin{tabular}{ | c | l | c | c | c | c | c | c | }
    \hline
Parameters/AIC & & Exponential & Dose Response & Asymptomatic & Gamma & Waning \\ \hline
$\beta_i$ & & 0.2706 & 0.271 & 0.2572 & 0.2789 & 0.2881 \\ \hline
$\beta_w$ & & 0.8178 & 0.31 & 0.3811 & 0.1802 & 0.6719 \\ \hline
$\xi$ & & 0.0036 & 0.0037 & 0.012 & 0.0038 & 0.0005 \\ \hline
$\alpha$ & & 0.0028 & 0.0028 & 0.0053 & 0.0026 & 0.0002 \\ \hline
$k$ & & 1.7532e-05 & 1.7274e-05 & 2.0013e-05 & 8.728e-06 & 8.285e-06 \\ \hline
$\Delta AIC$ & & 353 & 343 & 0 & 16973 & 612 \\ \hline
$\Ro$ & & 3.0864 & 2.6403 & 3.0393 & 14.199 & 5.5637 \\ \hline
\end{tabular}} \\
\subcaption{\textnormal{Gamma Data}\label{NormNNaive4}}{
    \tiny
   \centering
    \begin{tabular}{ | c | l | c | c | c | c | c | c | }
    \hline
Parameters/AIC & & Exponential & Dose Response & Asymptomatic & Gamma & Waning \\ \hline
$\beta_i$ & & 0.2854 & 0.2811 & 0.2787 & 0.2444 & 0.2413 \\ \hline
$\beta_w$ & & 0.0071 & 0.008 & 0.0273 & 0.4935 & 0.5058 \\ \hline
$\xi$ & & 0.0331 & 0.047 & 0.0487 & 0.0105 & 0.0108 \\ \hline
$\alpha$ & & 0.0047 & 0.0045 & 0.0045 & 0.0028 & 0.0027 \\ \hline
$k$ & & 7.728e-06 & 8.714e-06 & 9.333e-06 & 2.0136e-05 & 2.035e-05 \\ \hline
$\Delta AIC$ & & 2671 & 2582 & 2560 & 33 & 0 \\ \hline
$\Ro$ & & 5.7748 & 5.9765 & 5.7108 & 2.4269 & 3.6956 \\ \hline
\end{tabular}} \\
\subcaption{\textnormal{Waning Immunity Data}\label{NormNNaive5}}{
    \tiny
   \centering
       \begin{tabular}{ | c | l | c | c | c | c | c | c | }
    \hline
Parameters/AIC & & Exponential & Dose Response & Asymptomatic & Gamma & Waning \\ \hline
$\beta_i$ & & 0.2698 & 0.283 & 0.2679 & 0.2455 & 0.2382 \\ \hline
$\beta_w$ & & 0.053 & 0.0066 & 0.0575 & 0.5681 & 0.5225 \\ \hline
$\xi$ & & 0.0511 & 0.0408 & 0.052 & 0.0088 & 0.0109 \\ \hline
$\alpha$ & & 0.0047 & 0.005 & 0.0047 & 0.0028 & 0.0028 \\ \hline
$k$ & & 1.1313e-05 & 8.693e-06 & 1.1565e-05 & 2.024e-05 & 2.0648e-05 \\ \hline
$\Delta AIC$ & & 2337 & 2375 & 2352 & 168 & 0 \\ \hline
$\Ro$ & & 1.291 & 1.1978 & 1.3013 & 3.254 & 3.0422 \\ \hline
\end{tabular}} \\
\end{table}
\end{center}

\begin{center}
\begin{table}[H]
\centering
\caption{Poisson Noise with Informed Starting Parameters}
\label{PNIP3}
\subcaption{\textnormal{Exponential Data}\label{PNInformed1}}{
    \tiny
   \centering
 \begin{tabular}{ | c | l | c | c | c | c | c | c | }
 \hline
Parameters/AIC & & Exponential & Dose Response & Asymptomatic & Gamma & Waning \\ \hline
$\beta_i$ & & 0.2488 & 0.2474 & 0.2489 & 0.2844 & 0.2863 \\ \hline
$\beta_w$ & & 0.5133 & 0.206 & 0.4902 & 1.3858 & 0.2994 \\ \hline
$\xi$ & & 0.0099 & 0.0102 & 0.0104 & 0.0002 & 0.0007 \\ \hline
$\alpha$ & & 0.0027 & 0.0027 & 0.0028 & 0.0012 & 0.0011 \\ \hline
$k$ & & 2.0025e-05 & 1.9843e-05 & 1.9985e-05 & 7.016e-06 & 7.125e-06 \\ \hline
$\Delta AIC$ & & 0 & 1 & 7 & 5374 & 2609 \\ \hline
$\Ro$ & & 4.6086 & 4.6231 & 2.9614 & 1.8825 & 1.3443 \\ \hline
\end{tabular}} \\
\subcaption{\textnormal{Dose Response Data}\label{PNInformed2}}{
    \tiny
   \centering
       \begin{tabular}{ | c | l | c | c | c | c | c | c | }
    \hline
Parameters/AIC & & Exponential & Dose Response & Asymptomatic & Gamma & Waning \\ \hline
$\beta_i$ & & 0.2513 & 0.2464 & 0.2505 & 0.3169 & 0.3237 \\ \hline
$\beta_w$ & & 1.1872 & 0.4991 & 1.1563 & 0.3045 & 0.4892 \\ \hline
$\xi$ & & 0.0101 & 0.0102 & 0.0105 & 0.0021 & 0.0004 \\ \hline
$\alpha$ & & 0.0028 & 0.0028 & 0.0028 & 0.0022 & 0.0003 \\ \hline
$k$ & & 2.0087e-05 & 2.0057e-05 & 2.0122e-05 & 1.0264e-05 & 9.359e-06 \\ \hline
$\Delta AIC$ & & 11 & 0 & 11 & 20616 & 2665 \\ \hline
$\Ro$ & & 1.2258 & 1.2116 & 1.2007 & 3.0071 & 3.0295 \\ \hline
\end{tabular}} \\
\subcaption{\textnormal{Asymptomatic Data}\label{PNInformed3}}{
    \tiny
   \centering
    \begin{tabular}{ | c | l | c | c | c | c | c | c | }
    \hline
Parameters/AIC & & Exponential & Dose Response & Asymptomatic & Gamma & Waning \\ \hline
$\beta_i$ & & 0.266 & 0.2656 & 0.2512 & 0.2789 & 0.2836 \\ \hline
$\beta_w$ & & 0.8864 & 0.3562 & 0.4893 & 0.1918 & 0.0525 \\ \hline
$\xi$ & & 0.0037 & 0.0037 & 0.0102 & 0.0032 & 0.0064 \\ \hline
$\alpha$ & & 0.0028 & 0.0028 & 0.0047 & 0.0024 & 0.0027 \\ \hline
$k$ & & 1.7647e-05 & 1.7562e-05 & 2.0051e-05 & 8.117e-06 & 7.422e-06 \\ \hline
$\Delta AIC$ & & 394 & 378 & 0 & 18262 & 10941 \\ \hline
$\Ro$ & & 3.0476 & 3.0491 & 2.9556 & 6.6794 & 2.3422 \\ \hline
\end{tabular}} \\
\subcaption{\textnormal{Gamma Data}\label{PNInformed4}}{
    \tiny
   \centering
    \begin{tabular}{ | c | l | c | c | c | c | c | c | }
    \hline
Parameters/AIC & & Exponential & Dose Response & Asymptomatic & Gamma & Waning \\ \hline
$\beta_i$ & & 0.2825 & 0.2843 & 0.2858 & 0.249 & 0.2484 \\ \hline
$\beta_w$ & & 0.024 & 0.0075 & 0.0144 & 0.5029 & 0.5091 \\ \hline
$\xi$ & & 0.0432 & 0.0405 & 0.0367 & 0.01 & 0.01 \\ \hline
$\alpha$ & & 0.0044 & 0.0045 & 0.0045 & 0.0027 & 0.0026 \\ \hline
$k$ & & 9.236e-06 & 8.761e-06 & 8.503e-06 & 2.0042e-05 & 2.0147e-05 \\ \hline
$\Delta AIC$ & & 2170 & 2169 & 2164 & 0 & 2 \\ \hline
$\Ro$ & & 5.7528 & 5.9758 & 5.626 & 2.4851 & 3.2509 \\ \hline
\end{tabular}} \\
\subcaption{\textnormal{Waning Immunity Data}\label{PNInformed5}}{
    \tiny
   \centering
       \begin{tabular}{ | c | l | c | c | c | c | c | c | }
    \hline
Parameters/AIC & & Exponential & Dose Response & Asymptomatic & Gamma & Waning \\ \hline
$\beta_i$ & & 0.2851 & 0.2853 & 0.2816 & 0.2548 & 0.2518 \\ \hline
$\beta_w$ & & 0.026 & 0.0099 & 0.0359 & 0.5329 & 0.505 \\ \hline
$\xi$ & & 0.0386 & 0.0388 & 0.0429 & 0.0087 & 0.0098 \\ \hline
$\alpha$ & & 0.0048 & 0.0047 & 0.0047 & 0.0028 & 0.0027 \\ \hline
$k$ & & 9.763e-06 & 9.56e-06 & 1.0457e-05 & 1.979e-05 & 1.9935e-05 \\ \hline
$\Delta AIC$ & & 1928 & 1945 & 2050 & 42 & 0 \\ \hline
$\Ro$ & & 1.244 & 1.2401 & 1.2699 & 3.1503 & 3.0263 \\ \hline
\end{tabular}} \\
\end{table}
\end{center}

\begin{center}
\begin{table}[H]
\centering
\caption{Poisson Noise with Naive Starting Parameters}
\label{PNNP3}
\subcaption{\textnormal{Exponential Data}\label{PNNaive1}}{
    \tiny
   \centering
 \begin{tabular}{ | c | l | c | c | c | c | c | c | }
 \hline
Parameters/AIC & & Exponential & Dose Response & Asymptomatic & Gamma & Waning \\ \hline
$\beta_i$ & & 0.2552 & 0.2537 & 0.255 & 0.2859 & 0.2878 \\ \hline
$\beta_w$ & & 0.4554 & 0.1836 & 0.4444 & 0.7224 & 0.8403 \\ \hline
$\xi$ & & 0.0103 & 0.0106 & 0.0107 & 0.0004 & 0.0002 \\ \hline
$\alpha$ & & 0.0028 & 0.0028 & 0.0028 & 0.0013 & 0.0009 \\ \hline
$k$ & & 1.9629e-05 & 1.9456e-05 & 1.9629e-05 & 7.409e-06 & 7.356e-06 \\ \hline
$\Delta AIC$ & & 0 & 9 & 1 & 5977 & 2642 \\ \hline
$\Ro$ & & 4.9052 & 4.8738 & 3.1952 & 1.822 & 3.9554 \\ \hline
\end{tabular}} \\
\subcaption{\textnormal{Dose Response Data}\label{PNNaive2}}{
    \tiny
   \centering
       \begin{tabular}{ | c | l | c | c | c | c | c | c | }
    \hline
Parameters/AIC & & Exponential & Dose Response & Asymptomatic & Gamma & Waning \\ \hline
$\beta_i$ & & 0.2506 & 0.2464 & 0.2469 & 0.3163 & 0.3236 \\ \hline
$\beta_w$ & & 1.211 & 0.4997 & 1.0985 & 0.2978 & 0.1002 \\ \hline
$\xi$ & & 0.0099 & 0.0101 & 0.0115 & 0.0022 & 0.0046 \\ \hline
$\alpha$ & & 0.0028 & 0.0028 & 0.003 & 0.0023 & 0.0023 \\ \hline
$k$ & & 2.0143e-05 & 2.0092e-05 & 2.0322e-05 & 1.0274e-05 & 1.0532e-05 \\ \hline
$\Delta AIC$ & & 17 & 0 & 6 & 21009 & 7592 \\ \hline
$\Ro$ & & 1.2193 & 1.2534 & 1.2365 & 3.0394 & 3.0231 \\ \hline
\end{tabular}} \\
\subcaption{\textnormal{Asymptomatic Data}\label{PNNaive3}}{
    \tiny
   \centering
    \begin{tabular}{ | c | l | c | c | c | c | c | c | }
    \hline
Parameters/AIC & & Exponential & Dose Response & Asymptomatic & Gamma & Waning \\ \hline
$\beta_i$ & & 0.2601 & 0.2598 & 0.2444 & 0.2773 & 0.285 \\ \hline
$\beta_w$ & & 0.9665 & 0.3836 & 0.5545 & 0.1783 & 0.704 \\ \hline
$\xi$ & & 0.0037 & 0.0037 & 0.0096 & 0.0033 & 0.0004 \\ \hline
$\alpha$ & & 0.0028 & 0.0028 & 0.0046 & 0.0024 & 0.0003 \\ \hline
$k$ & & 1.7987e-05 & 1.7883e-05 & 2.0388e-05 & 7.794e-06 & 7.128e-06 \\ \hline
$\Delta AIC$ & & 553 & 533 & 0 & 17627 & 1233 \\ \hline
$\Ro$ & & 2.8416 & 2.8505 & 2.7973 & 4.0324 & 4.5113 \\ \hline
\end{tabular}} \\
\subcaption{\textnormal{Gamma Data}\label{PNNaive4}}{
    \tiny
   \centering
    \begin{tabular}{ | c | l | c | c | c | c | c | c | }
    \hline
Parameters/AIC & & Exponential & Dose Response & Asymptomatic & Gamma & Waning \\ \hline
$\beta_i$ & & 0.2844 & 0.2687 & 0.2819 & 0.2492 & 0.2483 \\ \hline
$\beta_w$ & & 0.0205 & 0.0179 & 0.0273 & 0.5109 & 0.5076 \\ \hline
$\xi$ & & 0.0403 & 0.0647 & 0.0437 & 0.0099 & 0.0101 \\ \hline
$\alpha$ & & 0.0044 & 0.0044 & 0.0044 & 0.0027 & 0.0026 \\ \hline
$k$ & & 9.013e-06 & 9.971e-06 & 9.507e-06 & 2.0071e-05 & 2.0105e-05 \\ \hline
$\Delta AIC$ & & 2263 & 3144 & 2301 & 0 & 6 \\ \hline
$\Ro$ & & 5.8454 & 5.9808 & 5.3806 & 2.4558 & 1.6952 \\ \hline
\end{tabular}} \\
\subcaption{\textnormal{Waning Immunity Data}\label{PNNaive5}}{
    \tiny
   \centering
       \begin{tabular}{ | c | l | c | c | c | c | c | c | }
    \hline
Parameters/AIC & & Exponential & Dose Response & Asymptomatic & Gamma & Waning \\ \hline
$\beta_i$ & & 0.2843 & 0.2864 & 0.283 & 0.2539 & 0.2508 \\ \hline
$\beta_w$ & & 0.0249 & 0.0069 & 0.027 & 0.5327 & 0.5013 \\ \hline
$\xi$ & & 0.039 & 0.0351 & 0.0408 & 0.0087 & 0.0098 \\ \hline
$\alpha$ & & 0.0048 & 0.0048 & 0.0049 & 0.0028 & 0.0027 \\ \hline
$k$ & & 9.613e-06 & 8.931e-06 & 9.775e-06 & 1.9806e-05 & 1.9943e-05 \\ \hline
$\Delta AIC$ & & 2051 & 2062 & 2112 & 42 & 0 \\ \hline
$\Ro$ & & 1.2365 & 1.2149 & 1.2398 & 3.1458 & 3.0077 \\ \hline
\end{tabular}} \\
\end{table}
\end{center}

\end{document}